\documentclass[mnsc,nonblindrev]{informs3-meng}

\OneAndAHalfSpacedXI 

\usepackage{natbib}
 \bibpunct[, ]{(}{)}{,}{a}{}{,}%
 \def\bibfont{\small}%
 \usepackage{tikz}
 \usepackage{booktabs}
\usetikzlibrary{positioning, decorations.pathreplacing}
\usepackage{amsmath} 
\usepackage{bbm}
\usepackage{amssymb}
\usepackage{mathtools}
\usepackage{soul}
\usepackage{dsfont}
\usepackage{color}
\usepackage{bm}
\usepackage{longtable}
\usepackage{threeparttable}
\usepackage{enumitem}
\usepackage[skip=0pt,font=scriptsize]{subcaption}
\usepackage{mathtools, nccmath}
\usepackage[skip=2pt,font=scriptsize]{caption}
\usepackage{fix-cm} 

\newtheorem{examplex}{Example}

\newenvironment{manualexample}[1]{%
  \renewcommand\theexamplex{#1}%
  \examplex
}{\endexamplex}

\usepackage{mwe}
\usepackage{subfig}
\usepackage{subfig}
\usepackage{graphicx}

\newcommand{\myeqmodel}[1]{\begin{equation} { 
\begin{aligned} #1 \end{aligned}}\end{equation}}

\newcommand{\myeq}[1]{$  {#1} $}
\newcommand{\myeql}[1]{\begin{equation}  {#1} \end{equation}}
\newcommand{\myeqln}[1]{\begin{equation*}  {#1} \end{equation*}}

\usepackage{algpseudocode}
\usepackage{enumitem}

\newtheorem{theorem}{Theorem}[section]
\newtheorem{corollary}{Corollary}[theorem] %
\newtheorem{lemma}[theorem]{Lemma} %
\newtheorem{assumption}[theorem]{Assumption}
\newtheorem{proposition}[theorem]{Proposition}

\newtheorem{axiom}{Axiom}
\newtheorem{definition}{Definition}[section]

\DeclareSymbolFont{boperators}   {OT1}{cmr} {bx}{n}
\DeclareSymbolFont{bletters}     {OML}{cmm} {b}{it}
\DeclareSymbolFont{bsymbols}     {OMS}{cmsy}{b}{n}
\DeclareMathSymbol{\BFa}{\mathalpha}{boperators}{`a}
\DeclareMathSymbol{\BFb}{\mathalpha}{boperators}{`b}
\DeclareMathSymbol{\BFc}{\mathalpha}{boperators}{`c}
\DeclareMathSymbol{\BFd}{\mathalpha}{boperators}{`d}
\DeclareMathSymbol{\BFe}{\mathalpha}{boperators}{`e}
\DeclareMathSymbol{\BFf}{\mathalpha}{boperators}{`f}
\DeclareMathSymbol{\BFg}{\mathalpha}{boperators}{`g}
\DeclareMathSymbol{\BFh}{\mathalpha}{boperators}{`h}
\DeclareMathSymbol{\BFi}{\mathalpha}{boperators}{`i}
\DeclareMathSymbol{\BFj}{\mathalpha}{boperators}{`j}
\DeclareMathSymbol{\BFk}{\mathalpha}{boperators}{`k}
\DeclareMathSymbol{\BFl}{\mathalpha}{boperators}{`l}
\DeclareMathSymbol{\BFm}{\mathalpha}{boperators}{`m}
\DeclareMathSymbol{\BFn}{\mathalpha}{boperators}{`n}
\DeclareMathSymbol{\BFo}{\mathalpha}{boperators}{`o}
\DeclareMathSymbol{\BFp}{\mathalpha}{boperators}{`p}
\DeclareMathSymbol{\BFq}{\mathalpha}{boperators}{`q}
\DeclareMathSymbol{\BFr}{\mathalpha}{boperators}{`r}
\DeclareMathSymbol{\BFs}{\mathalpha}{boperators}{`s}
\DeclareMathSymbol{\BFt}{\mathalpha}{boperators}{`t}
\DeclareMathSymbol{\BFu}{\mathalpha}{boperators}{`u}
\DeclareMathSymbol{\BFv}{\mathalpha}{boperators}{`v}
\DeclareMathSymbol{\BFw}{\mathalpha}{boperators}{`w}
\DeclareMathSymbol{\BFx}{\mathalpha}{boperators}{`x}
\DeclareMathSymbol{\BFy}{\mathalpha}{boperators}{`y}
\DeclareMathSymbol{\BFz}{\mathalpha}{boperators}{`z}
\DeclareMathSymbol{\BFA}{\mathalpha}{boperators}{`A}
\DeclareMathSymbol{\BFB}{\mathalpha}{boperators}{`B}
\DeclareMathSymbol{\BFC}{\mathalpha}{boperators}{`C}
\DeclareMathSymbol{\BFD}{\mathalpha}{boperators}{`D}
\DeclareMathSymbol{\BFE}{\mathalpha}{boperators}{`E}
\DeclareMathSymbol{\BFF}{\mathalpha}{boperators}{`F}
\DeclareMathSymbol{\BFG}{\mathalpha}{boperators}{`G}
\DeclareMathSymbol{\BFH}{\mathalpha}{boperators}{`H}
\DeclareMathSymbol{\BFI}{\mathalpha}{boperators}{`I}
\DeclareMathSymbol{\BFJ}{\mathalpha}{boperators}{`J}
\DeclareMathSymbol{\BFK}{\mathalpha}{boperators}{`K}
\DeclareMathSymbol{\BFL}{\mathalpha}{boperators}{`L}
\DeclareMathSymbol{\BFM}{\mathalpha}{boperators}{`M}
\DeclareMathSymbol{\BFN}{\mathalpha}{boperators}{`N}
\DeclareMathSymbol{\BFO}{\mathalpha}{boperators}{`O}
\DeclareMathSymbol{\BFP}{\mathalpha}{boperators}{`P}
\DeclareMathSymbol{\BFQ}{\mathalpha}{boperators}{`Q}
\DeclareMathSymbol{\BFR}{\mathalpha}{boperators}{`R}
\DeclareMathSymbol{\BFS}{\mathalpha}{boperators}{`S}
\DeclareMathSymbol{\BFT}{\mathalpha}{boperators}{`T}
\DeclareMathSymbol{\BFU}{\mathalpha}{boperators}{`U}
\DeclareMathSymbol{\BFV}{\mathalpha}{boperators}{`V}
\DeclareMathSymbol{\BFW}{\mathalpha}{boperators}{`W}
\DeclareMathSymbol{\BFX}{\mathalpha}{boperators}{`X}
\DeclareMathSymbol{\BFY}{\mathalpha}{boperators}{`Y}
\DeclareMathSymbol{\BFZ}{\mathalpha}{boperators}{`Z}
\DeclareMathSymbol{\BFzero}{\mathalpha}{boperators}{`0}
\DeclareMathSymbol{\BFone}{\mathalpha}{boperators}{`1}
\DeclareMathSymbol{\BFtwo}{\mathalpha}{boperators}{`2}
\DeclareMathSymbol{\BFthree}{\mathalpha}{boperators}{`3}
\DeclareMathSymbol{\BFfour}{\mathalpha}{boperators}{`4}
\DeclareMathSymbol{\BFfive}{\mathalpha}{boperators}{`5}
\DeclareMathSymbol{\BFsix}{\mathalpha}{boperators}{`6}
\DeclareMathSymbol{\BFseven}{\mathalpha}{boperators}{`7}
\DeclareMathSymbol{\BFeight}{\mathalpha}{boperators}{`8}
\DeclareMathSymbol{\BFnine}{\mathalpha}{boperators}{`9}
\DeclareMathSymbol{\BFalpha}{\mathord}{bletters}{"0B}
\DeclareMathSymbol{\BFbeta}{\mathord}{bletters}{"0C}
\DeclareMathSymbol{\BFgamma}{\mathord}{bletters}{"0D}
\DeclareMathSymbol{\BFdelta}{\mathord}{bletters}{"0E}
\DeclareMathSymbol{\BFepsilon}{\mathord}{bletters}{"0F}
\DeclareMathSymbol{\BFzeta}{\mathord}{bletters}{"10}
\DeclareMathSymbol{\BFeta}{\mathord}{bletters}{"11}
\DeclareMathSymbol{\BFtheta}{\mathord}{bletters}{"12}
\DeclareMathSymbol{\BFiota}{\mathord}{bletters}{"13}
\DeclareMathSymbol{\BFkappa}{\mathord}{bletters}{"14}
\DeclareMathSymbol{\BFlambda}{\mathord}{bletters}{"15}
\DeclareMathSymbol{\BFmu}{\mathord}{bletters}{"16}
\DeclareMathSymbol{\BFnu}{\mathord}{bletters}{"17}
\DeclareMathSymbol{\BFxi}{\mathord}{bletters}{"18}
\DeclareMathSymbol{\BFpi}{\mathord}{bletters}{"19}
\DeclareMathSymbol{\BFrho}{\mathord}{bletters}{"1A}
\DeclareMathSymbol{\BFsigma}{\mathord}{bletters}{"1B}
\DeclareMathSymbol{\BFtau}{\mathord}{bletters}{"1C}
\DeclareMathSymbol{\BFupsilon}{\mathord}{bletters}{"1D}
\DeclareMathSymbol{\BFphi}{\mathord}{bletters}{"1E}
\DeclareMathSymbol{\BFchi}{\mathord}{bletters}{"1F}
\DeclareMathSymbol{\BFpsi}{\mathord}{bletters}{"20}
\DeclareMathSymbol{\BFomega}{\mathord}{bletters}{"21}
\DeclareMathSymbol{\BFvarepsilon}{\mathord}{bletters}{"22}
\DeclareMathSymbol{\BFvartheta}{\mathord}{bletters}{"23}
\DeclareMathSymbol{\BFvarpi}{\mathord}{bletters}{"24}
\DeclareMathSymbol{\BFvarrho}{\mathord}{bletters}{"25}
\DeclareMathSymbol{\BFvarsigma}{\mathord}{bletters}{"26}
\DeclareMathSymbol{\BFvarphi}{\mathord}{bletters}{"27}
\DeclareMathSymbol{\BFGamma}{\mathalpha}{boperators}{"00}
\DeclareMathSymbol{\BFDelta}{\mathalpha}{boperators}{"01}
\DeclareMathSymbol{\BFTheta}{\mathalpha}{boperators}{"02}
\DeclareMathSymbol{\BFLambda}{\mathalpha}{boperators}{"03}
\DeclareMathSymbol{\BFXi}{\mathalpha}{boperators}{"04}
\DeclareMathSymbol{\BFPi}{\mathalpha}{boperators}{"05}
\DeclareMathSymbol{\BFSigma}{\mathalpha}{boperators}{"06}
\DeclareMathSymbol{\BFUpsilon}{\mathalpha}{boperators}{"07}
\DeclareMathSymbol{\BFPhi}{\mathalpha}{boperators}{"08}
\DeclareMathSymbol{\BFPsi}{\mathalpha}{boperators}{"09}
\DeclareMathSymbol{\BFOmega}{\mathalpha}{boperators}{"0A}
\DeclareMathSymbol{\BFimath}{\mathord}{bletters}{"7B}
\DeclareMathSymbol{\BFjmath}{\mathord}{bletters}{"7C}
\DeclareMathSymbol{\BFell}{\mathord}{bletters}{"60}
\DeclareMathSymbol{\BFwp}{\mathord}{bletters}{"7D}
\DeclareMathSymbol{\BFnabla}{\mathord}{bsymbols}{"72}
\DeclareSymbolFontAlphabet{\BFcal}{bsymbols}
\DeclareMathSymbol{\BFcalA}{\mathalpha}{bsymbols}{`A}
\DeclareMathSymbol{\BFcalB}{\mathalpha}{bsymbols}{`B}
\DeclareMathSymbol{\BFcalC}{\mathalpha}{bsymbols}{`C}
\DeclareMathSymbol{\BFcalD}{\mathalpha}{bsymbols}{`D}
\DeclareMathSymbol{\BFcalE}{\mathalpha}{bsymbols}{`E}
\DeclareMathSymbol{\BFcalF}{\mathalpha}{bsymbols}{`F}
\DeclareMathSymbol{\BFcalG}{\mathalpha}{bsymbols}{`G}
\DeclareMathSymbol{\BFcalH}{\mathalpha}{bsymbols}{`H}
\DeclareMathSymbol{\BFcalI}{\mathalpha}{bsymbols}{`I}
\DeclareMathSymbol{\BFcalJ}{\mathalpha}{bsymbols}{`J}
\DeclareMathSymbol{\BFcalK}{\mathalpha}{bsymbols}{`K}
\DeclareMathSymbol{\BFcalL}{\mathalpha}{bsymbols}{`L}
\DeclareMathSymbol{\BFcalM}{\mathalpha}{bsymbols}{`M}
\DeclareMathSymbol{\BFcalN}{\mathalpha}{bsymbols}{`N}
\DeclareMathSymbol{\BFcalO}{\mathalpha}{bsymbols}{`O}
\DeclareMathSymbol{\BFcalP}{\mathalpha}{bsymbols}{`P}
\DeclareMathSymbol{\BFcalQ}{\mathalpha}{bsymbols}{`Q}
\DeclareMathSymbol{\BFcalR}{\mathalpha}{bsymbols}{`R}
\DeclareMathSymbol{\BFcalS}{\mathalpha}{bsymbols}{`S}
\DeclareMathSymbol{\BFcalT}{\mathalpha}{bsymbols}{`T}
\DeclareMathSymbol{\BFcalU}{\mathalpha}{bsymbols}{`U}
\DeclareMathSymbol{\BFcalV}{\mathalpha}{bsymbols}{`V}
\DeclareMathSymbol{\BFcalW}{\mathalpha}{bsymbols}{`W}
\DeclareMathSymbol{\BFcalX}{\mathalpha}{bsymbols}{`X}
\DeclareMathSymbol{\BFcalY}{\mathalpha}{bsymbols}{`Y}
\DeclareMathSymbol{\BFcalZ}{\mathalpha}{bsymbols}{`Z}

\newcommand{\Tau}{\mathcal{T}}

\newcommand{\bmvec}{\BFm_{[2:K]}}

\usepackage[linesnumbered,ruled,vlined]{algorithm2e} %

\usepackage{algpseudocode}
\usepackage{graphicx}
\usepackage{caption}
\usepackage{subcaption}

\ECRepeatTheorems

\EquationsNumberedThrough    

\MANUSCRIPTNO{}

\begin{document}


\RUNAUTHOR{Qi, Zhu}

\RUNTITLE{Pitfalls of Shapley Values in Collaborative Federated Learning}

\TITLE{Mechanism for Collaborative Federated Learning: Pitfalls of Shapley Values}

\ARTICLEAUTHORS{%
\AUTHOR{Meng Qi}
\AFF{SC Johnson College of Business, Cornell University, Ithaca, NY, 14850, \\ \EMAIL{mq56@cornell.edu}}
\AUTHOR{Mingxi Zhu}
\AFF{Scheller College of Business, Georgia Institute of Technology\\ \EMAIL{mzhu359@gatech.edu}}
} 

\ABSTRACT{
This paper investigates the impact of mechanism design on collaborative learning systems enabled by federated learning (FL). We propose a multi-action collaborative federated learning (MCFL) framework, capturing the interplay between agent strategies, platform mechanisms, and FL algorithms—a ``three-body problem" in collaborative learning. This work 
demonstrates 
how the convergence rate and computational efficiency of FL are endogenously determined by the agent participation equilibrium that is induced by the mechanism. By doing so, we establish a direct link between incentive design in collaborative learning systems and the performance of the underlying optimization algorithms, a connection that has been largely overlooked in the existing literature.

Specifically, we characterize the equilibrium of agent participation under two prominent mechanisms: the Shapley Value (SV) and Marginal Contribution (MC) mechanisms. Although SV is fair in surplus allocation and budget balanced,  it has a vital pitfall: agents are incentivized to split their data across newly created fake identities. This is critical especially in the MCFL setting as it leads to slow convergence of FL optimization, which increases the number of required synchronization/communication rounds even when the per-round cost is fixed. In contrast, while MC is not budget-balanced, it is robust to such strategic manipulation and is able to induce an equilibrium that maximizes the MCFL system efficiency.

Overall, our study lays a foundation for jointly designing incentives and algorithms in MCFL systems. We provide insights on pitfalls of SV: it induces a system equilibrium that leads to tremendous training cost and slower convergence, ultimately undermining the effectiveness of collaborative learning.
}%




\maketitle


%


\section{Introduction}
\label{sec:intro}
In the digital economy, companies have data assets, often leading to data silos where information is isolated and stored locally. In the meantime, advances in large-scale machine learning models highlight the growing need for collaboration in developing machine learning methods that enable individuals and organizations to contribute their own data for mutual benefit. However, there are often concerns of privacy and regulation that are against directly sharing raw data. In such contexts, federated learning (FL) offers a promising solution to facilitate collaborative learning, allowing multiple agents to collaboratively train a machine learning model without directly sharing raw data \citep{mcmahan2017communication}. We model such emerging collaborative learning ecosystem as a system consisting of one central platform and multiple federated learning agents with local data. The following figure illustrates the collaborative learning ecosystem.
\begin{figure}[htb!]
\centering
\includegraphics[width=0.9\textwidth]{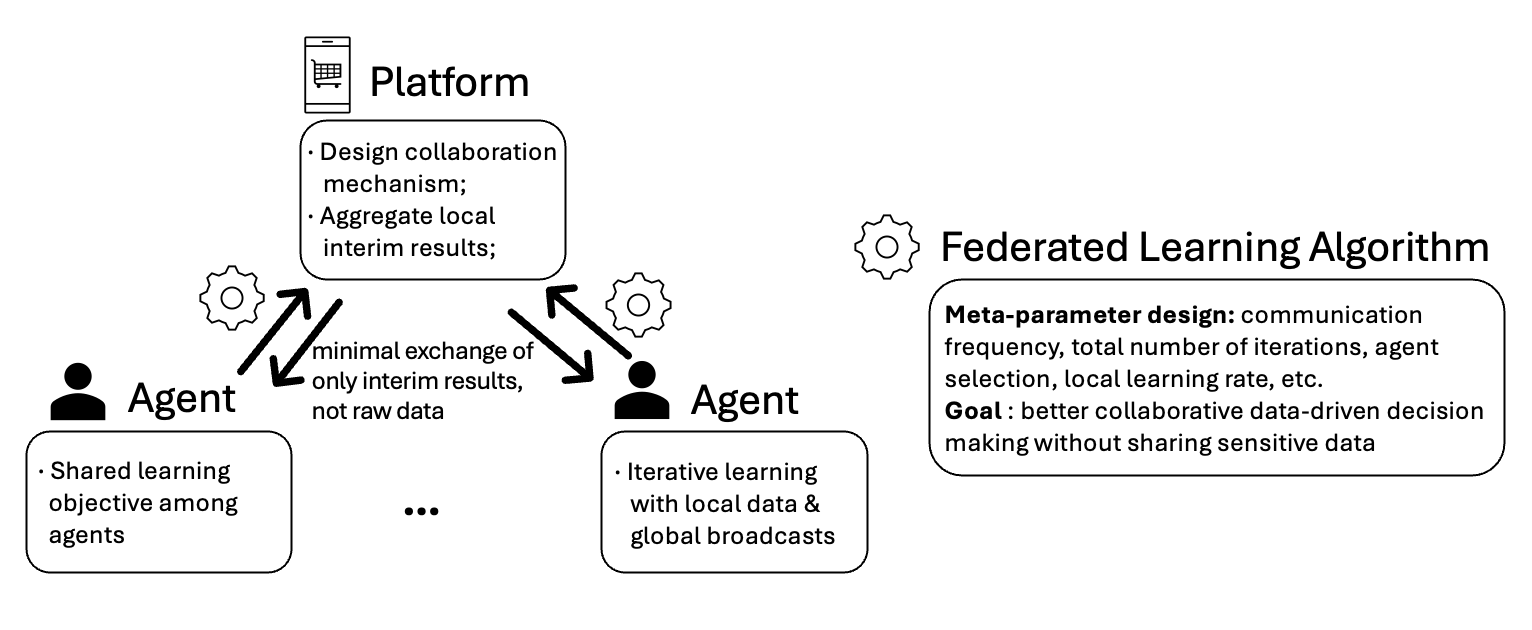}
\caption{Illustration of the Collaborative Learning Ecosystem}
\end{figure}
The role of the central platform is not to pool data or centralize learning tasks, but to provide FL infrastructure and coordinate collaboration among agents. In particular, the platform uses mechanism design as a lever to incentivize the participation of the agents. Agents, who share a common learning objective, but each with their own local data, will strategically decide their level of participation depending on the platform-announced mechanism.%
Such strategic behaviors may include contributing only a subset of their data or creating fake identities and splitting datasets to participate using multiple (fake) identities. Although such behaviors are often not considered explicitly malicious or attacking, as splitting datasets does not change the aggregate set of training samples available to the platform, yet we will show later in this work that they may lead to significant inefficiencies and degrade the performance of the FL algorithm in the system. 

Notably, the issue of potential data splitting across fake identities has become increasingly salient in the emerging multi-agent AI training and data marketplace industry, where participation is often account-based and remote. Business Insider \citep{GoelRolletMann2025ShadowMarket} reports that, in collaborative AI training and data marketplaces, platforms have identified cases where multiple data-provider accounts were, in fact, controlled by the same underlying individual. Moreover, a single entity can split its participation across multiple identities through partitioning its contributions across accounts, to potentially increase the total payment received. 
Reflecting this concern, major data/ML contribution platforms explicitly prohibit multi-account participation. For example, the Kaggle competition rules \citep{KaggleDigitRecognizerRules} state ``one account per participant" and disallow submissions from multiple accounts. However, rule-based prohibitions alone can be difficult to enforce reliably in large-scale distributed settings without managing agents' incentives. Therefore, the platform must design a robust mechanism that encourages the full participation of agents while mitigating the impact of strategic behavior.

In this work, we model this collaborative learning system, highlighting the intricate interplay between mechanism design, agent participation equilibrium, and the performance of FL algorithms. Specifically, we explore the following questions.
\textit{(1) Given an FL learning outcome, how can we design a mechanism that fairly allocates the associated learning surplus to agents? (2) What equilibrium of agent participation will this mechanism induce? (3) How does this induced equilibrium affect the convergence of the FL algorithm and the system efficiency?}

This complex interaction between the three parties (mechanism, agent participation, and learning efficiency) has been overlooked in existing studies. Some prior research focuses mainly on encouraging local agent participation, yet often neglects its impact on algorithm performance. Other works instead emphasize improving FL algorithm efficiency and reducing communication costs, but mostly disregard the crucial role of mechanism design in improving algorithm performance. 

In Appendix \ref{app:example}, we present two motivating examples in supply-chain and portfolio management, which serve as running examples throughout the paper. In Section \ref{subsec:mcfl-real}, we also extend to a real-world healthcare application, the MELLODDY project \citep{oldenhof2023industry,heyndrickx2023melloddy}, a cross-company federated learning platform where large pharmaceutical firms jointly train predictive models for drug discovery without sharing proprietary data, directly matching our collaborative learning setting.

\subsection{Main Contributions and Outline}
Our study proposes a framework that balances efficiency, fairness, and strategic behavior in collaborative
learning systems, offering insights on mechanism design and the efficiency of FL algorithms in collaborative learning systems. Specifically, this paper introduces the multi-action collaborative federated learning (MCFL) framework, modeling the interplay between agents, platforms, and learning algorithms to address the ``three-body problem" in collaborative learning. Since the agent participation in MCFL system can be modeled as a cooperative game, we focus on two mechanisms: Shapley values (SV) and marginal contribution (MC), and extend the traditional Shapley models by considering a multi-action agent strategy space, where agents not only make binary decisions of whether or not to participate, but also decide the quantity of data they would like to contribute. %
To our knowledge, this is among the first works to apply a multi-action (multi-choice) Shapley framework to collaborative federated learning, where agents can contribute at multiple data-contribution levels rather than make only binary participation decisions. Although SV is fair in surplus allocation and budget balanced, it has a vital pitfall: agents are incentivized to split their data across newly created fake identities. It is especially vital in the MCFL setting as it leads to slow convergence of FL optimization with higher number of synchronizations and communication cost. It is worth pointing out that, this pitfall of SV is technically not trivial to demonstrate. Unlike the standard binary-action Shapley value, the MCFL Shapley value requires comparing payoff differences over a combinatorially complex space of participation profiles. To handle this, we develop new proof techniques that decompose the MCFL Shapley value into a weighted sum of incremental learning surpluses, and then establish positivity through a Beta function integral representation together with a stochastic dominance argument over the profile distribution. This work is the first to connect the combinatorial nature of multi-action Shapley value with Beta function and the distribution over participation profiles. We would also like to emphasize that agents' participation equilibrium affects learning efficiency not simply through the amount of data contributed or number of agents to communicate with. In fact, it has a more profound and complicated effect on the convergence rate of the federated learning algorithm performed in the system, bridging mechanism design and algorithm performance in collaborative learning systems. To connect mechanism design to FL optimization (Theorem~\ref{thm:convergence}), it requires deriving explicit convergence bounds that reveal a polynomial gap in synchronization cost---$\mathcal{O}((T|\mathbf{m}|)^{3/4})$ under SV versus $\mathcal{O}((TK)^{3/4})$  under MC --- arising purely from the mechanism's equilibrium effect, providing the first formal quantitative bridge between incentive design and federated learning efficiency.

The remainder of the paper is organized as follows. In Section \ref{sec: model}, we introduce a multi-action collaborative federated learning (MCFL) framework with strategic agents, modeling the agent–platform–algorithm triad. Agents choose whether, how much, and how to participate, including the option to split data. In Section \ref{sec: mechanism}, we study surplus allocation via Shapley Value (SV) \citep{shapley1953value} and Marginal Contribution (MC), extend SV to multi-action agents, characterize participation equilibria under data splitting, and show that SV is not robust to splitting data in equilibrium. In Section \ref{sec: algorithm}, we define the efficiency of the MCFL system by jointly considering the surplus of the coalition generated from learning outcome and the training cost which largely depends on the number of synchronization rounds \citep{Kairouz2021}, and show how identity splitting directly reduces system efficiency, revealing a pitfall of SV. Section \ref{sec: numerical} provides a numerical demonstration of the pitfall of SV under newsvendor problem, portfolio optimization, and a real-world healthcare image-classification application. Section \ref{sec: extensions} discusses extensions of the MCFL framework, including alternative platform objectives, comparison to traditional anonymity-proof SV \citep{ohta2008anonymity}, and mechanisms beyond SV and MC.

\section{Literature Review}
\label{ref: literature}

We study a collaborative learning ecosystem in which multiple agents train a shared machine learning model under the orchestration of a platform. A closely related empirical setting is documented by \citet{chen2024role}, who study Taobao platform and examine how data and machine learning analytics sharing across marketplace retailers, orchestrated by the Taobao Platform, benefits the market participants. Using a quasi-experimental design with a synthetic difference-in-differences approach, they find that data sharing increases retailers' sales revenue by about 30\%. Importantly, \citet{chen2024role} also highlights privacy concerns in data/model sharing and addresses them by sharing analytics outputs (rather than disclosing identifiable raw data), so that the released results cannot be used to identify any individual customer or firm. Privacy concerns have long posed a central challenge in collaborative learning and data-sharing ecosystems \citep{goldfarb2012shifts,menon2016privacy,martin2020ethical}. This challenge is further amplified by current data regulations, such as the EU General Data Protection Regulation (GDPR) \citep{gdpr2016} and the California Consumer Privacy Act (CCPA) \citep{ccpa2018}, which impose stringent constraints on collecting, processing, and sharing personal data, making direct raw-data sharing infeasible in many settings. In our problem, we mitigate the privacy challenge through the emerging technology of federated learning (FL). We study how mechanism design of FL platform shapes the collaborative learning system outputs, including both the multi-agent equilibrium participation decisions and FL learning algorithm efficiency. Our paper, hence, is directly related to the following literature streams: strategic multi-agent interaction in business environments; mechanism design and optimization in platforms; federated learning (FL) and FL incentives.

\paragraph{\textbf{Cooperative Games and Shapley Value.}} Our paper is closely related to the literature that studies collaborations among multiple agents in business operations. In our setting, we model this collaboration as a cooperative game that studies the coalitions formed by the players and their cooperative actions \citep{branzei2008models}. A canonical allocation rule in this framework is the Shapley value \citep{shapley1953value}, which allocates the rewards based on the marginal contribution of each player in the coalitions, and is widely adopted as it ensures fairness in the allocation of surpluses. Our paper differs from the above cooperative-game formulations in that we study a data-sharing collaborative game. The classical Shapley value is defined for binary participation decisions. However, in our setting, an agent's decision is richer: beyond whether to participate in a coalition, each agent also chooses how to participate (potentially with multiple identities) and how much data to share.
\citet{hsiao1993shapley} extends the concept to \emph{multi-choice} (multi-action/effort-level) cooperative games. Our proposed MCFL Shapley value can be viewed as an application of the multi-choice Shapley value proposed in \citet{hsiao1993shapley} in the collaborative learning context.
Beyond collaborative learning with operational decisions, cooperative games have been adopted in other contexts in operations management, for example, \citet{he2012polymatroid} studies cooperative joint replenishment games through a polymatroid optimization lens, linking equilibrium allocations to submodularity and separable concave objectives. Moreover, a complementary stream studies agents’ non-cooperative interactions through Nash equilibrium \citep{wang2009epidemic, he2017noncooperative,kwon2022common}, which is beyond our cooperative-game focus. This work differs from the above cooperative and noncooperative settings for operations because we take a unique perspective on the Shapley value. Although the Shapley value is well-known as fair and budget-balanced as an allocation rule, we identify a vital pitfall of the Shapley value in the decentralized collaborative learning context, as it may reduce the overall efficiency of the federated learning ecosystem by inducing agent mis-behaviors that deteriorate the convergence of the underlying FL training algorithm. 

\paragraph{\textbf{Shapley Value in Platform Mechanism Design.}} Our work is closely related to mechanism design and optimization in platforms. Despite its computational complexity, the Shapley value is still widely adopted as a platform mechanism to incentivize collaboration among agents. Specifically, in supply chain management, \cite{leng2009allocation} explores the Shapley value to distribute the surplus generated from the information on shared demand among a manufacturer, a distributor and a retailer. \cite{kemahliouglu2011centralizing} applies the Shapley mechanism for allocating surplus generated from inventory pooling among retailers. \cite{gopalakrishnan2021incentives} use the Shapley value to allocate carbon emission responsibilities among firms in a supply chain. Beyond supply chain management, \cite{anily2010cooperation} employs the Shapley mechanism to divide the surplus from pooling service capacities in service systems.  \cite{singal2019shapley} tackles the challenge of allocating eventual conversion among online advertisers  through a modified counterfactual adjusted Shapley value. \cite{bergantinos2020sharing} considers the equal-split rule, aligned with the Shapley values in this specific context of splitting revenues from broadcasting sports events. \cite{leng2021multiplayer} investigates a game class with diminishing marginal contributions and analyzes the properties of the Shapley value mechanism. \cite{gopalakrishnan2023cooperative} considers a Shapley mechanism variant for firm security cost-sharing arrangements. %

\paragraph{\textbf{Platform Mechanism Design and Optimization Beyond Shapley Value.}}
Beyond the Shapley value, the literature has studied alternative mechanisms and related optimization problems for surplus/cost allocation in operations.
For example, \citet{aswani2019data} proposes a data-driven incentive design in a principal--agent setting (e.g. Medicare shared savings), where the mechanism is learned from data rather than derived solely from primitives; \citet{balseiro2019multiagent} studies multi-agent mechanism design without monetary transfers for dynamic resource allocation;  \citet{chen2024managing} designs a demand mechanism to incentivize multiple newsvendors to make platform-favorable inventory decisions;  \citet{chen2024incentivizing} analyzes the incentives to participate in decentralized dynamic matching markets, and \citet{chen2025incentivizing} studies mechanisms that incentivize resource pooling between self-interested agents in service systems. 
In contrast to the platform mechanism design literature, which focuses on how to use Shapley value for surplus allocation, we go one step beyond this and discuss how the mechanism shapes both the agent equilibrium and machine learning efficiency. We also investigate agents' dishonest behavior that is unique in the MCFL setting of creating duplicated identities and data splitting.
Lastly, while much of the existing literature studies how to use optimization to solve
mechanism design problems (see, e.g., \cite{chan2025inverse} for using inverse optimization in platform mechanism design as an example), in this work, we instead focus on the reversed process: how the mechanism itself impacts optimization, particularly the convergence of gradient-based algorithms while solving for the best machine learning outcome. The mechanism shapes agents' participation equilibrium in the collaborative learning setting, which in turn affects the convergence and efficiency of the learning outcome. 

\paragraph{\textbf{Federated Learning and its Communication Cost.}}
Federated learning (FL) trains models across decentralized agents without sharing raw data \cite{mcmahan2017communication,kairouz2021advances}. In the FL setting, the canonical algorithm is FedAvg (local SGD) \citep{stich2018local, yu2019parallel, khaled2019first}, rooted in parallel SGD \cite{Zinkevich2010}. Because FL costs are often dominated by communication \cite{kairouz2021advances}, previous work studies reducing communication cost through faster convergence algorithms \cite{shamir2014communication, Yuan2020} or bandwidth compression \cite{konevcny2016federated,chraibi2019distributed, hamer2020fedboost}, typically taking the number of participating agents as exogenous. A closely related work, \citep{qu2023unified}, characterizes how FedAvg convergence scales with the number of agents. To our knowledge, our work is the first that considers the impact of mechanism design on the equilibrium number of participating agents, which in turn impacts the communication cost of FL algorithms. %

\paragraph{\textbf{Mechanism Design and Incentive Alignment in FL.}}
\cite{zhan2021survey} and \cite{zeng2021comprehensive} provide surveys of recent work on incentives and mechanism design in FL. \cite{donahue2021optimality} models FL as a game among agents who can share updates, characterizing stable and socially efficient collaboration outcomes and how heterogeneity affects coalition formation and stability. \cite{donahue2021model} formalizes ``model-sharing games" with voluntary participation, analyzing equilibrium participation/sharing behavior, and the welfare loss from strategic under-contribution in FL. \cite{karimireddy2022mechanisms} considers a model with partial participation due to the cost of data provision, and constructs a mechanism to incentivize collaborative learning. \cite{zhang2022data} analyzes the incentive mechanism design while considering partial participation and the cost of computation for FL algorithms. \cite{haghtalab2025platforms} studies a platform-design problem for incentive-aware collaboration, proposing mechanisms that induce efficient participation and information sharing when agents strategically decide whether and how much to collaborate. For non-cooperative agents, \cite{gafni2022long} considers an FL platform with non-collaborative agents and investigates how to manage the conflicting incentives, where we focus on a collaborative environment. \cite{bi2024understanding}, on the other hand, model federated learning partnership formation and repeated data contributions as a dynamic noncooperative game, characterizing when collaboration can be sustained over time.  
This work is distinguished from existing literature on mechanism and incentive alignment in FL. First, existing work typically posits exogenous functional forms for learning benefits and FL costs, whereas we endogenize FL cost through equilibrium participation induced by the mechanism. 
Second, to the best of our knowledge, this is the first work to study dishonest behavior of the agents in Shapley-value-based mechanism and how it impacts the performance of federated learning algorithms.

It is worth noting that a separate line of work uses the Shapley value primarily as a contribution-evaluation and explainability tool in federated learning and collaborative learning to quantify the value of the data \citep{ghorbani2019data,jia2019towards,sim2020collaborative,rozemberczki2022shapley,ai2024instrumental}. This stream differs from our setting: in these studies, the Shapley value is not employed as an incentive mechanism, but rather as a data-valuation metric. In contrast, we study a platform mechanism design problem with decision-aware agents, focusing on equilibrium participation and the efficiency of the resulting collaborative learning process.

\section{The Multi-Action Collaborative Federated Learning (MCFL) Framework}
\label{sec: model}
In this section, we describe the ecosystem of collaborative federated learning. The system consists of (i) multiple agents who share a common learning objective and aim to make better informed decisions by participating in the collaborative learning; (ii) a digital platform that is the coordinator that provides an infrastructure that enables cross-agent learning and seeks to design a mechanism that leads to the most efficient outcome; (iii) a federated learning algorithm that performs the collaborative decentralized learning. 
In this section, we introduce the three factors of the MCFL system and discuss the complicated interplay between the three factors.

\subsection{Agent}
In this part, we introduce the agent's possessed dataset, incentive to collaborate, goal, and possible actions. We assume that there are potentially $K$ agents trying to form a coalition for FL. 
Each agent holds its own proprietary data samples, possibly with a different number of samples. The vector $\mathbf{m}=[m_1,\dots,m_K] \in \mathbb{R}^K$ denotes the amount of data each agent possesses, with $\BFm=\sum^{K}_{1}m_k$ denoting the total data samples of all agents. We also let $S_k:= \{\mathbf{z}_j := (\mathbf{x}_j, y_j), \quad \forall j = 1, \dots, m_k \}$ denote the data samples possessed by agent $k$, with each observation $\mathbf{z}_j := (\mathbf{x}_j, y_j)\in \mathbb{R}^{p+1}$ representing a data sample. All data samples are generated independently from an unknown ground-truth distribution $\mathbb{P}_{\BFz}$. We consider a family of hypothesis classes $\mathcal{H}_\BFtheta := \{h_\BFtheta | \BFtheta \in \Theta\}$ where $h_\BFtheta : \mathbf{x} \rightarrow y$, and the true hypothesis parameter $\BFtheta^\ast$ is unknown. Here we assume the model-specified case, where $\BFtheta^\ast \in \Theta $.  In the following two examples, we provide the data set that agents may possess in a specific business context.

The incentive of the agents for collaborative learning is aligned with a common objective: to estimate the true unknown parameter $\boldsymbol{\theta}^*$ and maximize the associated performance measure $\pi(\hat{\boldsymbol{\theta}}, \boldsymbol{\theta}^*)$. This function $\pi$ could capture either estimation accuracy or the value of a downstream decision based on $\hat{\boldsymbol{\theta}}$. We assume that the function $\pi(\hat{\boldsymbol{\theta}}, \boldsymbol{\theta}^*)$ satisfies the following assumption. We also refer the reader to the Appendix~\ref{app:pi} for a detailed justification that the stated conditions are satisfied by a broad class of data-driven decision-making problems.

\begin{assumption}
\label{assump:pi}
$\pi(\hat{\boldsymbol{\theta}}, \boldsymbol{\theta}^*)$ is $L_{\pi}$-Lipschitz continuous in $\hat{\boldsymbol{\theta}}$ for any $\BFtheta^*$, and there exists a unique maximizer to $\max_{\hat{\boldsymbol{\theta}}} \ \pi(\hat{\boldsymbol{\theta}}, \boldsymbol{\theta}^*)$, with $\arg\max_{\hat{\boldsymbol{\theta}}} \ \pi(\hat{\boldsymbol{\theta}}, \boldsymbol{\theta}^*):=\BFtheta^*$.
\end{assumption}

While the incentive to collaborate is to jointly maximize the common objective $\pi$, an individual agent's goal within the platform is to maximize the surplus it receives under the platform-specified mechanism. This mechanism, announced by the platform, evaluates each agent's contributed datasets and redistributes surplus accordingly. Its formal specification is provided in Section~\ref{sec: mechanism}. Thus, although the incentive for collaboration is globally aligned, each agent is still strategic, aiming to maximize its received allocation as a function of its actions.

The agent's possible actions include partial data provision and data splitting across duplicated fake identities. To be more specific, for an agent $k$ with data samples  $S_k$, let $s_k \subseteq S_k$ represent a subset of $S_k$, and let $\mathcal{P}(s_k)$ denote the set of all possible partitions of $s_k$ with $p(s_k)\in \mathcal{P}(s_k)$ being one specific partition of $s_k$. The strategy space of agent $k$ is then defined as: $\{\mathcal{P}(s_k)\ \mid  \forall \  s_k \subseteq S_k\}$, which encompasses all possible ways an agent can contribute to the coalition, potentially by splitting their data across multiple identities.  All agents, including possible fake identities, that decide to participate form a coalition $\mathcal{A}$. Note that the total number of (split) agents contained in the coalition $\mathcal{A}$ is denoted by $|\mathcal{A}| := \sum_{k=1}^K |p(s_k)|$ where $|p(s_k)|$ denotes the number of partitions for $p(s_k)$. $|\mathcal{A}|$ characterizes the total number of identities, including the fake ones, in the system. Then we let the participation decision profile $\BFtau_\mathcal{A} \in \mathbb{R}^{|\mathcal{A}|}$ denote the participation decision for agents in the coalition $\mathcal{A}$. It has $|\mathcal{A}|$ elements, each of which denotes the contributed volume of data (number of data samples) from each of the (could be fake) $|\mathcal{A}|$ identities. Specifically, we let $\tau^k_{\mathcal{A}}$ denote the number of data samples provided by agent $k$ in coalition $\mathcal{A}$. Then $|\BFtau_\mathcal{A}|$ denotes the total sample size of the data within the coalition. 

This misconduct behavior of splitting data samples across duplicated fake identities is related to, but distinct from, the well-known false-name manipulation and the setting of anonymity-proof Shapley Values \citep{iwasaki2010worst, conitzer2010using, aziz2011false}. A detailed comparison and further explanation can be found in Section \ref{sec: compare_falsename}. In addition, we would like to clarify that we do not consider modification or pollution of data provided for the collaborative learning. We also leave attacking behaviors, which involve disrupting the system, damaging learning outcomes, or inferring sensitive data, outside the scope of our study. Examples include data poisoning, Byzantine attacks, Sybil attacks, and inference attacks. Attacking behaviors have been extensively studied in previous research. Our work instead focuses on investigating the impact of misconduct that is not attacking but only abusing rules of the well-established mechanisms that have been demonstrated to be effective in cooperative games. The main insight from our result is that the presence of potential misconduct of data-splitting will dramatically harm the collaborative federated learning system.

\subsection{Platform}

In MCFL Framework, we consider an altruistic platform that has two objectives: maximize the aggregate performance measure of participating agents; and fairly redistribute such performance measure. We will discuss the platform with other objectives in Section \ref{sec: extensions_social_surplus}. The platform faces two core challenges: (i) agents must be willing to join despite outcome uncertainty; (ii) the platform must commit to a redistribution rule before accessing coalition data, which requires an ex-ante valuation of the coalition surplus.

The platform can resolve these challenges by committing to a high-probability performance guarantee that brings each coalition’s performance measure $\pi(\hat{\boldsymbol{\theta}}_{\boldsymbol{\tau}_{\mathcal{A}}}, \boldsymbol{\theta}^*)$ close to the oracle value \(v^*:=\pi(\boldsymbol{\theta}^*,\boldsymbol{\theta}^*)\). Specifically, let \(\epsilon:\mathbb{N}\times(0,1)\to\mathbb{R}_{\ge 0}\) be an error function and \(\delta\in(0,1)\) a pre-specified confidence level. Then for any coalition \(\mathcal A\) with total sample size \(|\boldsymbol{\tau}_{\mathcal A}|\),
\begin{equation}\label{eq:prob_bound_revenue}
\mathbb{P}\left( \pi(\hat{\boldsymbol{\theta}}_{\boldsymbol{\tau}_{\mathcal{A}}}, \boldsymbol{\theta}^*) \geq v^*-\epsilon(|\boldsymbol{\tau}_{\mathcal{A}}|, \delta) \right) \geq 1-\delta.
\end{equation}

We define the class of admissible guarantees as
\(\mathbb{G} := \left\{  (\epsilon, \delta) \;\middle|\; \epsilon: \mathbb{N} \times (0,1) \to \mathbb{R}_{\geq 0},\; \delta \in (0,1) \right\},\)
and say that a performance guarantee $\mathcal{G} \in \mathbb{G}$ is \emph{satisfied} if \eqref{eq:prob_bound_revenue} holds for all coalitions $\mathcal{A}$. In this work, we focus on a specific functional form for the error bound, given by
\(\epsilon(|\boldsymbol{\tau}_{\mathcal{A}}|, \delta) := L_\pi \sqrt{ \frac{1}{C_1 |\boldsymbol{\tau}_{\mathcal{A}}|} \log\left( \frac{C_2}{\delta} \right) },\)
which is commonly adopted in data-driven optimization analysis. This bound captures two desirable statistical properties: the width of the confidence interval shrinks at a rate of \( O(1/\sqrt{n}) \) as the number of samples increases, and the probability of large deviations decays exponentially in the number of samples. We refer the reader to Appendix~\ref{app:pg} for a detailed justification of this probabilistic bound.

This bound allows the platform to offer a guaranteed value that incentivizes participation by ensuring a small, quantifiable gap to the oracle surplus ex-ante, 
\begin{equation}\label{eq:def_vtaua}
v(|\boldsymbol{\tau}_{\mathcal{A}}|) := v^* - \varepsilon(|\boldsymbol{\tau}_{\mathcal{A}}|, \delta_0).
\end{equation}

The guaranteed value function $v:\mathbb{N}\to \mathbb{R}$ satisfies:
(i) \emph{Monotonicity}: $v(\cdot)$ is non-decreasing;
(ii) \emph{Diminishing returns}: $v(\cdot)$ is concave (the guaranteed gain per additional sample shrinks);
(iii) \emph{Dependence only on the data volume}: for any coalition $\mathcal A$, $v(|\boldsymbol{\tau}_{\mathcal A}|)$ depends only on the total sample size;
(iv) \emph{Consistency}: $\lim_{|\boldsymbol{\tau}_{\mathcal{A}}|\to\infty}v(|\boldsymbol{\tau}_{\mathcal{A}}|)=v^*$, $v(|\boldsymbol{\tau}_{\mathcal{A}}|)$ converges from below to $v^*$ as $|\boldsymbol{\tau}_{\mathcal{A}}|\to\infty$.

Given the agents' action space, it is natural to form the collaboration as a cooperative game. Naturally, $v(|\boldsymbol{\tau}_{\mathcal{A}}|)$ can also serve as a characteristic function of collaborative games. In a cooperative game, agents are the players and $v$ is the characteristic function. 
The mechanism specified by the platform announces the allocation rule $\mathbf{\psi}(v) \in \mathbb{R}^K$ which specifies the share of the payoff allocated to each player $k = 1, \dots, K$. Thus, agents decide the coalition $\mathcal{A}$ and the participation profile $\BFtau_{\mathcal{A}}$ after observing the allocation rule $\mathbf{\psi}(v)$.

\subsection{Federated Learning Algorithm}

After forming a coalition, the platform utilizes Federated Learning (FL) algorithms to perform the collaborative learning without raw data sharing. A widely adopted FL algorithm is the Federated Averaging algorithm (FedAvg) \citep{kairouz2021advances}, where agents keep their raw data locally to preserve privacy and perform local training on their own data. Periodically, the platform collects interim results from each agent and synchronizes all agents by distributing the average of these local outcomes. 

The platform, after observing the participation profile $\BFtau_{\mathcal{A}}$, orchestrates FL, and specifies the FL design parameter set $\Phi$, which includes relevant configuration details of learning rates, total training epochs, number of synchronizations, to achieve the target performance guarantee $\mathcal{G}$.
The details of the algorithm are provided in Section \ref{sec: algorithm}.

We let $\hat{\BFtheta}^{FL}_{\BFtau_\mathcal{A}}$ denote the estimator produced by the platform conducting FL under participation profile $\BFtau_\mathcal{A}$. FL requires multiple rounds of synchronization to obtain an estimator $\hat{\BFtheta}^{FL}_{\BFtau_\mathcal{A}}$ that satisfies the performance guarantee $\mathcal{G}$ in \eqref{eq:prob_bound_revenue}. The majority cost of performing FL lies in this synchronization process where the platform is required to aggregate and communicate the results across agents \citep{kairouz2021advances}. 

A measure of this cost is determined by the total number of synchronizations needed to achieve an estimator meeting the performance criterion, denoted as $N_{sync}$, which is a function of \textit{platform performance guarantee $\mathcal{G}$, other FL design parameters in $\Phi$, and the announced mechanism $\mathcal{M}$}, highlighting the challenge of interdependency between agent, platform and algorithm.  

Specifically, when the platform’s guarantee $\mathcal{G}$ requires higher confidence or a tighter error bound, the accuracy target is stricter and $N_{\mathrm{sync}}$ must increase. Poor FL design (e.g., cold initialization, suboptimal learning rates) likewise raises $N_{\mathrm{sync}}$. Crucially, and often overlooked in previous literature, the announced mechanism $\mathcal{M}$ also shapes $N_{\mathrm{sync}}$: ill-designed incentives can induce adversarial, highly fragmented participation (e.g., many small identities), increasing gradient variance during local training and slowing convergence to the target specified by $\mathcal{G}$, hence, raises $N_{sync}$.

The following figure summarizes the interplay between the agent, the platform and the algorithm. 
\begin{figure}[htb!]
\centering
\includegraphics[width=0.6\textwidth]{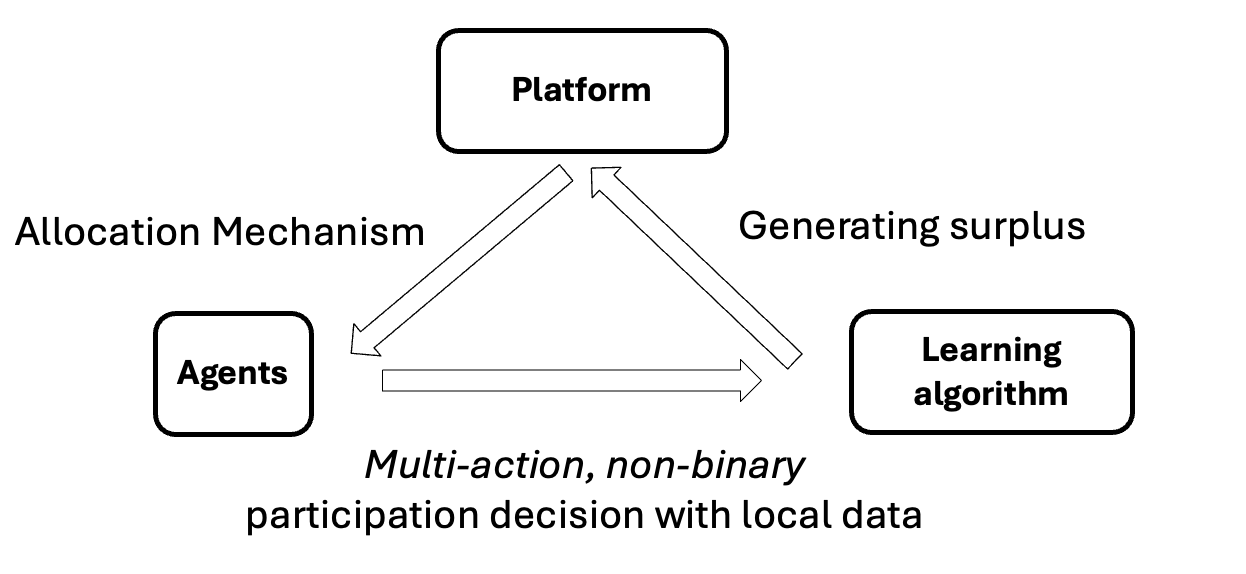}
\caption{Illustration of the MCFL framework}
\label{fig:mcfl}
\end{figure}

The MCFL system, described in Figure \ref{fig:mcfl}, can therefore be described as follows:

\begin{itemize}
\item The platform specifies the performance guarantee $\mathcal{G}$ with a guaranteed value $v(|\BFtau_{\mathcal{A}}|)$, and announces the mechanism $\mathcal{M}$, specifying the allocation rule $\psi(v)$ for all possible participation profiles $\BFtau_\mathcal{A}$.
\item Observing $v$, $\mathcal{G}$ and $\psi$, the agents participate by deciding how to share, and how much to share under announced mechanism $\mathcal{M}$. The coalition $\mathcal{A}$ is formed with the associated participation profile $\BFtau_{\mathcal{A}}$.
\item The platform conducts the learning task through federated learning, designs FL parameters $\Phi$ to achieve target precision level specified in performance guarantee $\mathcal{G}$. 
\item  Once the guarantee is met, the platform announces the estimator produced by FL, $\hat{\boldsymbol{\theta}}^{FL}_{\boldsymbol{\tau}_\mathcal{A}}$, and the coalition surplus is realized as $\pi(\hat{\boldsymbol{\theta}}^{FL}_{\boldsymbol{\tau}_\mathcal{A}}, \boldsymbol{\theta}^*)$. With high probability according to performance guarantee $\mathcal{G}$, the realized surplus is greater than the guaranteed value $v(|\BFtau_{\mathcal{A}}|)$. The platform then redistributes the guaranteed value $v(|\BFtau_{\mathcal{A}}|)$ among the agents in $\mathcal{A}$ according to the announced mechanism $\mathcal{M}$.
\end{itemize}

\section{Mechanisms and Participation Equilibrium}
\label{sec: mechanism}

In this section, we discuss the surplus mechanism announced by the platform in the MCFL framework and the agent participation equilibrium induced by the mechanism. We start by describing the axioms that characterize the properties of ideal mechanisms in the MCFL context. Then we focus on two popular mechanisms based on Shapley value and marginal contribution, respectively, as both of them have desirable properties under MCFL framework. We will discuss other popular mechanisms, which are less preferable under MCFL framework, and their performance under MCFL framework in Appendix \ref{sec: extensions_mechanism}.

As mentioned in Section \ref{sec: model}, we model the decision-aware collaboration problem under MCFL which allows agents to have various levels of participation decisions by selecting the number of samples they would like to contribute to the coalition, as quantified by the participation profile.  

The mechanism allocation rule $\psi(v) \in \mathbb{R}^K$, which characterizes a mechanism $\mathcal{M}$, specifies the payoff allocated to each player $k = 1, \dots, K$, given their contribution of data samples. Thus, agents decide the participation profiles $\BFtau_\mathcal{A}$ after observing the allocation rule $\mathbf{\psi}(v)$. Specifically, $\psi_{\BFtau_\mathcal{A}}$ denotes the allocated payoff received by agent $k$ in a coalition $\mathcal{A}$, where agent participation is characterized by the participation profile $\BFtau_\mathcal{A}$.

\subsection{Mechanisms in the MCFL System}
The standard definition of marginal contribution mechanism applies directly in our setting, as stated in the following:
\begin{definition}[Marginal Contribution Mechanism in MCFL]
\label{def: MC}In a cooperative game, given a formed coalition $\mathcal{A}$, the participation profile $\BFtau_{\mathcal{A}}$, and the characteristic function $v(|\BFtau_{\mathcal{A}}|)$, the payoff allocated to each agent $k \in \mathcal{A}$ is defined as
\begin{equation*}
\psi^{MC}_{\tau_{\mathcal{A}},k}(v) = v(|\BFtau_{\mathcal{A}}|)-v(|\BFtau_{\mathcal{A}}|-\tau^k_{\mathcal{A}}).
\end{equation*} 
\end{definition}
The allocation rule of marginal contribution is straightforward, which is the marginal surplus generated by comparing the surplus of the data from the entire coalition, $v(|\BFtau_{\mathcal{A}}|)$ subtracted by the surplus generated by data except for the samples contributed by agent $k$.

Another popular mechanism for fair and efficient pay-off allocation is the renowned Shapley value. The original definition of the Shapley value can be traced back to \cite{shapley1953value}. The standard definition is applicable for cooperative games with binary participation, as stated below.
\begin{definition}[Shapley value \cite{shapley1953value}] 
In the MCFL cooperative game, given a coalition $\mathcal{A}$, and the characteristic function $v: 2^{|\mathcal{A}|}\to \mathbb{R}$. Then the payoff allocated to agent $k \in \mathcal{A}$ is defined as
\begin{equation*}
\psi_k(v) = \sum_{\mathcal{S}\subseteq \mathcal{A} \backslash\{k\}} \frac{|\mathcal{S}|!(K-|\mathcal{S}|-1)!}{K!} (v(\mathcal{S}\cup \{k\}) - v(\mathcal{S})).
\end{equation*}
\end{definition}

Intuitively, the Shapley value assigns each player their expected marginal contribution by averaging the extra value they add when joining across all possible coalition orders, thereby ensuring fair and efficient allocation among agents. However, the standard Shapley value does not accommodate multiple levels of participation. Therefore,  based on the multi-choice Shapley value proposed in \cite{hsiao1993shapley,hsiao2004power}, we introduce the Shapley value for the MCFL context, which specifies the allocation rule of the surplus. 
\begin{definition}[MCFL Shapley Value]
\label{def: MCLG-SV}
In the MCFL cooperative game, given any agent participation decision profile $\BFtau_\mathcal{A}$, the allocated payoff for agent $k$ is 

\begin{equation*}
\psi^{SV, \alpha}_{\BFtau_\mathcal{A},k}(v) = 
\sum^{\tau^k_{\mathcal{A}}}_{j=1} \sum_{\substack{\BFtau \in \{  \tau_k = j, \\ \BFtau \neq \mathbf{0}| \forall \BFtau \in \mathcal{\Tau}\}}  } \left[\sum_{\substack{T\subseteq M_{k}(\BFtau_\mathcal{A})}}(-1)^{|T|}\frac{\alpha(j)}{||\BFtau_\mathcal{A}||_{\alpha}+\sum\limits_{\substack{r\in T }}[\alpha(\tau_r+1)-\alpha(\tau_r)]}\right] [v(|\BFtau_\mathcal{A}|)-v(|\BFtau_\mathcal{A}-\BFb(k)|)],
\end{equation*}

where $\mathcal{\Tau} := \Pi_{k=1}^K\{0, 1, \dots, m_k\}$ denotes the Cartesian product of $\{0, 1, \dots, m_k\}$ for $k=1, \dots, K$. For any $\BFtau \in \mathcal{\Tau}$, $\|\BFtau\|_{\alpha} := \sum_{k=1}^K \alpha(\tau_k)$. $M_{k}(\BFtau)=\{v|\tau_v\neq m_v, v\neq k\}$ as the set of players that is not an agent $k$ and does not share all the data. Let $\BFb(k)=[0,0\dots,1,0,\dots,0]\in R^{K}$ where the $k^{th}$ element of $\BFb(k)$ is equal to 1. 
\end{definition}

Similarly to the original Shapley value, the generalized Shapley value allowing multi-action assigns each agent the expected marginal gain of each incremental action level, averaging over all others’ action profiles, weighting the contributions by function $\alpha(\cdot)$, so that the allocation reflects how much the agent contributes.

Note that to define the multichoice Shapley value, a weight function must be defined prior to the Shapley value (\cite{hsiao1993shapley,hsiao2004power}). Particularly, the weight function maps any possible action to a non-negative number and satisfies $\alpha(0) = 0$, and $\alpha(i) \leq \alpha(i+1)$ for any $i = 1, \dots, K-1$. In the problem context of data sharing, there often exists a unit effort to obtain data samples \citep{karimireddy2022mechanisms}. Thus, it is natural to consider linear weights function, specifically, sharing a sample of size $i$ has a weight as a linear function of $i$.

For simplicity, we set $\alpha(\tau_k) = \tau_k$\footnote{ This is without loss of generality as any proportional function $\alpha(\tau_k) = \alpha \tau_k$ could be degenerate into $\alpha(\tau_k) = \tau_k$.}, for all $\tau_k \in \mathbb{N}^+$. Thus, the Shapley value defined in \ref{def: MCLG-SV} can be simplified as the following 
\begin{definition}[MCFL Shapley Value with Linear Weights]\label{def: MCLG-SV-linear}
For any agent participation decision profile $\BFtau_{\mathcal{A}}$, for an agent $k$ sharing $\tau_k$ observations, the allocated payoff is given by
\begin{equation*}
\psi^{SV}_{\BFtau_\mathcal{A},k}(v) = \sum^{\tau^k_{\mathcal{A}}}_{j=1} \sum_{\BFtau: \tau_k=j, \BFtau\neq 0 } \left[\sum_{T\subseteq M_{k}(\BFtau)}(-1)^{|T|}\frac{j}{|\BFtau_{\mathcal{A}}|+|T|}\right] [v(|\BFtau_{\mathcal{A}}|)-v(|\BFtau_{\mathcal{A}}-\BFb(k)|)].
\end{equation*}
\end{definition}

 Beyond its use in our setting, it is worth mentioning that, the above definitions of MCFL Shapley values offer a simple and tractable way to evaluate a set of samples contributions to the learning objective.
 In contrast to most prior ML applications of Shapley value, which focus on binary inclusion/exclusion of features (or other ``single-action” players), our setup values the contribution of a specified set of data samples to the global objective under multi-unit participation, extending existing literature in utilizing SV for data evaluation in machine learning tasks. Moreover, our characterization extends to general (potentially non-linear) weighting functions $\alpha(\tau_k)$, providing a unified approach to data valuation under machine learning.

\subsection{Properties of the Mechanism}

Firstly, as the platform goal is to maximize the coalition performance measure, which translates to the platform announced performance guarantee $v(|\BFtau_{\mathcal{A}}|)$. An ideal mechanism should induce full participation, since $v(|\BFtau_{\mathcal{A}}|)$ is an increasing function in the number of data samples provided by the coalition $\BFtau_{\mathcal{A}}$. In the following, we first demonstrate that both marginal contribution mechanism and MCFL Shapley value induce the full participation of the agents.

\begin{proposition}
\label{prop: full-part}
Both the MCFL Shapley value and marginal contribution mechanism motivate agents' full participation. In other words, let $\mathcal{A}^{\text{SV}}$ and $\mathcal{A}^{\text{MC}}$ denote the coalition induced by the MCFL Shapley value respectively, then the total number of samples contributed within the coalition equals the total number of samples possessed by the agents, $|\BFtau_{\mathcal{A}^{\text{SV}}}| = |\BFtau_{\mathcal{A}^{\text{MC}}}| = \sum_{k=1}^K m_k$.
\end{proposition}

Besides full participation, there are other desirable axioms that an ideal surplus allocation mechanism should satisfy, as summarized below:
\begin{axiom}[Null player.]\label{axiom: null-player}
A player who contributes no value always gets zero surplus. In other words, for any agent participation profile $\BFtau_{\mathcal{A}}$, we set $\BFtau'_{\mathcal{A}}$, such that $\tau'_j =\tau_j$ for all $j \neq k\in\mathcal{A}$ and $\tau'_k =0$. If $v(\BFtau_{\mathcal{A}}) = v(\BFtau'_{\mathcal{A}})$, then $\psi_{\tau_\mathcal{A},k} (v) = 0$.
\end{axiom}
The null player axiom is a common fundamental property desired in allocation mechanisms in cooperative games. It requires that a player who does not contribute to the overall value of the coalition receives zero payoff. This axiom is included to discourage free-riding and encourage participation.

\begin{axiom}[Symmetry.]
\label{axiom: symmetry}
Two players that contribute equally get same surplus. To formally state this, for any $\BFtau_{\mathcal{A}} := [\tau_1, \dots, \tau_i, 0,\dots, \tau_K]$ and $\BFtau'_{\mathcal{A}} := [\tau_1, \dots, 0, \tau_j,\dots, \tau_K]$, if $v(\BFtau_{\mathcal{A}}) = v(\BFtau'_{\mathcal{A}})$, then $\psi_{\BFtau,i} (v) = \psi_{\BFtau',j} (v)$.
\end{axiom}
The symmetry axiom is another fundamental requirement of the payoff allocation vector. A mechanism satisfies the symmetry axiom if it allocates the same amount to two players that contribute equally to the coalition's value.

\begin{axiom}[Additivity.]
\label{axiom: additivity}
For two characteristic functions $v$ and $u$, $\psi_{\BFtau,k}(v+u) = \psi_{\BFtau,k}(u)+ \psi_{\BFtau,k}(v)$. 
\end{axiom}
The additivity axiom requires that, for two different characteristic functions $v$ and $u$, $\psi_{i,k}(v+u) = \psi_{i,k}(u)+ \psi_{i,k}(v)$, for all $i, k$. Additivity ensures allocation consistency when simple collaborative learning tasks are combined, making the mechanism robust across multiple learning tasks.

\begin{axiom}[Efficiency/Budget Balance.]
$\sum_{k \in \mathcal{A}} \psi_{\BFtau_\mathcal{A},k}(v) = v(|\BFtau_\mathcal{A}|)$.\label{axiom: efficiency}
\end{axiom}
The efficiency axiom, also referred to as the budget balanced axiom, requires that all of the surplus generated by the coalition be allocated to the participants. If $\sum_{k \in \mathcal{A}} \psi_{\BFtau_\mathcal{A},k}(v) < v(|\BFtau_\mathcal{A}|)$, then part of the surplus is kept by the platform.

\begin{proposition}
\label{prop: SV-axiom}
The MCFL Shapley value, as defined in Definition \ref{def: MCLG-SV}, is the only mechanism that satisfies Axioms \ref{axiom: null-player}-\ref{axiom: efficiency}.
\end{proposition}

The marginal contribution mechanism, on the other hand, does not satisfy Axiom \ref{axiom: efficiency} if the value function $v(\cdot)$ is strictly concave. Note that $v(\cdot)$ is strictly concave in the context of collaborative learning, as the marginal gain of having additional data decreases. This result is formally stated in the following theorem.
\begin{theorem}
\label{thm: MC-BB} In the MCFL framework, the total surplus allocated to participants, denoted as $\sum_{k\in \mathcal{A}} \psi^{\text{MC}}_{\BFtau_{\mathcal{A}}, k}(v) $, has the following properties
\begin{itemize}
\item[(i)] $\sum_{k\in \mathcal{A}} \psi^{\text{MC}}_{\BFtau_{\mathcal{A}}, k}(v)<v(|\BFtau_{\mathcal{A}}|)$ if $v(\cdot)$ is a strictly concave function,
\item[(ii)] $\sum_{k\in \mathcal{A}} \psi^{\text{MC}}_{\BFtau_{\mathcal{A}}, k}(v)>v(|\BFtau_{\mathcal{A}}|)$ if $v(\cdot)$ is a strictly convex function, 
\item[(iii)] $\sum_{k\in \mathcal{A}} \psi^{\text{MC}}_{\BFtau_{\mathcal{A}}, k}(v)=v(|\BFtau_{\mathcal{A}}|)$ if $v(\cdot)$ is a linear function,
\end{itemize}
under assumption that $v(0)=0$. 
\end{theorem}

Theorem \ref{thm: MC-BB} implies that, if the platform adopts the MC mechanism, under the MCFL framework where $v(\cdot)$ is usually a strictly concave function, it will leave the surplus on the table instead of allocating all. Proposition \ref{prop: SV-axiom} and Theorem \ref{thm: MC-BB} suggest that, SV is the only mechanism that maintains fair allocation satisfying Axioms 1-3, while keeping budget-balance with redistributing all surplus generated by coalition back to the coalition member. MC, in contrast, leaves part of the surplus unassigned.

\subsection{Vulnerability Against Data Splitting}

According to Proposition \ref{prop: SV-axiom}, the Shapley value is the only mechanism that satisfies all axioms. However, in addition to satisfying these axioms, it is also important for a mechanism to remain robust with respect to dishonest behaviors conducted by the participating agents that abuse the mechanism. One common abuse is the data splitting behavior across fake identities. To be more specific, an agent might create fake multiple identities and then splits their data among these fake identities. In other words, if the player $k$ splits into $m$ identities, then the original sample $S_k$ is split into $m$ small sub-samples, each containing at least one data sample, and the player $k$ pretends that the $m$ small sub-samples come from $m$ different (fake) agents. Note that, in this work, we do not discuss dishonest behaviors that are adversarial, malicious, or attacking, as it involves safety and security issues in learning and cyber system, which is beyond the scope of this paper.

In the following, we further discuss the participation equilibrium induced by the two mechanisms -- MCFL Shapley value and Marginal contribution, considering potential risk of data splitting across duplicated fake identities.

We first discuss the equilibrium induced by the MCFL Shapley value. In the following theorem, we demonstrate that, splitting data always leads to more allocated surplus.  
\begin{theorem}[Vulnerability of MCFL Shapley Against Dishonest Data Splitting] 
\label{thm: main-SV}
For any agent within the coalition, $k \in \mathcal{A}$, suppose that agent $k$ participates with $\tau_k$ data samples. They may create two fake identities, denoted as (fake) agent $k_1$ and $k_2$, and split $\tau_k$ samples among identities $k_1$ and $k_2$, each possesses $t_1$ and $t_2$, respectively. Then for all $t_1, t_2 >0$ and $t_1+t_2 = \tau_k$, we have
\begin{itemize}
\item[(i)] $\psi^{\text{SV}}_{\BFtau_{\mathcal{A}},k}(v)<\psi^{\text{SV}}_{\BFtau_{\mathcal{A}},k_1}(v)+\psi^{\text{SV}}_{\BFtau_{\mathcal{A}},k_2}(v)$ if $v(\cdot)$ is a strictly concave function,
\item[(ii)] $\psi^{\text{SV}}_{\BFtau_{\mathcal{A}},k}(v)=\psi^{\text{SV}}_{\BFtau_{\mathcal{A}},k_1}(v)+\psi^{\text{SV}}_{\BFtau_{\mathcal{A}},k_2}(v)$ if $v(\cdot)$ is a linear function,
\end{itemize}
where $\psi^{\text{SV}}_{\BFtau_{\mathcal{A}},k}$ denotes the allocated surplus received by agent $k$ participated using its true original identity with $\tau_k$ data samples, and $\psi^{\text{SV}}_{\BFtau_{\mathcal{A}},k_1}$ and $\psi^{\text{SV}}_{\BFtau_{\mathcal{A}},k_2}$ denote the surplus received by (fake) agent $k_1$ and $k_2$, with $t_1$ and $t_2$ data samples split from the original $\tau_k$ samples.
\end{theorem}

Theorem \ref{thm: main-SV} indicates that, if the characteristic function $v(\cdot)$ is concave, which is the case in learning, an agent would receive more allocated surplus if they create duplicated fake identities and split their original data set.

Achieving the above result requires developing new proof techniques. The main challenge is that, unlike the standard Shapley value where agents' actions are binary, the MCFL Shapley value operates in a multi-action setting, and thus requires comparing payoff differences across a combinatorially complex space of participation profiles. The proof idea is to first decompose the MCFL Shapley value to reduce the comparison to a weighted sum of incremental learning surpluses. The key step is then to establish the positivity of this sum through a Beta function integral representation and a stochastic dominance argument over the profile distribution. The detailed proof is outlined in Appendix~\ref{subsec: proof-main-thm}.

Combining the results from Theorem \ref{thm: main-SV} and Proposition \ref{prop: full-part}, we are able to characterize the agent participation equilibrium as stated in the following corollary under MCFL framework, where $v(\cdot)$ is a strictly increasing and concave function:
\begin{corollary}
\label{cor: equi-SV}
In the MCFL system, with $K$ real agents possessing $m_1, \dots, m_K$ data samples respectively, the MCFL Shapley value leads to the following agent participation equilibrium with a coalition consisting of $|\mathcal{A}| = \sum_{k=1}^K m_k$ (fake) agents, and the participation profile: $\BFtau_{\mathcal{A}} = \mathbf{1}\in\mathbb{R}^{|\BFm|}$. In other words, real agent $k$ would participate using $m_k$ fake identities and fully split the original $m_k$ data samples across the fake identities until each identity only contribute one data sample.
\end{corollary}

More importantly, building on Proposition~\ref{prop: SV-axiom}, we obtain an impossibility result for mechanism design: no mechanism can simultaneously satisfy the standard fairness and efficiency axioms (Axioms~\ref{axiom: null-player}–\ref{axiom: efficiency}), while remaining robust to data splitting. This impossibility points to a broader tradeoff in MCFL: mitigating data splitting can improve FL training efficiency (e.g., fewer communication rounds to reach a target accuracy), but may require sacrificing allocation efficiency in the mechanism. Hence, FL algorithm efficiency is not free under MCFL, and there is no “free lunch.”

\begin{corollary}\label{cor:impossible}
No mechanism jointly satisfies Axioms~\ref{axiom: null-player}--\ref{axiom: efficiency} while being robust to data splitting.
\end{corollary}

In contrast, the marginal contribution mechanism, although not budget balanced, is robust to the dishonest behavior of data splitting and thus induces a completely different equilibrium of user participation profile.
\begin{theorem}[Robustness of Marginal Contribution Against Dishonest Data Splitting] 
\label{thm: main-MC}
For any agent within the coalition, $k \in \mathcal{A}$, suppose that agent $k$ participates with $\tau_k$ data samples. They may create two fake identities, denoted as (fake) agent $k_1$ and $k_2$, and split $\tau_k$ samples among identities $k_1$ and $k_2$, each possessing $t_1$ and $t_2$, respectively. Then for all $t_1, t_2 >0$ and $t_1+t_2 = \tau_k$, we have
\begin{itemize}
\item[(i)] $\psi^{\text{MC}}_{\BFtau_{\mathcal{A}},k}(v)>\psi^{\text{MC}}_{\BFtau_{\mathcal{A}},k_1}(v)+\psi^{\text{MC}}_{\BFtau_{\mathcal{A}},k_2}(v)$ if $v(\cdot)$ is a strictly concave and increasing function,
\item[(ii)] $\psi^{\text{MC}}_{\BFtau_{\mathcal{A}},k}(v)=\psi^{\text{MC}}_{\BFtau_{\mathcal{A}},k_1}(v)+\psi^{\text{MC}}_{\BFtau_{\mathcal{A}},k_2}(v)$ if $v(\cdot)$ is a linear function,
\end{itemize}
where $\psi^{\text{MC}}_{\BFtau_{\mathcal{A}},k}$ denoted the allocated surplus received by agent $k$ participated using its true original identity with $\tau_k$ data samples, and $\psi^{\text{MC}}_{\BFtau_{\mathcal{A}},k_1}$ and $\psi^{\text{MC}}_{\BFtau_{\mathcal{A}},k_2}$ denote the surplus received by (fake) agent $k_1$ and $k_2$, with $t_1$ and $t_2$ data samples split from the original $\tau_k$ samples.
\end{theorem}
Similar to the MCFL Shapley setting, the following corollary describes the agent participation equilibrium. 

\begin{corollary}
\label{cor: equi-MC}
In the MCFL system, with $K$ real agents possessing $m_1, \dots, m_K$ data samples respectively, the Marginal Contribution mechanism leads to the agent participation equilibrium with a coalition consisted of $|\mathcal{A}| = K$ real agents, and the participation profile: $\BFtau_{\mathcal{A}} = [m_1, \dots, m_K]$.  In other words, real agent $k$ would participate using their all possessed $m_k$ data samples without an incentive of creating fake identities and splitting data across them. 
\end{corollary}

Corollary \ref{cor: equi-SV} suggests that while the Shapley value satisfies all the desirable properties and encourages full participation in the provision of all data samples, it inherently incentivizes agents to split data with fake identities. Although this dishonest behavior still incentivizes agents to contribute to all data samples, it leads to an equilibrium with a massive number of fake agent identities in the coalition.

Why is dishonest data splitting particularly critical in the MCFL framework? This is because, under traditional collaborative games, splitting across identities does not hurt the overall coalition efficiency, nor does it change the resulting allocation under splitting equilibrium.  However, under MCFL framework, data splitting will directly hurt FL algorithm performance, which, in turn, harms the overall coalition efficiency. We will formally analyze how this data splitting behavior hurts FL performance, in the next section.
\section{Federated Learning Algorithms: Convergence and Cost}
\label{sec: algorithm}

In Section \ref{sec: mechanism}, we discuss the MCFL Shapley value and marginal contribution mechanisms that incentivize full participation of the agents. We also discuss the potential vulnerability of the MCFL Shapley value under dishonest data splitting behaviors. In this section, we elaborate on how this vulnerability would impact the performance of FL algorithms and communication costs.  Building on the  foundation in Section \ref{sec: mechanism}, Section~\ref{subsec:fl_efficiency} introduces FL algorithms, develops metrics for system efficiency in FL, based on coalition surplus and cost of FL training. Section~\ref{subsec:sv_pitfall1} demonstrates a first pitfall: Shapley value allocations can be misaligned with FL efficiency objectives and tuning FL hyperparameters cannot systematically restore this alignment. 

\subsection{Algorithm Convergence and System Efficiency}
\label{subsec:fl_efficiency}
The goal of the platform is to estimate $\BFtheta^*$ to achieve the performance guarantee described in (\ref{eq:prob_bound_revenue}), as promised when announcing the characteristic function $v$ of the mechanism. As discussed earlier in this paper, the estimation is obtained through FL algorithms. We first introduce the FL setting in the MCFL system.

Given the data samples provided by the coalition $\mathcal{A}$ with the participation profile $\BFtau_{\mathcal{A}}$, the target of the FL algorithm is to learn the estimator
\myeqln{\hat{\boldsymbol{\theta}}^*_{FL} := \arg\min_{\boldsymbol{\theta}} L(\boldsymbol{\theta}), } where the loss function $L$ to minimize is decomposable:
\myeql{\label{eq:FL_general}
L({\boldsymbol{\theta}}):=\frac{1}{|\mathcal{A}|}\sum_{k\in\mathcal{A}}L_k(\BFtheta) = \frac{1}{|\mathcal{A}|} \sum_{k \in \mathcal{A}}\left(\frac{|\mathcal{A}|}{|\BFtau_{\mathcal{A}}|}\sum^{\BFtau^k_{\mathcal{A}}}_{j=1}l({\boldsymbol{\theta}}; \mathbf{z}_j)\right),
}
with $L_k=\frac{|\mathcal{A}|}{|\BFtau_{\mathcal{A}}|}\sum^{\BFtau^k_{\mathcal{A}}}_{j=1}l({\boldsymbol{\theta}}; \mathbf{z}_j)$\footnote{We use uniform sample weights. The factor $|\mathcal{A}|/|\boldsymbol{\tau}_{\mathcal{A}}|$ ensures that the global objective is a \emph{sample average} that is invariant to how samples are partitioned across participants. Although the $|\mathcal{A}|$ terms cancel algebraically, writing $L(\boldsymbol{\theta})=\frac{1}{|\mathcal{A}|}\sum_{k\in\mathcal{A}}L_k(\boldsymbol{\theta})$ matches the standard FedAvg/local-GD form (agent-wise averaging) while keeping $L$ unchanged under data splitting/merging, as long as the total number of samples $|\boldsymbol{\tau}_{\mathcal{A}}|$ is fixed.}. If the performance metric $\pi(\hat{\boldsymbol{\theta}}, \boldsymbol{\theta}^*)$ satisfies the required properties for FL algorithms (e.g., decomposability among local agents), then $L$ may be taken as negative of $\pi$, so the estimator directly targets the performance measure.

Even when the platform objective $\pi(\boldsymbol{\theta})$ is not identical to the training loss $L(\boldsymbol{\theta})$, we assume that $L$ is a well-designed surrogate whose population minimizer coincides with the target parameter $\boldsymbol{\theta}^*$, i.e., $\boldsymbol{\theta}^*=\arg\min_{\boldsymbol{\theta}}\mathbb{E}_{z\sim\mathbb{P}_z} \, l(\boldsymbol{\theta};z)$. Lemma~\ref{lemma: high-prob-consistent} then guarantees that the FL output $\hat{\boldsymbol{\theta}}^*_{FL}$ obtained by (approximately) minimizing the empirical loss concentrates around $\boldsymbol{\theta}^*$ with high probability. Together with Assumption~\ref{assump:pi}, which ensures that $\pi$ is uniquely maximized at $\boldsymbol{\theta}^*$, this justifies using FL minimization of $L$ as an indirect method to (approximately) maximize $\pi$.

\begin{lemma}[High-Probability Convergence of FL Estimator]
\label{lemma: high-prob-consistent}
There exist constants \( C_1 > 0 \), \( C_2 > 0 \), and radius \( r > 0 \) such that for any confidence level \( \delta \in (0, C_2] \), with $|\BFtau_{\mathcal{A}}|\geq\frac{1}{C_1 r^2}\log\!\Bigl(\frac{C_2}{\delta}\Bigr)$, define \(\epsilon_L(|\BFtau_{\mathcal{A}}|, \delta) := \sqrt{ \frac{1}{C_1 |\BFtau_{\mathcal{A}}|} \log\left( \frac{C_2}{\delta} \right) }
\), the following holds
\[
\mathbb{P}\left( \left\| \hat{\BFtheta}^*_{FL}- {\BFtheta}^* \right\| \geq \epsilon_L(|\BFtau_{\mathcal{A}}|, \delta) \right) \leq \delta.
\]
Here, ${\BFtheta}^*$ is defined as the unknown true parameter. In the meanwhile, it is also the minimizer of the population expected loss, in other words, ${\BFtheta}^* = \argmin_{\BFtheta} \mathbb{E}_{z\sim \mathbb{P}_z} l(\BFtheta; \BFz)$.
\end{lemma}
This lemma requires a high-probability bound on the difference between the optimal in-sample estimator and the ground truth parameter, which minimizes the expected loss.

Under Assumption~\ref{assump:pi}, which ensures that \( \pi \) is uniquely maximized at \( \boldsymbol{\theta}^* \), and Lemma~\ref{lemma: high-prob-consistent}, minimizing \( L \) yields an estimator that asymptotically maximizes \( \pi \) as the data size grows. This separation between estimation and optimization enables the use of federated algorithms to efficiently estimate \( \boldsymbol{\theta}^* \) while still providing rigorous guarantees on performance under specific business context.

In practice, obtaining $\hat{\BFtheta}^*_{FL}$ typically relies on FL algorithms. Although often the convergence of the FL algorithms can be guaranteed, iterations often stop when the solution is close enough within a certain tolerable optimization error. This results in an inherent gap between $\hat{\BFtheta}^*_{FL}$ and the output of the FL algorithm, denoted as $\hat{\BFtheta}^{FL}_{\BFtau_\mathcal{A}}$. 
FL algorithms follow a decentralized iterative update mechanism: the platform broadcasts the current model, agents perform local optimization, and their updates are aggregated to form the new global model.
The algorithm is governed by a set of hyper-parameters $\Phi=\{K^{FL}, \rho, \BFtheta^0, T, H\}$, where $K^{FL}, \rho, \BFtheta^0, T, H$ represent the number of selected agents per round, learning rate, starting point, total number of iterations (including federated individual iteration steps) and number of local update iterations (which, together with $T$, pins down the number of synchronization required for FL), respectively. 

\begin{figure}[htb!]
\centering
\includegraphics[width=0.85\textwidth]{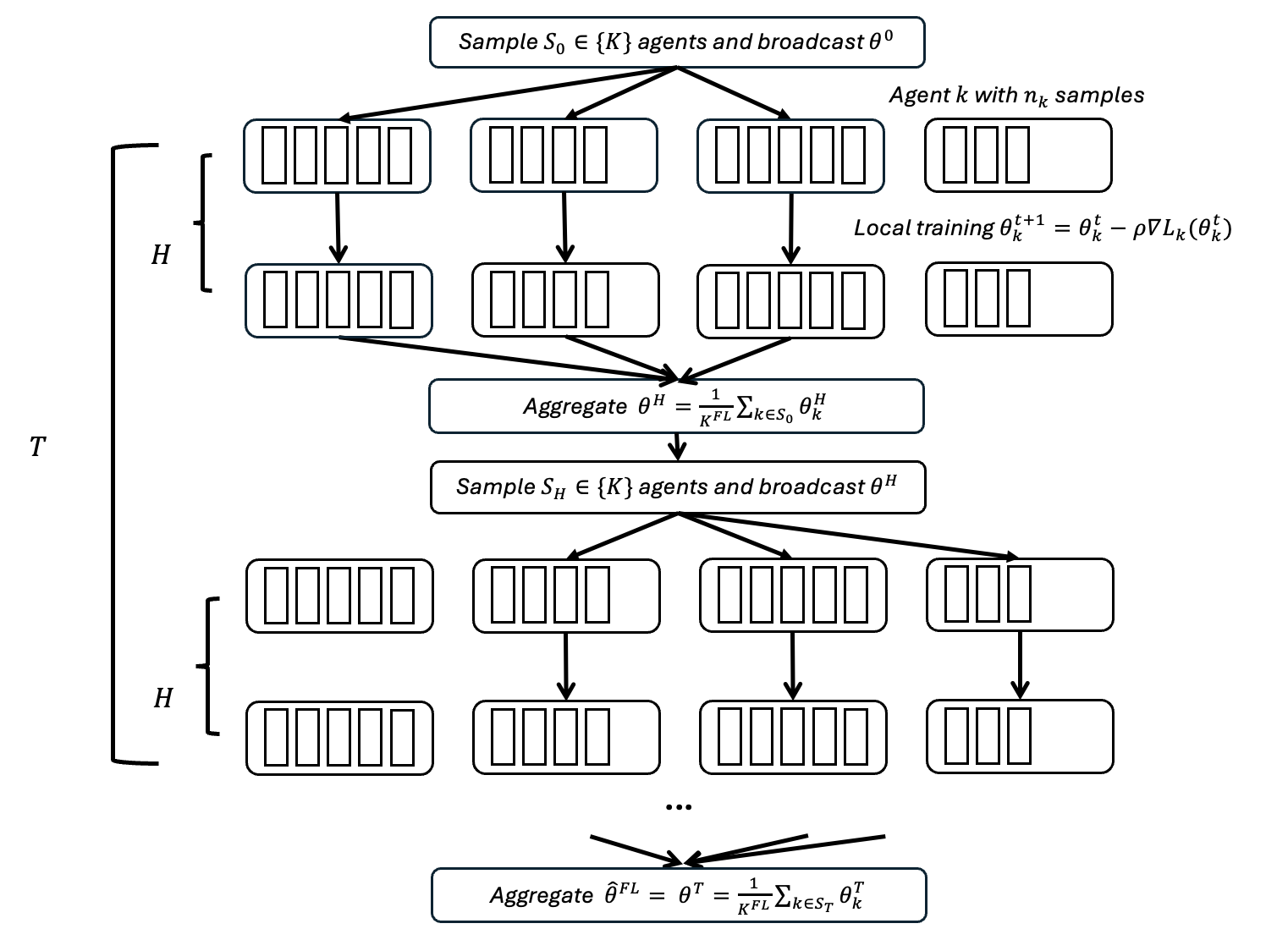}
\caption{Optimization Algorithm Framework for MCFL, with key decision parameters $\Phi=\{K^{FL}, \rho, \BFtheta^0, T, H\}$.}
\label{fig:FL_process}
\end{figure}

Figure \ref{fig:FL_process} illustrates the roles of these hyper-parameters in the federated learning algorithm. In each round, the platform selects $K^{FL}$ agents to receive the global parameters broadcast. Each agent performs $H$ steps of local gradient descent with step-size $\rho$, starting from the received parameters. The platform then aggregates the local updates, typically by averaging, to obtain the next global model. A full description of the algorithm appears in Appendix \ref{app:FL_alg}.
For simplicity of the analysis, we assume that each agent performs gradient descent (GD) rather than stochastic gradient descent (SGD) during local training, which is a practical approach when agents do not possess a large volume of data. For simplicity of the analysis and to focus on the main implication, we also assume that \( K^{FL} = |\mathcal{A}| \), implying that at each synchronization, the platform queries all participating agents, a standard assumption in cross-silo federated learning \citep{kairouz2021advances}. The results and implications that we provided can be relaxed for SGD, and the case where the platform queries only a subset of agents with $K^{FL}<|\mathcal{A}|$ at each synchronization step.

While performing FL algorithm, the MCFL platform sets hyperparameters to minimize training cost while meeting a prescribed performance guarantee and the dominant cost typically comes from synchronization/communication rounds \citep{kairouz2021advances}. Accordingly, we model total training cost as primarily driven by the number of synchronization rounds. Given a target performance guarantee level, we use $N_{sync}$ to denote the \textit{minimum number of synchronization} rounds required for a set of hyperparameter choice to achieve the target guarantee. In general, more aggressive synchronization often leads to faster convergences, but synchronizations are expensive. 
Therefore, the platform’s hyperparameter choice trades off optimization performance against synchronization cost by affecting $N_{sync}$.

In typical FL setting, increasing the number of participating agents often raises the per-round training cost, since it requires additional communication and coordination effort for each additional client \citep{kairouz2021advances}. In this work, we would like to highlight that, we do not directly assume that the per-round synchronization cost scales with the number of participating agents, even though this is true to some extent. Instead, we took a more sophisticated model that, in each global synchronization round,  there is a \textit{fixed unit cost $c$}, so the total training cost scales linearly with number of synchronization. This is because we would like to highlight how the number of agents in the system affects the convergence of the FL algorithm.
Note that, this model of communication cost is actually in favor of SV, as it underestimates the communication cost, yet there is still a vital pitfall of adopting SV in the MCFL system. We take $N_{\text{sync}}$ to be the minimum number of synchronization rounds required for the FL procedure to meet the target performance guarantee under the chosen hyperparameters.

\begin{assumption}
The FL training cost scales linearly with the minimum number of synchronization rounds required, denoted by $N_{\text{sync}}$. Specifically, the training cost is $c\,N_{\text{sync}}$, where $c>0$ is the fixed unit cost per synchronization round.
\end{assumption}

We then define the \emph{system efficiency} as
\myeql{\label{eq : system_efficiency}
\textbf{System Efficiency}\quad 
\Pi^{sys}=\sum_{k\in\mathcal{A}}\psi_{\tau_k, k}(v( |\boldsymbol{\tau}_{\mathcal{A}}| ))-cN_{sync}.
}
Here, the first term is the total surplus accrued by participating agents under the announced mechanism, and $cN_{\text{sync}}$ is the (amortized) cost of federated training.

\subsection{Pitfall of Shapley Values in System Efficiency}
\label{subsec:sv_pitfall1}
The number of synchronization is inherently impacted by the announced mechanism because different mechanisms cause different equilibrium of agent participation. In this part, we analyze the minimum number of synchronization rounds required to achieve the announced performance guarantee.
To guarantee the convergence of the  FL algorithm, we assume that the loss function $l(\cdot; \cdot)$ satisfies the following properties:
\begin{assumption}
\label{assump: asmp_Lsmooth_convex}
The function $l(\cdot; \BFz)$ satisfies the following properties for any $\BFz\in \mathbb{R}^{p+1}$:
\begin{enumerate}[label=\alph*.]
    \item \textbf{Smoothness and strong convexity.} $l(\cdot; \BFz)$ is $\nu$-smooth and $\mu$-strongly convex.
    \item \textbf{Uniform Lipschitzness.} there exists a nonempty compact convex set $\BFTheta$ such that all FL iterates remain in $\BFTheta$, and $\| l(\BFtheta_1; \BFz) - l(\BFtheta_2; \BFz) \| \leq
    L \|\BFtheta_1 -\BFtheta_2\|$ for all $\BFtheta\in \BFTheta$ and $\BFz$.
\end{enumerate}
\end{assumption}

These assumptions are common regularity conditions for loss functions in machine learning. 

To analyze the convergence of the FL algorithm under the MCFL framework, we first need to specify the key parameters $\Phi$ that the platform adopts. Note that the platform could potentially fine tune the parameters and choose different $\Phi$ in different mechanisms, based on different equilibriums induced by those mechanisms. 

In this section, we adopt the FL parameter that is tuned to optimize the convergence under the SV mechanism. That said, the platform will select the optimal combination of $N_{sync}$ and $\rho$ with respect to the participation profile $\BFtau_\mathcal{A}$ induced by SV under fixed epoch of training $T$. The comparison of $N_{\mathrm{sync}}^{SV}$ and $N_{\mathrm{sync}}^{MC}$ is conducted for a fixed total training epoch $T$, consistent with both our definition of system efficiency, and the standard convention in federated learning convergence analysis \citep{khaled2019first, stich2018local, qu2023unified}. This convention reflects the practical setting in which the platform's computational budget in total epoch of training, determined by agent hardware capacity, data volume, and project timeline, is fixed independently of the announced mechanism. The mechanism's effect on FL efficiency then operates purely through the number of synchronization rounds required within this budget. Later, we will show that, the system is less efficient with SV even with the hyper-parameters are set in favor of it.

\begin{theorem}\label{thm:convergence}
Suppose Assumptions~\ref{assump:pi} and~\ref{assump: asmp_Lsmooth_convex} hold, with any given $\delta\in(0,1)$, and let
$\epsilon(|\boldsymbol{\tau}_{\mathcal A}|,\delta)$ be the target tolerance in~\eqref{eq:thm:convergence}.
Then FL converges to the desired performance guarantee
\myeql{
\mathbb{P}\!\left( v^* - \pi(\hat{\BFtheta}^{FL}_{\BFtau_\mathcal{A}}, \boldsymbol{\theta}^*) \geq \epsilon(|\boldsymbol{\tau}_{\mathcal{A}}|, \delta) \right) \leq \delta,
\quad \forall\,\mathcal{M}\in\{SV,MC\}.
\label{eq:thm:convergence}
}
under both $SV$ and $MC$ for $T\geq\underline{T}$, where $\underline{T}$ is a fixed constant defined in Appendix \ref{app:FL_alg}, equation (\ref{eq-barT}).

Moreover, the number of synchronization required $N_{sync}$ satisfies the following properties.
\begin{enumerate}
\item \textbf{For Shapley Value (SV)}: There exists a constant $C_{\mathrm{SV}}>0$ and a choice of FL parameters such that, the number of synchronization required that achieves the optimal dependence on both $T$ and $|\BFm|$ is
\myeql{
N^{SV}_{\mathrm{sync}}  \;=\; C_{\mathrm{SV}}(T|\BFm|)^{3/4}.
}
\item \textbf{For Marginal Contribution (MC)}: There exists a constant $C_{\mathrm{MC}}>0$ and a configuration of FL parameters $\Phi$ such that,
\begin{equation}
N^{MC}_{\mathrm{sync}}  \;=\; C_{\mathrm{MC}}(TK)^{3/4}.
\label{eq:MC_Nsync}
\end{equation}
\end{enumerate}
\end{theorem}

The contribution of Theorem \ref{thm:convergence} is two-fold. First, while the convergence of FL is well-established, Theorem \ref{thm:convergence} provides the convergence of FL for decision-aware agent with general business objectives, highlighting the potential of FL to business applications. Second, Theorem \ref{thm:convergence} states that SV requires a greater number of synchronization to converge to the desired performance guarantee, than MC. Note that, as mentioned earlier, we set the hyperparameters in favor to SV, so that number of synchronization required that achieves the best balanced leading-order dependence on both $T$ and $|\BFm|$ under SV. 

Since SV is budget-balanced, it redistributes all the surplus generated from the learning outcome back to participating agents, while MC is not budget balanced and will leave surplus on table when the objective is concave, as demonstrated in Theorem \ref{thm: MC-BB}. Intuitively, one would think that SV thus has an advantage in system efficiency. However, in the decentralized collaborative learning context, the results is the opposite of the intuition. Considering the substantial increment of the training costs, SV is dominated by MC in terms of system efficiency, as demonstrated in the following theorem.  Since the exact constant is not material here, we assume $C_{SV}|\BFm|^{3/4}>C_{MC}K^{3/4}$, which is satisfied whenever the total number of data samples is sufficiently large relative to the number of agents, as is typically the case in practice.

\begin{corollary}
\label{cor:system_analysis_MS}
Under the conditions specified in Theorem~\ref{thm:convergence}, 

\myeql{\Pi^{sys}(\mathcal{M}=SV)=v(|\BFm|)-c\cdot C_{SV}(T|\BFm|)^{3/4}}
and 
\myeql{\Pi^{sys}(\mathcal{M}=MC)\geq \sum_k (v(|\BFm|)-v(|\BFm|-m_k))-c\cdot C_{MC}(TK)^{3/4}}
And under $C_{SV}|\BFm|^{3/4}-C_{MC}K^{3/4}>0$, for $c>\frac{\sum_{k=1}^{K}v(|\BFm|-m_k)-(K-1)v(|\BFm|)}
{T^{3/4}\bigl(C_{SV}|\BFm|^{3/4}-C_{MC}K^{3/4}\bigr)}$,
\myeql{\Pi^{sys}(\mathcal{M}=SV)<\Pi^{sys}(\mathcal{M}=MC).\label{ineq:sys_eff}}
Moreover, for any $c>0$, there exists a $v(\cdot)$ such that the inequality (\ref{ineq:sys_eff}) holds for all data profiles, and SV leads to worse system efficiency compared with MC.
\end{corollary}

The following figure represents the total system efficiency under equilibrium participation profile for Shapley mechanism and marginal contribution mechanism. 

\begin{figure}[htb!]
\begin{center}
\includegraphics[scale=.20]{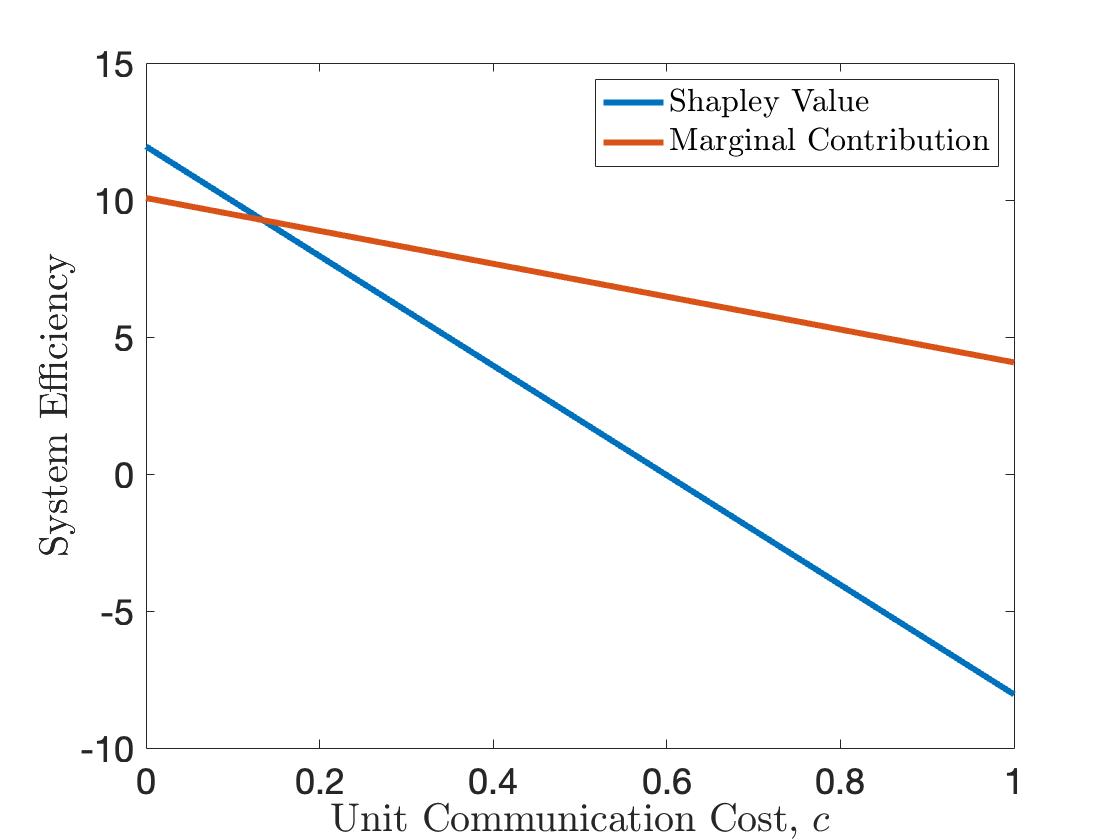}
\end{center}

\caption{System efficiency of different mechanisms. The x-axis represents the cost per synchronization across all participating agent identities. The y-axis represents the system efficiency under different mechanisms equilibrium participation profile. }
        \label{fig:sys_eff}
\end{figure}

While much attention has been devoted to the allocation of surplus while designing a collaborative learning mechanism, the training cost incurred during the learning process with the induced equilibrium is often overlooked. This omission is critical, as training cost--particularly under decentralized or federated settings--can significantly affect the overall system efficiency. Results presented in this section show that, although SV ensures budget balance and achieves higher allocation efficiency relative to MC, SV unintentionally incentivizes agents to split their data across multiple identities to increase their perceived marginal contribution, thereby inflating synchronization and training costs.  When training cost is incorporated into the system-level evaluation—especially under scenarios with expensive synchronization—MC can outperform SV in terms of total system efficiency. These results underscore the importance of jointly considering both surplus allocation and training costs in mechanism design for collaborative learning systems.

\section{Numerical Experiments}
\label{sec: numerical}

We further validate our framework using numerical experiments with synthetic data and a case study using real-world data. In numerical experiment with synthetic data, we consider two business operation examples of the MCFL framework, the newsvendor problem and portfolio optimization problem. For the case study, we consider the collaborative learning task for skin lesions diagnosis.

In numerical experiments, we aim to demonstrate the performance of the MCFL framework, especially focusing on the inefficiency caused by dishonest data splitting among fake identities. Recall that the agent participation equilibrium under SV is that all agents split to an extreme with many (fake) identities, each possessing one data sample. In this section, we do not examine the convergence of the FL algorithm in the full splitting equilibrium, instead, we demonstrate how data splitting, even not to the full splitting equilibrium, still significantly hurts the convergence of the learning algorithm.

\subsection{MCFL for Newsvendor Problem}
\label{sec: num_nv}
Following up on Example 1.1 in Appendix \ref{app:example}, we introduce the MCFL setting for the newsvendor problem. 
 The agents aim to collaboratively learn an estimator $\hat{\BFtheta}$ by minimizing the objective function 
$(w-y_i)^{+}=\max\{0, w-y_i\}$, and $(y_i-w)^{+}=\max\{0,y_i-w\}$, the FL learning objective is to obtain an estimator $\hat{\BFtheta}$ that minimizes the following NV objective with an additional term of $L_2$-regularization to prevent overfitting through FL, with
\myeql{
\hat{\BFtheta}^*_{FL}=\arg\min_{\BFtheta}\ L(\BFtheta)=\sum^{K}_{k=1}\left(\sum_{\BFz_j\in S_k} \ l_j(\BFx^T_j\BFtheta,y_j)\right).
}

In this numerical example, we focus on the newsvendor problem as presented in Example \ref{example: NV}. Specifically, we assume the demand data held by each agent independently follows the distribution of $y_i=\BFx^T_i\BFtheta^*+\epsilon_i$, and $\BFz_i := (\BFx_i, y_i)$, where $\BFx_i\in \mathbb{R}^{p}$ is the contextual information, and $\epsilon_i$ follows the Normal distribution with zero mean and variance $\sigma^2$.  Here, $\BFtheta^*\in \mathbb{R}^{p}$ is the unknown estimator we try to obtain from FL. According to the newsvendor problem formulation, $l_j(w,y_j)=h(w-y_j)^{+}+b(y_j-w)^{+}$, 
where $(w-y_i)^{+}=\max\{0, w-y_i\}$, and $(y_i-w)^{+}=\max\{0,y_i-w\}$, the platform objective is to obtain an estimator $\hat{\BFtheta}^*_{FL}$ that minimizes the following $l_2$-regularized NV objective 
\myeql{
\label{eq: NV-loss}
\hat{\BFtheta}^*_{FL}=\arg\min_{\BFtheta}\ L(\BFtheta)=\sum^{K}_{k=1}\sum_{\BFz_j\in S_k} \ l_j(\BFx^T_j\BFtheta,y_j)+\lambda \|\BFtheta\|^2_2.
}

In this numerical experiment, we set $h=0.1$, $b=0.9$, $\lambda = 1$, and $\epsilon_i$ follow $N(0,\sigma^2)$ with $\sigma^2 = 1$. The feature vectors, $\BFx_j \in \mathbb{R}^1$, are generated from $N(0,\sigma^2_{x})$ with $\sigma_{x}=2$. 

To demonstrate the harm to the convergence if agents commit dishonest data splitting, we consider the setting with two agents, denoted as agent $k$ and $j$, each possesses $8$ data samples. Note that here we consider a very small sample size as a toy example to better illustrate the local and global convergence of the estimator through visualization of the estimator convergence process. Agent $k$ (or $j$) could create one fake identity by equally split their data samples, and participate using two identities $k_1$ (or $j_1$), denoting the real one, and $k_2$ (or $j_2$), denoting the fake one. Then the agent split the samples evenly among two identities. We set FL parameter \(\Phi\) as \(\{K^{FL}=K, \rho=0.1, \hat{\BFtheta}^{FL}_0=3, T = 55\}\).

\begin{figure}[htb!]
    \centering
    \begin{subfigure}{.5\linewidth}
        \centering
        \includegraphics[scale=.18]{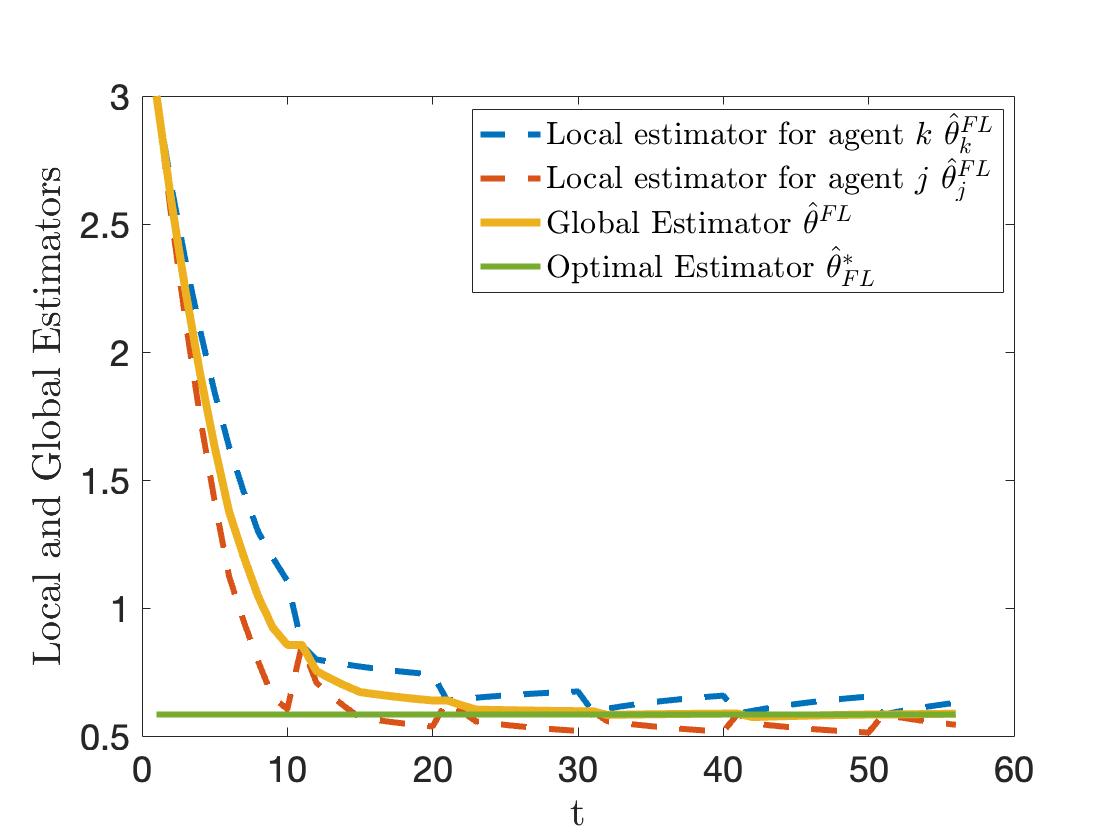}
        \caption{Performance of FL, no data splitting for 2 agents, \\
        with $N_{sync}=6$.}
        \label{fig:NV2}
    \end{subfigure}%
    \begin{subfigure}{.5\linewidth}
        \centering
        \includegraphics[scale=.18]{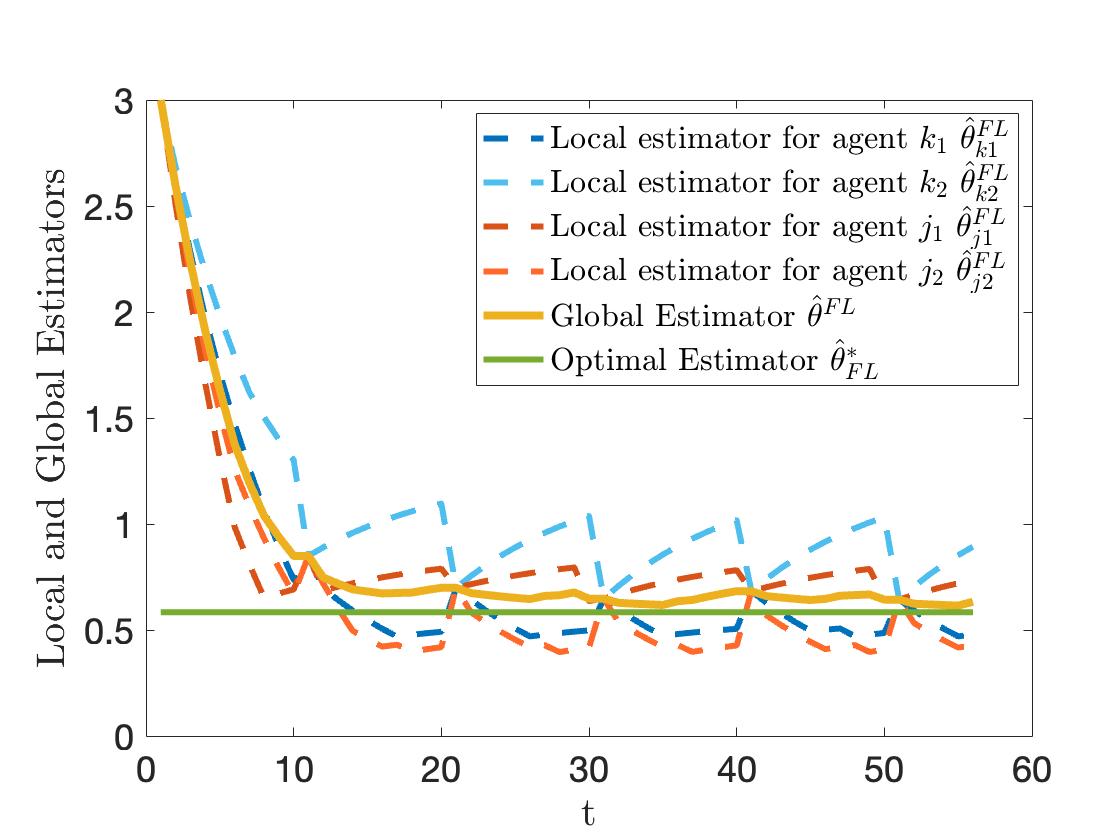}
        \caption{Data splitting, fixing \(N_{\text{sync}}\) the same as Fig.\ref{fig:NV2}, \\
        with $ N_{sync}=6$.}
        \label{fig:NV4_a}
    \end{subfigure}\par\medskip
    \centering
    \begin{subfigure}{\linewidth}
        \centering
        \includegraphics[scale=.18]{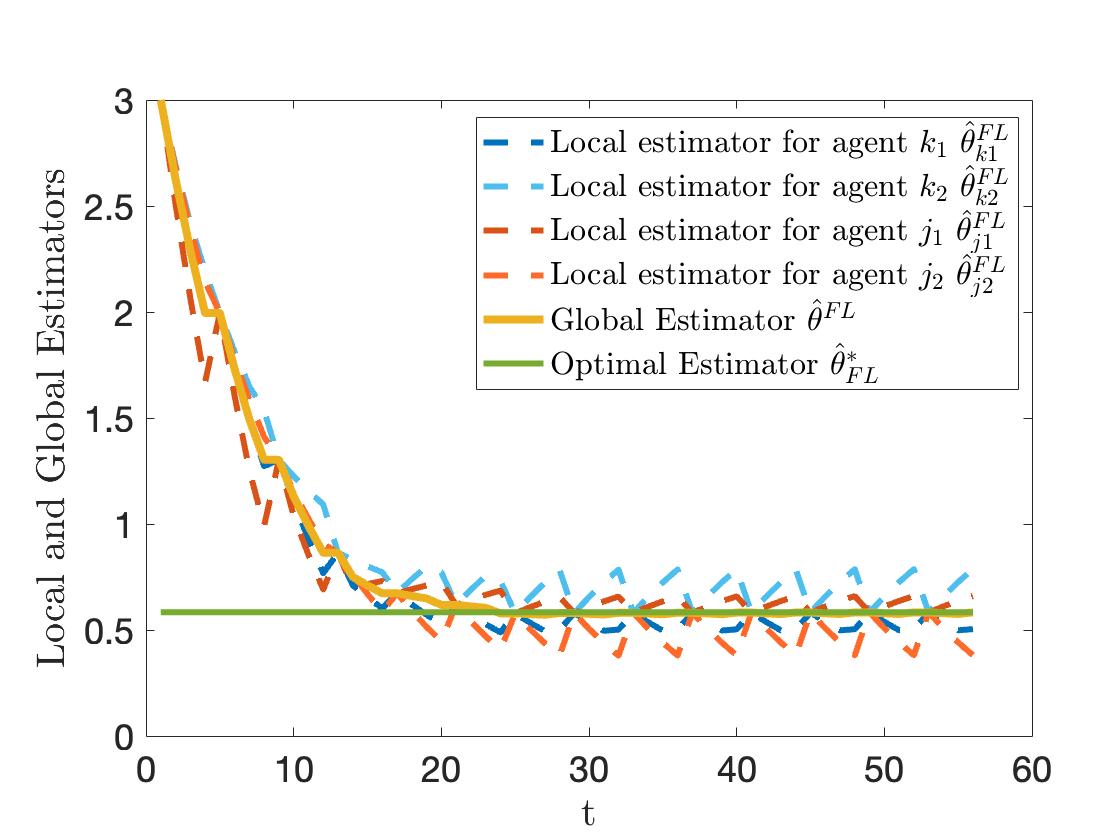}
        \caption{Performance of FL under data splitting, increasing number of synchronization \(N_{\text{sync}}\) such that output \(\hat{\BFtheta}^{FL}\) is within \(10^{-1}\) distance to \(\hat{\BFtheta}^{FL}\) obtained in Fig.\ref{fig:NV2} after fixed \(T\) epochs, with $N_{sync}=19$.}
        \label{fig:NV4_b}
    \end{subfigure}
    \caption{Performance of FL under (potential) data splitting.}
    \label{fig:FLNV}
\end{figure}

\begin{figure}[htb!]
    \centering
    \begin{subfigure}{.5\linewidth}
        \centering
        \includegraphics[scale=.18]{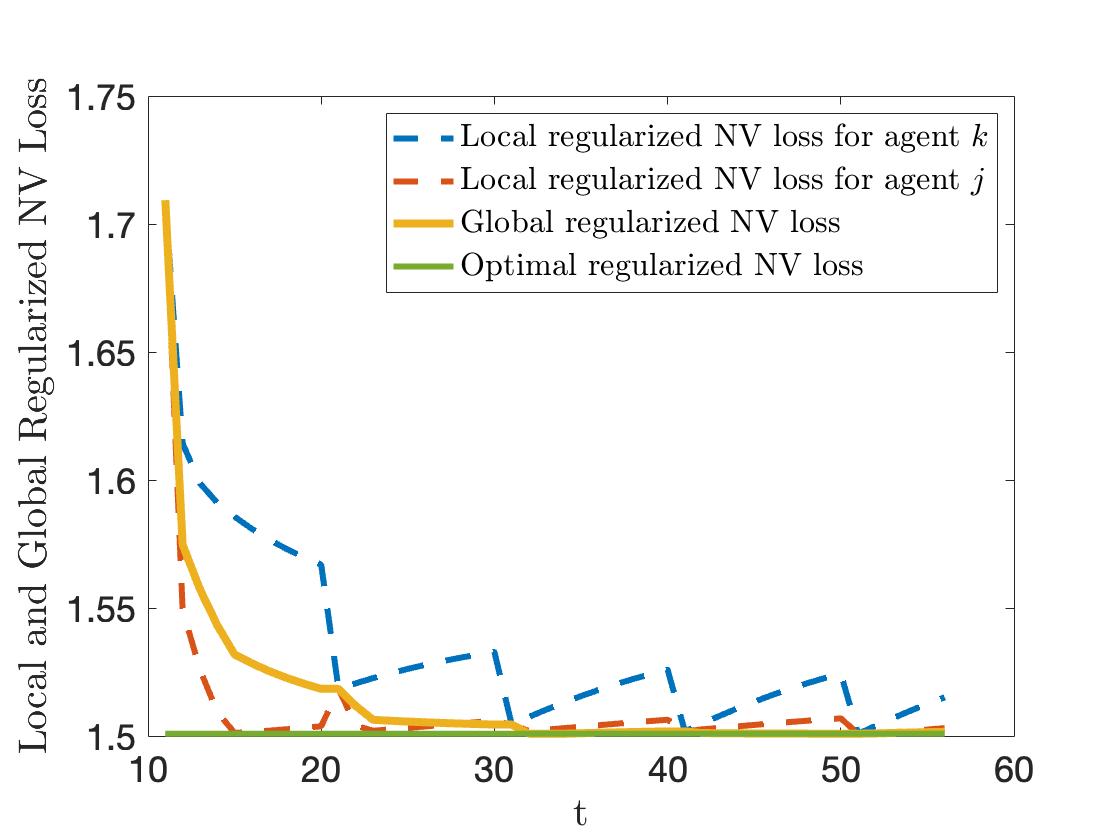}
        \caption{Performance of FL, no data splitting,\\ with $N_{sync}=6$.}
        \label{fig:NVL2}
    \end{subfigure}%
    \begin{subfigure}{.5\linewidth}
        \centering
        \includegraphics[scale=.18]{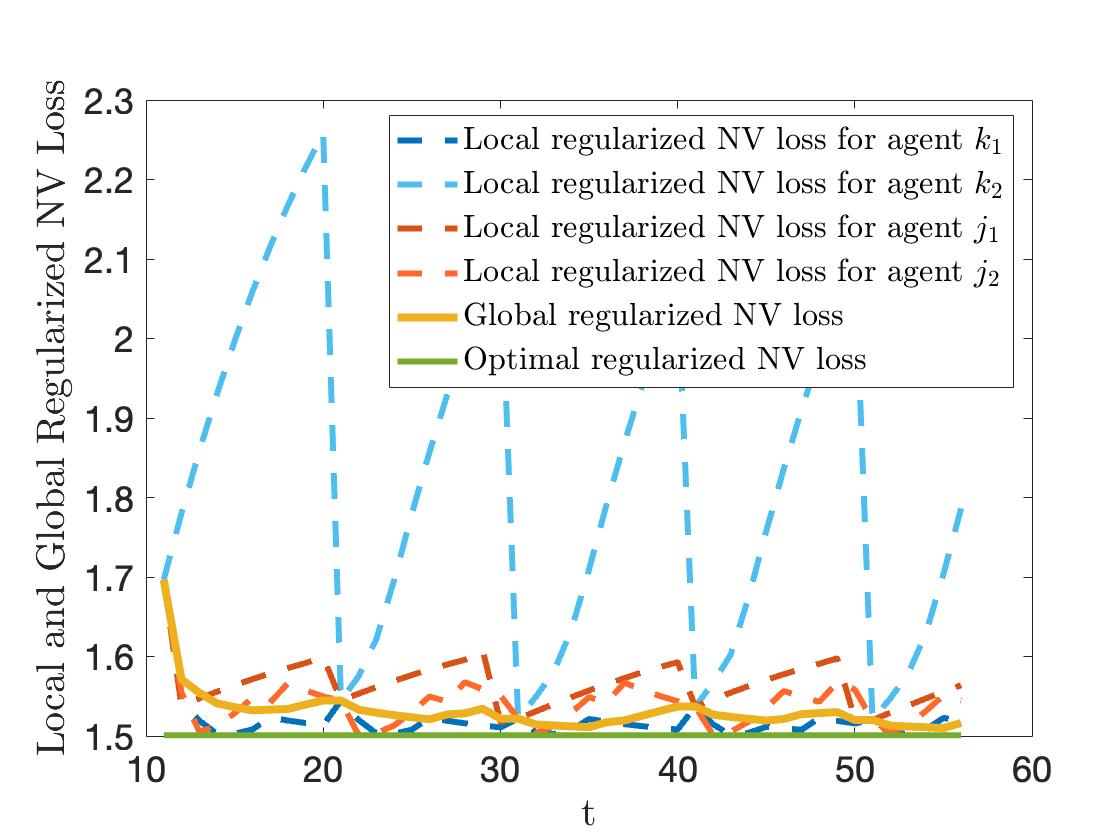}
        \caption{Data splitting, fixing \(N_{\text{sync}}\) the same as Fig.\ref{fig:NVL2},\\ with  $N_{sync}=6$.}
        \label{fig:NVL4_a}
    \end{subfigure}\par\medskip
    \centering
    \begin{subfigure}{\linewidth}
        \centering
        \includegraphics[scale=.18]{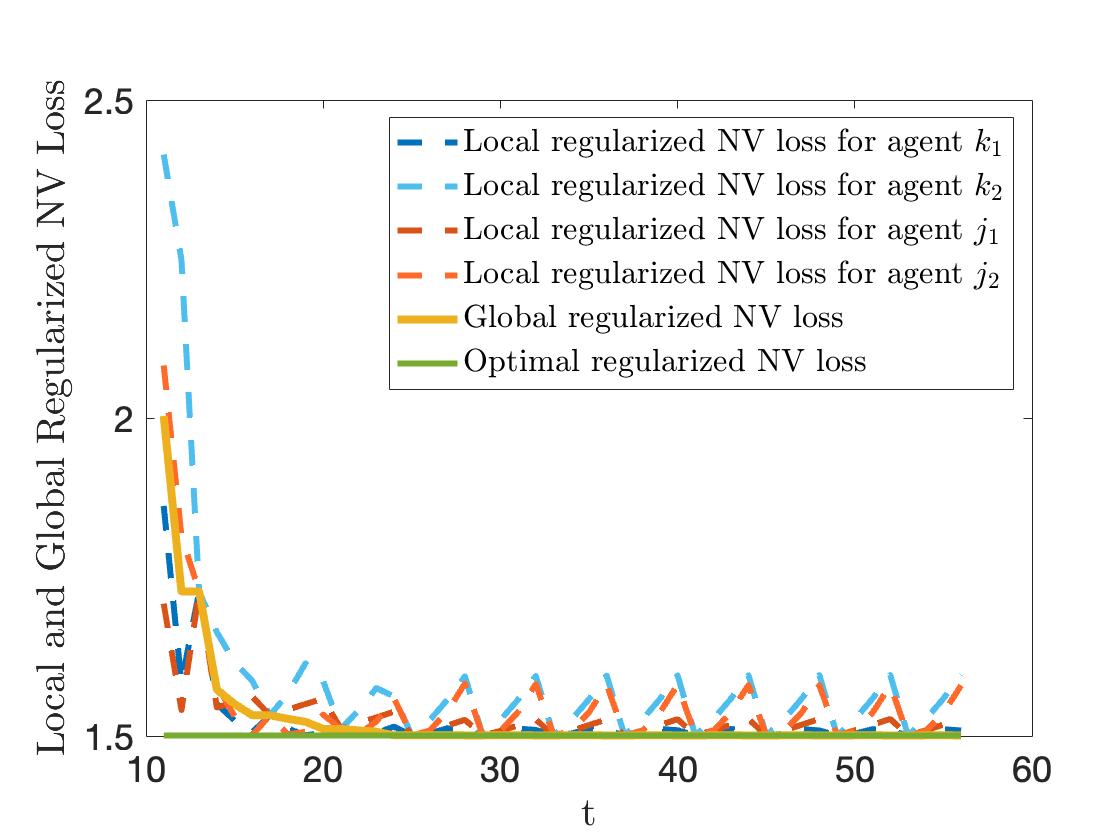}
        \caption{Performance of FL under data splitting, increasing number of synchronization \(N_{\text{sync}}\) such that output \(\hat{\BFtheta}^{FL}\) is within \(10^{-1}\) distance to \(\hat{\BFtheta}^{FL}\) obtained in Fig.\ref{fig:NVL2} after \(T\) epochs, with $N_{sync}=19$.}
        \label{fig:NVL4_b}
    \end{subfigure}
    \caption{Performance of FL under (potential) data splitting.}
    \label{fig:NVLFL}
\end{figure}

Figures \ref{fig:FLNV} and \ref{fig:NVLFL} demonstrate the performance of the estimator learned using FL algorithm. Figure \ref{fig:FLNV} plots the value of the estimator while in Figure \ref{fig:NVLFL} the performance is quantified by the loss function specified in Equation \ref{eq: NV-loss}.  Each plot demonstrates the performance of the global FL estimator, the local estimators of each (fake) agent, compared with the optimal in-sample FL estimator. 
Figures \ref{fig:NV2} and \ref{fig:NVL2} examine the setting without data splitting while Figures \ref{fig:NV4_a} and \ref{fig:NVL4_a} examine the convergence when agent conduct data splitting, under the number of synchronization $N_{sync} =6$ and other FL algorithm hyper-parameters. Comparing these two settings, we observe that when agents create more fake identities and split data among them, the convergence is significantly slower. This is because of the higher variation of each local estimator after splitting. The global estimator is not able to converge within $T=55$ epochs, which prevents the platform  from providing an estimator that satisfies the performance guarantee announced in MCFL.
Therefore, to obtain faster convergence, one has to increase number of synchronization. 
By increasing the number of synchronization from $N_{sync} =6$ to $N_{sync}=19$, the desired performance guarantee can be reached.  Figures \ref{fig:NV4_b} and \ref{fig:NVL4_b} illustrate how increasing the number of synchronization in data splitting setting leads to similar convergence speed with fewer synchronizations when there is no data splitting. However, this convergence comes with a cost since an increasing number of synchronization will lead to significantly higher communication cost.

\subsection{MCFL for Portfolio Optimization}
Following up on Example 2.1 in Appendix \ref{app:example}, we examine the MCFL setting for the risk-averse portfolio optimization problem. Suppose $\BFxi \in \mathbb{R}^d$ denotes the random return of assets, and that $\xi_i = \BFx^T_i\BFtheta^{*}   + N(0, \sigma)$, where $\BFx_i\in R^{1\times p }$ is fixed local feature data.  Here,  $\BFtheta^*\in \mathbb{R}^{p\times 1}$ is the unknown estimator we try to obtain from FL. In the MCFL framework, the agent first collaboratively predict the return of each assets using the mean square error (MSE) loss function $l_{MSE}(\BFxi, \BFxi'):= \|\BFxi- \BFxi'\|^2$ for any $\BFxi$, $\BFxi'$, with the following FL learning objective: \myeq{
\min_\BFtheta L(\BFtheta) = \sum_{k=1}^K \sum_{\BFy_j\in S_{k}} l_{MSE}(\BFx^T_j\BFtheta, \BFxi_j)}, and 
\myeq{
\hat{\BFtheta}^*_{FL}=\arg\min_\BFtheta L(\BFtheta).}
Then with the estimated $\hat{\BFtheta}^{FL}$, each agent first predicts the return of assets $\hat{\BFxi} = \BFx^T\hat{\BFtheta}^{FL} $ for any given $\BFx$. Then make investment decision $\BFw\in \mathbb{R}^{d}$ and $w_0\in \mathbb{R}$ to maximize the return of assets  by solving the following portfolio optimization problem
\begin{align}
\max_{\mathbf{w}} & \quad \alpha (\sum_{l=1}^d w^l \hat{\BFxi} ^l -w_0)^2 -\sum_{l=1}^d w^l \hat{\BFxi} ^l \label{eq: portfolio-problem}\\
s.t. &\quad  \sum_lw^l+w_0=1,\nonumber\\
&\quad  w_l,w_0\geq0. \nonumber
\end{align}
where $w^l$ and $\xi^l$ denote the $l$-th component.

Similar to the newsvendor example, we still consider a toy model with two agents, \(k\) and \(j\), who might each create a fake identity and evenly split their data. Each agent holds 8 data samples, similar to the setting in Section \ref{sec: num_nv}. The samples $\{x_i\}_{i=1}^n$ are generated following i.i.d. standard normal distribution while \(\theta^*=1\) and \(\varepsilon\sim \mathcal{N}(0,\sigma =0.01)\). We set \(\alpha=\tfrac{1}{2}\) and the FL parameters \(\Phi=\{K^{FL}=K, \rho=0.1, \hat{\BFtheta}^{FL}_0=2, T = 55\}\).

\begin{figure}[htb!]
    \centering
    \begin{subfigure}{.5\linewidth}
        \centering
        \includegraphics[scale=.18]{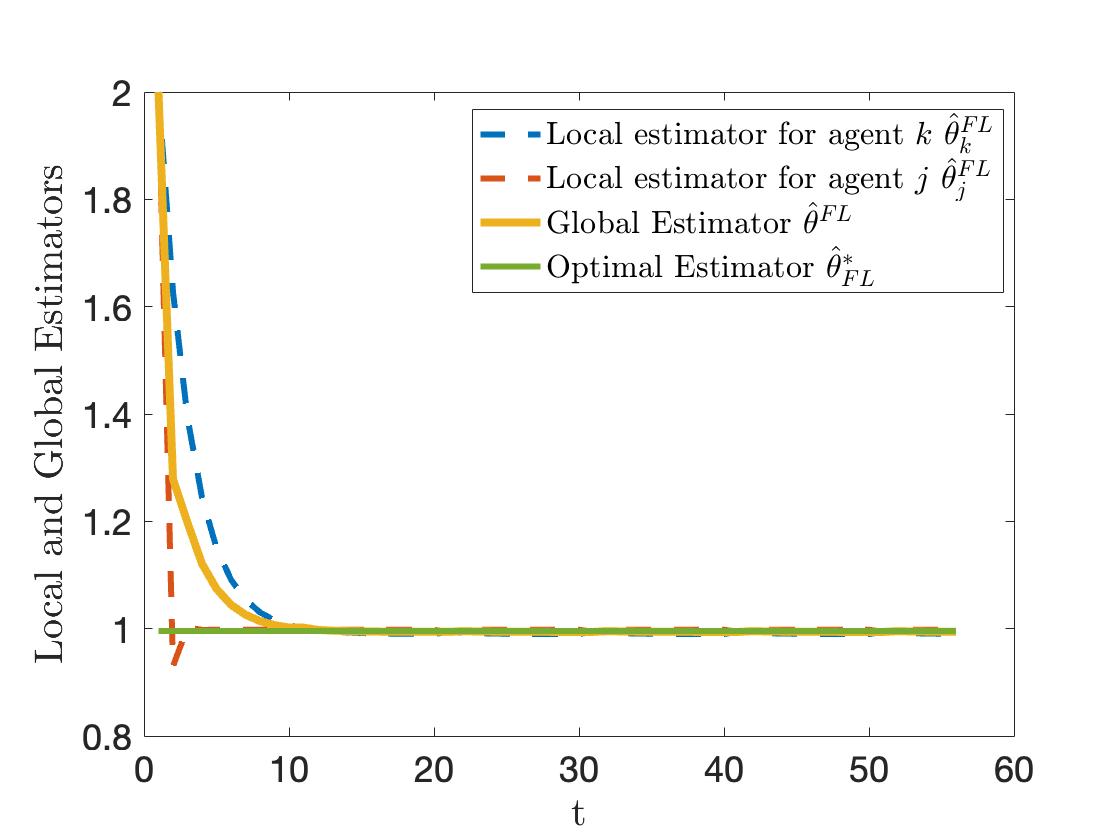}
        \caption{Performance of FL, no data splitting,\\ with $ N_{sync}=6$.}
        \label{fig:p2}
    \end{subfigure}%
    \begin{subfigure}{.5\linewidth}
        \centering
        \includegraphics[scale=.18]{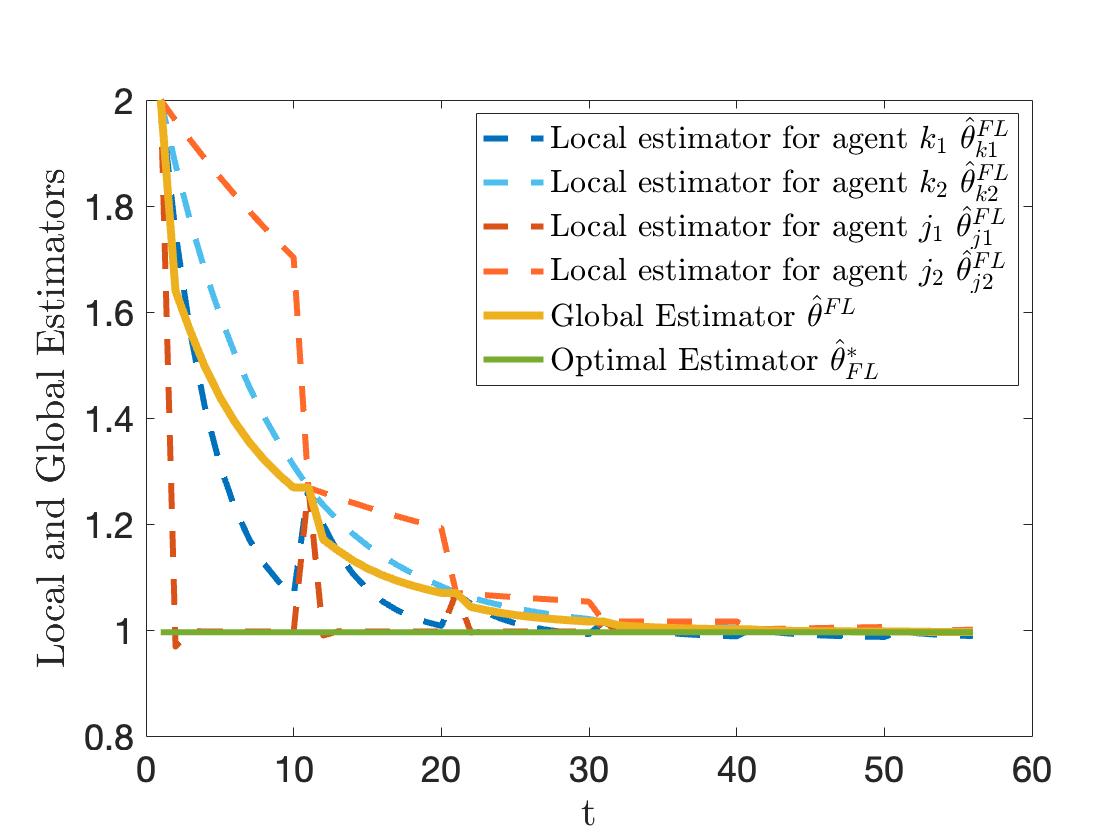}
        \caption{Data splitting, fixing \(N_{\text{sync}}\) the same as Fig.\ref{fig:p2},\\ with $ N_{sync}=6$.}
        \label{fig:p4_a}
    \end{subfigure}\par\medskip
    \centering
    \begin{subfigure}{\linewidth}
        \centering
        \includegraphics[scale=.18]{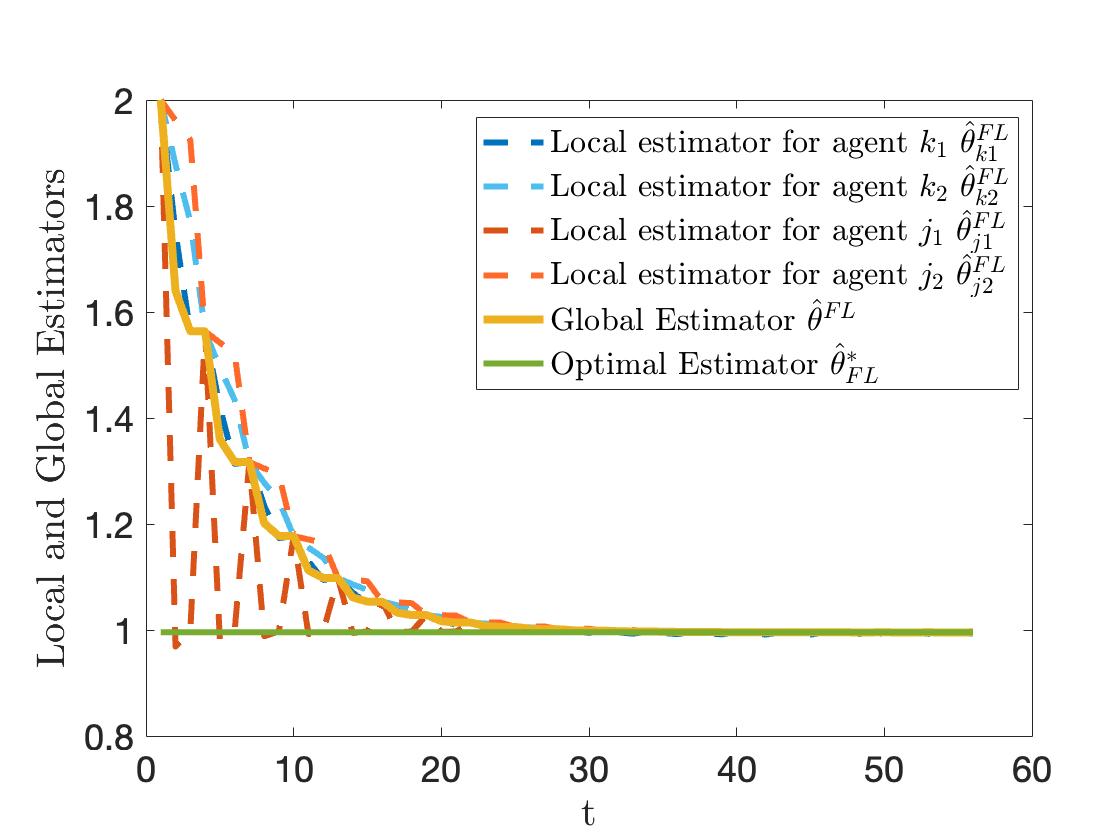}
        \caption{Performance of FL under data splitting, increasing number of synchronization \(N_{\text{sync}}\) such that output \(\hat{\theta}^{FL}\) is within \(10^{-3}\) distance to \(\hat{\theta}^{FL}\) obtained in Fig.\ref{fig:p2} after \(T\) epochs,  with $ N_{sync}=19$.}
        \label{fig:p4_b}
    \end{subfigure}
    \caption{Performance of FL under (potential) data splitting.}
    \label{fig:FL_portfolio}
\end{figure}

\begin{figure}[htb!]
    \centering
    \begin{subfigure}{.5\linewidth}
        \centering
        \includegraphics[scale=.18]{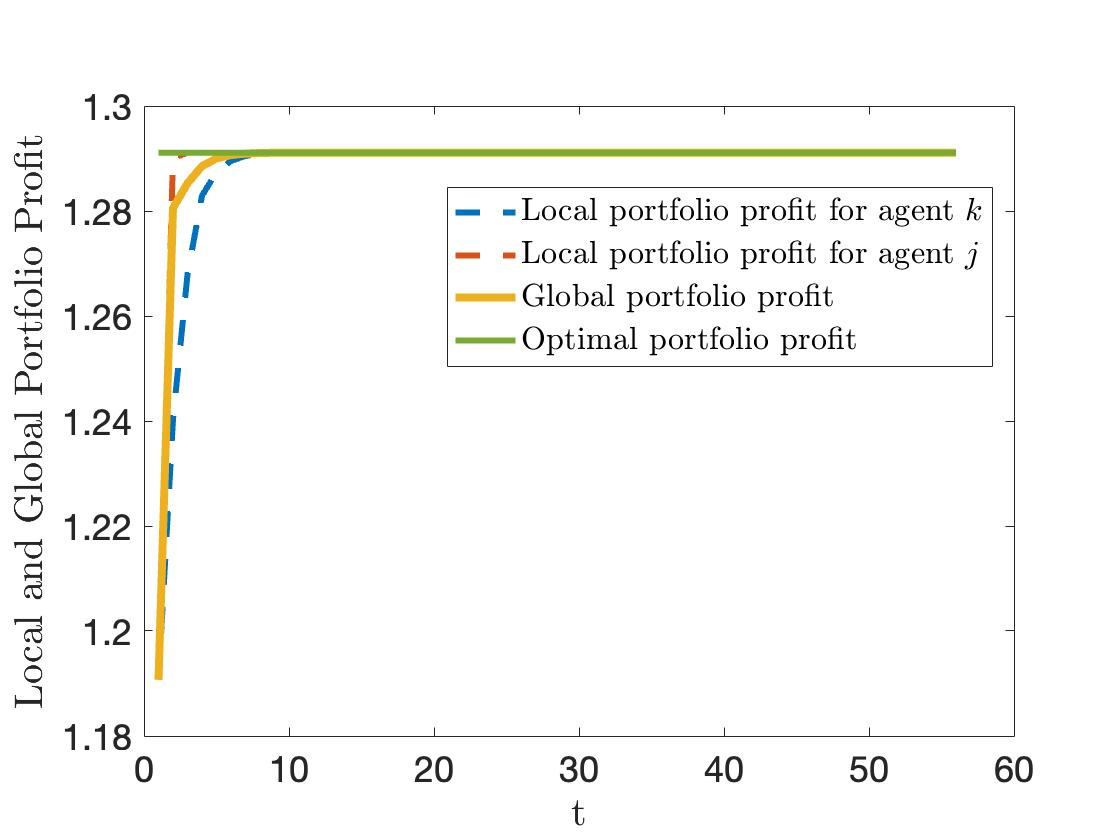}
        \caption{Performance of FL, no data splitting,\\ with $N_{sync}=6$.}
        \label{fig:p2_profit}
    \end{subfigure}%
    \begin{subfigure}{.5\linewidth}
        \centering
        \includegraphics[scale=.18]{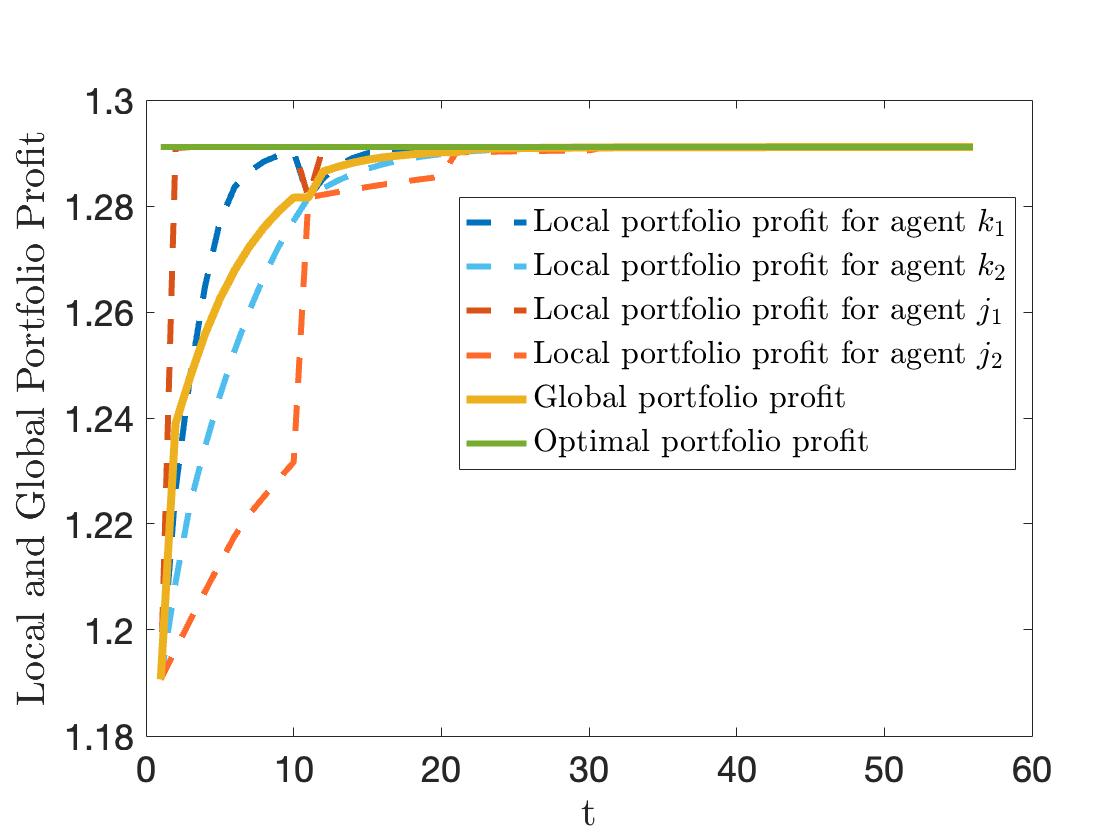}
        \caption{Data splitting, fixing \(N_{\text{sync}}\) the same as Fig.\ref{fig:p2_profit},\\ with $ N_{sync}=6$.}
        \label{fig:p4_a_profit}
    \end{subfigure}\par\medskip
    \centering
    \begin{subfigure}{\linewidth}
        \centering
        \includegraphics[scale=.18]{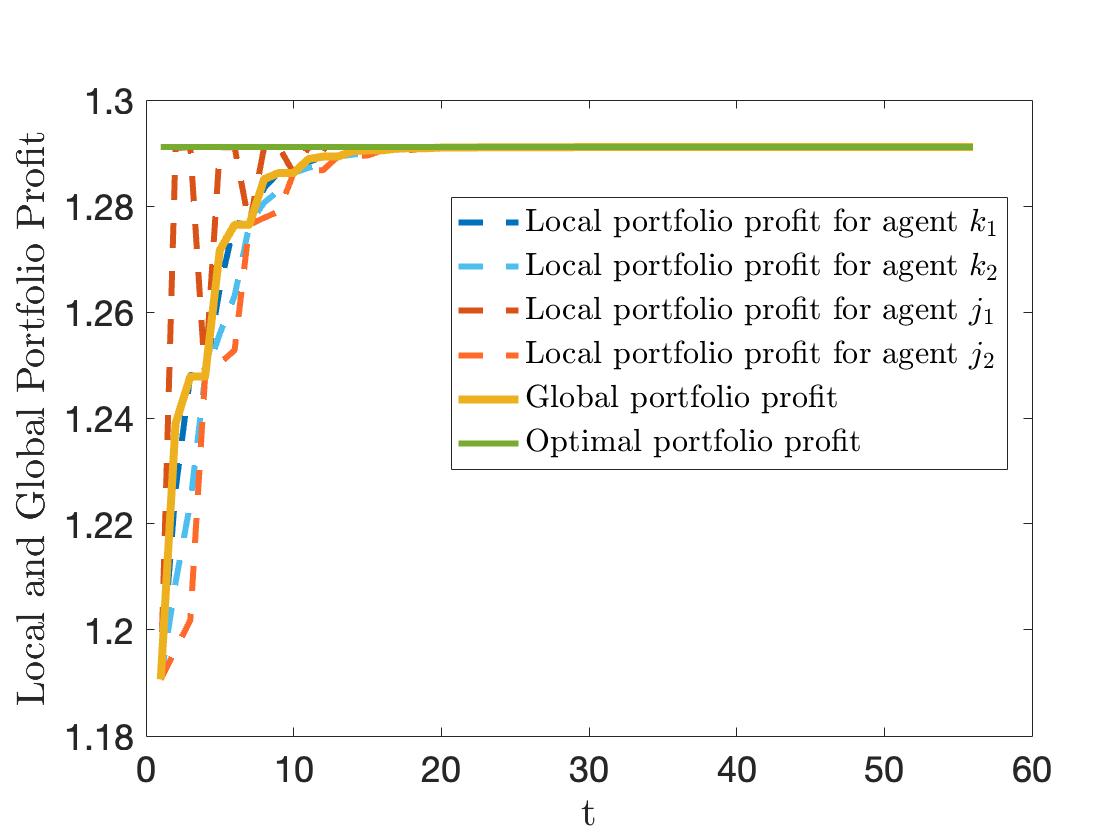}
        \caption{Performance of FL under data splitting, increasing number of synchronization \(N_{\text{sync}}\) such that output \(\hat{\theta}^{FL}\) is within \(10^{-3}\) distance to \(\hat{\theta}^{FL}\) obtained in Fig.\ref{fig:p2_profit} after \(T\) epochs, with $N_{sync}=19$.}
        \label{fig:p4_b_profit}
    \end{subfigure}
    \caption{Performance of FL under (potential) data splitting.}
    \label{fig:FL_portfolio_profit}
\end{figure}

Similarly, Figures \ref{fig:FL_portfolio} and \ref{fig:FL_portfolio_profit}  demonstrate the performance of the estimators. Figure \ref{fig:FL_portfolio} shows the value of the estimator while   \ref{fig:FL_portfolio_profit} quantifies the performance using the optimal portfolio return obtained after solving (\ref{eq: portfolio-problem}). We observe that data splitting directly leads to a potential decrease in the profits that the agents would gain. Each plot demonstrates the performance of the global FL estimator, the local estimators of each (fake) agent, compared with the optimal in-sample FL estimator. 
Figures \ref{fig:p2} and \ref{fig:p2_profit} examine the setting without data splitting while Figures \ref{fig:p4_a} and \ref{fig:p4_a_profit} examine the convergence when agents conduct data splitting, under the same number of synchronization $N_{sync} =6$ and other FL algorithm hyper-parameters. Comparing these two settings, we again observe that when agents create more fake identities and split data among them, the convergence is significantly slower and the global estimator is not able to converge within $T=55$ epochs.
By increasing the number of synchronization from $N_{sync} =6$ to $N_{sync}=19$, the desired performance guarantee can be reached, as demonstrated in Figures \ref{fig:p4_b} and \ref{fig:p4_b_profit}.

\subsection{MCFL for Healthcare Application with Real-World Data}\label{subsec:mcfl-real}
In this section, we examine the MCFL framework in the case of dermatoglyphics skin lesion detection. We consider \texttt{Fed-ISIC2019} benchmark with $23{,}247$ dermatoscopic images partitioned across \textbf{six} hospitals/data centers. In total, the datasets contain \(12{,}413\), \(3{,}954\), \(3{,}363\), \(2{,}259\), \(819\), and \(439\) samples for each hospital respectively. The task is multi-class lesion diagnosis, with 8 classes of skin  lesions detection \citep{flambyFedISIC2019,flwrDatasets,ham10000,isic2017,bcn20000}. We use a federated image-classification pipeline:  \texttt{Flower} with more details provided in Appendix \ref{app:flwr-parameter}. In the experiment, we fix the total number of training epochs, $T=48$ and experiment with various number of synchronizations $N_{sync}=6, 8, 12, 24, 48$. We consider each agent splitting into $2$ or $10$ identities to examine different system efficiency outcome under different agent splitting actions.

To characterize the MCFL system efficiency (\ref{eq : system_efficiency}), we first specify the learning surplus.
In the lesion diagnosis setting, we set the coalition learning surplus to be a function of the accuracy of the diagnosis. Specifically, let $\mathrm{Acc}^{FL}(\BFtau_{\mathcal{A}})\in[0,1]$ denote the accuracy of the MCFL learning outcome for skin lesions classification. Note that this learning outcome is a function of participation profile $\BFtau_{\mathcal{A}}$.  Following this,  we model the MCFL coalition surplus as 
\(
v(\BFtau_{\mathcal{A}}) \;=\; r \cdot N \cdot \bigl(\mathrm{Acc}^{FL}(\BFtau_{\mathcal{A}}) - \mathrm{Acc}_{\text{base}}\bigr),
\)
where $N$ is the number of skin-lesion cases handled per month and the baseline accuracy is set as $\mathrm{Acc}_{\text{base}}=0.50$. We let $r$ denote the marginal benefit (in $\$$) per accurate case and estimate it using the average cost of a benign skin lesion exam, which is \$18.56 USD according to \cite{anderson2021cost}. This is because that higher acuracy of the diagnosis reduces unnecessary follow-ups and avoids missed malignant lesions. We proceed with the assumption that, one accurate diagnosis saves a cost equivalent to $20\%$ of one benign skin lesion exam cost, thus we set $r=3.712$. We further set $N=100$, assuming $5$ skin tests received by a healthcare center per day and a total of $20$ working days in a month. 

Note that the second component of system efficiency consists of the FL learning cost. We consider a linear cost of the number of synchronization \( c N_{sync}\)
where $c$ (in \$ per round) aggregates communication, agent compute, and orchestration costs per aggregation round. Note that this is an underestimate of the communication cost, as it is not directly multiplied by the number of (fake) agents in agent participation equilibrium. Instead, it implicitly depends on the number of agents through the number of synchronization required to reach a certain accuracy. While the exact cost per synchronization in FL training process might be hard to determine and differ from case to case, we refer to the current pricing scheme of FL \citep{FedMLPricing2025} which indicates that \(c\;\approx\; 1.584 \ \$/\text{round}.\)

Then we run each simulation for 20 repetitions and report the mean and standard deviation of three metrics: the diagnostic accuracy of the MCFL learning outcome, the cross-entropy loss function which is related to the likelihood of the diagnosis classification, and the system efficiency $\Pi^{\text{sys}}$, specified in (\ref{eq : system_efficiency}) and evaluated using the above process. We examine different values of $K$: $K=6$ denotes no splitting and no fake identity created,  $K=12$ represents the setting where each agent split to two identities, and finally $K=60$ denotes the setting where each agent splits to 10 identities. In each setting, we investigate the MCFL outcomes under different numbers of synchronization $N_{\text{sync}} = 6, 12, 48$. Table~\ref{tab:results} summarizes the results.

\begin{table}[h!]
\centering
\caption{Simulation results: average accuracy, loss, and system surplus $\Pi^{Sys}$ over 20 runs.}
\label{tab:results}
\begin{tabular}{l c c c c}
\toprule
$K$ & $N_{\text{sync}}$ & Accuracy (mean $\pm$ std) & Cross-Entropy Loss (mean $\pm$ std) & $\Pi^{Sys}$ \\
\midrule
6  &  6  & 0.7259 $\pm$ 0.0191 & 1.1278 $\pm$ 0.0881 & 74.35 \\
\midrule
12 &  6  & 0.6275 $\pm$ 0.0189 & 1.4059 $\pm$ 0.1153 & 37.82 \\
12 & 12  & 0.6336 $\pm$ 0.0186 & 1.1403 $\pm$ 0.0606 & 30.58 \\
12 & 48  & 0.5929 $\pm$ 0.0127 & 1.1848 $\pm$ 0.0294 & -41.55 \\
\midrule
60 &  6  & 0.5428 $\pm$ 0.0140 & 1.4066 $\pm$ 0.0581 & 6.38 \\
60 & 12  & 0.5501 $\pm$ 0.0102 & 1.2533 $\pm$ 0.0192 & -0.41 \\
60 & 48  & 0.5584 $\pm$ 0.0097 & 1.2851 $\pm$ 0.0247 & -54.35 \\
\bottomrule
\end{tabular}
\end{table}

According to the results summarized in Table~\ref{tab:results}, we observe the following findings. First, under the same number of synchronization $N_{\text{sync}} =6$, splitting data among fake identities consistently reduces the learning outcome, as indicated by accuracy and cross-entropy loss. 
Table~\ref{tab:results} illustrates two key findings.  
Second, this performance loss cannot be fully mitigated by increasing the number of synchronizations in practice. Specifically, in practice, instead of full-batch local gradient descent (GD), platform often adopts stochastic gradient descent (SGD) training for neural networks with FL, because local full-batch GD is computationally infeasible, and federated learning must rely on SGD. In addition, when applying SGD in FL, the effect of synchronization is non-monotone: excessive rounds of synchronization may even harm convergence, as documented in \citet{woodworth2020local} (e.g., in Table \ref{tab:results}, when $K=12$, increasing $N_{sync}$ from $12$ to $48$ even hurts overall accuracy). Consequently, while the earlier toy examples suggest that the platform can improve convergence accuracy by incurring the cost of more frequent synchronization in local gradient descent, the same does not necessarily hold for SGD: even with additional synchronizations, the adverse effect of data splitting may not be fully mitigated. This suggests that data splitting among fake identities might introduce a structural inefficiency that cannot be remedied simply by adjusting the synchronization frequency.
\section{Extensions}
\label{sec: extensions}
\subsection{The Second Pitfall of SV in Maximizing Social Surplus}
\label{sec: extensions_social_surplus}

In previous sections, we focused on a platform whose objective was to maximize the coalition’s learning surplus and to redistribute this surplus fairly, while overlooking the cost of FL training. In this section, we instead consider a platform that values both the learning surplus and the FL training cost, and show that Shapley value still suffers from an additional pitfall: it induces data splitting even when the platform adopts this more comprehensive objective.

Formally, we assume that the platform maximizes social efficiency, which we define as
\myeql{\label{eq : social_efficiency}
\textbf{Social Efficiency}\quad
\Pi^{\text{social}}= v\bigl(|\boldsymbol{\tau}_{\mathcal{A}}|\bigr) - c N_{sync}.
}

Social efficiency refers to the total surplus generated under the MCFL framework, minus the overall cost of training. Unlike system efficiency, it does not account for potential inefficiencies arising from how the surplus is distributed among agents under different mechanisms. Social efficiency focuses on the overall coalition surplus, taking into account the cost of FL training. Hence, under this broader notion of efficiency, allocation efficiency is no longer important. 

It is clear that, under the full participation requirement, compared to SV, MC leads to higher social efficiency once training cost is taken into account. This is because, as both SV and MC satisfy the desirable property of encouraging full participation, the surplus gained from learning remains the same across mechanisms. However, since MC leads to lower training cost compared to SV, it leads to higher social efficiency.

However, in order to maximize social efficiency, full participation is no longer necessarily desirable from a policy-maker perspective to maximize social efficiency. In fact, increasing the size of the coalition can raise training costs to the point where total system efficiency deteriorates. This leads us to explore two key questions: (1) What is the optimal participation profile to maximize social efficiency $\Pi^{social}(\delta_0,\Phi,\mathcal{M})$, given $K$ players, each with $m_k$ samples? (2) Does there exist a mechanism that incentivizes such an optimal participation profile?

The following proposition states the optimal participation profile under mild assumptions, and we also show that the marginal contribution mechanism, while not efficient for participants' surplus, indeed maximizes the social surplus. 

 \begin{proposition}
  \label{prop_opt_participation}
 Suppose that the coalition surplus $v(|\BFtau_{\mathcal{A}}|)$ is strictly increasing and concave in $|\BFtau_{\mathcal{A}}|$, and
 the cost of synchronization $cN_{\text{sync}}(|\mathcal{A}|):=C|\mathcal{A}|^{3/4}$ for some constant $C>0$. Then:
  \begin{enumerate}
  \item[(i)] The optimal participation profile that maximizes social surplus $\Pi^{\text{social}} = v(|\BFtau_{\mathcal{A}}|) -
  C|\mathcal{A}|^{3/4}$ follows a threshold policy: there exists $m^*$ such that it is optimal for all agents with $m_k \geq m^*$ to
  participate with full data provision $\tau_k = m_k$, and for all agents with $m_k < m^*$ to abstain.

  \item[(ii)] The marginal contribution mechanism with characteristic function $v^{S}(\BFtau_{\mathcal{A}}) = v(|\BFtau_{\mathcal{A}}|) -
  C|\mathcal{A}|^{3/4}$ induces the optimal participation profile as an equilibrium.

  \item[(iii)] The Shapley value mechanism under the same characteristic function $v^S$ leads to data splitting.
  \end{enumerate}
  \end{proposition}

Under a social welfare objective, the central question becomes identifying the optimal participation profile that maximizes overall social efficiency—that is, the net gain from learning after accounting for training costs. When the coalition surplus is a concave function of the number of data samples—as is typically the case in collaborative learning, SV-based mechanisms inherently incentivize a certain degree of data splitting. This tendency leads to inflated training costs and represents another critical pitfall of SV.

In contrast, the MC-based mechanism can induce more socially efficient outcomes. Specifically, agents are discouraged from splitting their data, and participation is limited to those whose inclusion increases net social efficiency. Agents with insufficient data, whose participation would raise training costs without contributing significantly to the surplus, are incentivized to opt out voluntarily under the MC-based mechanism.

\subsection{Comparison with Anonymity-proof Shapley Value}
\label{sec: compare_falsename}

Possible misconduct through duplicate identities is similar to the well-known false-name manipulation in voting \citep{conitzer2010using}. However, it is distinct from the typical false-name manipulation that the misconduct does not only involve a duplication of identity, but also the splitting of the possessed data across the identities. Moreover, it is also identified in the federated learning literature that additional massive number of agents could be harmful to the FL algorithm \citep{fung2020limitations}.

Anonymity-proof SV ensures that a participant's Shapley-based payoff is invariant to relabeling or hiding behind multiple anonymous identities, which blocks false-name manipulation (creation of additional ``agents") and is conceptually related but distinct from data split incentives, where a single real participant strategically partitions its dataset into multiple pseudo-clients to alter marginal contributions.
To further compare our MCFL framework with the anonymity-proof Shapley value proposed in (\cite{ohta2008anonymity}), 
the anonymity-proof Shapley value, when applied to the MCFL framework, treats each agent \textit{data sample $\BFy_j$} as an individual skill, and the mechanism announces $\pi(\BFy_j, S_{\BFtau_{\mathcal{A}}})$ for all data samples $\BFy_j$ and data samples $S_{\BFtau_{\mathcal{A}}}$ from the coalition $\mathcal{A}$ with the participation decision profile $\BFtau_{\mathcal{A}}$ where $\BFy_j\in S_{\BFtau_{\mathcal{A}}}$. And the payoff to the agent $k$ who contributes data samples $S_k$ is determined as $\sum_{\BFy_j\in S_k} \pi(\BFy_j, S_{\BFtau_{\mathcal{A}}})$. Under this formulation, the false name manipulation no longer becomes a concern since it does not change the payoffs to the skills.

However, there are potential issues in directly applying the anonymity-proof Shapley value mechanism to the MCFL framework. 

The first concern is the computational complexity. It is well-known that, even for the plain SV mechanism, it already has high computation complexity. The anonymity-proof Shapley value is given by 
$\pi(\BFy_j, S_{\BFtau_{\mathcal{A}}})=v(S_{\BFtau_{\mathcal{A}}})\frac{sh_v(\BFy_j, T)}{\sum_{\BFy_i\in S_{\BFtau_{\mathcal{A}}}}sh_v(\BFy_i, T)}$, where $sh_v(\BFy_j, T)$ is the Shapley value of data sample $\BFy_j$ in the grand coalition $T=\cup_{k} S_k$. 
Hence the complexity of anonymity-proof Shapley value is $O(|\BFm|2^{|\BFm|})$ \citep{ohta2008anonymity, conitzer2010using},%
and the complexity of calculating our algorithm is $O(\Pi_{k}m_k|\BFm|2^{K-1})$. 
This is because to directly apply the anonymity-proof Shapley value, one has to treat each data sample as an individual skill, thus significantly increases the dimension. Note that in the collaborative learning setting, the number of (real) agents is often much smaller than the total number of samples all agents possessed, i.e., $K<<|\BFm|$.  Using our proposed MCFL framework reduces the exponential dependency from $2^{|\BFm|}$ to $2^K$. 

Another concern is the additivity axiom. Our proposed MCFL framework satisfies the additivity axiom (Axiom \ref{axiom: additivity}), which requires that, for two different characteristic functions $v$ and $u$, $\psi_{i,k}(v+u) = \psi_{i,k}(u)+ \psi_{i,k}(v)$, for all $i, k$. 
Recall that in the MCFL setting, the value function is defined as $v(\BFtau_{\mathcal{A}}) = \pi(\BFw^*(\hat{\BFtheta}_{\BFtau_{\mathcal{A}}}), \BFtheta^\ast)$, where $\pi(\BFw, F_{\BFtheta^\ast}) := \mathbb{E}_{\BFy \sim F_{\BFtheta^\ast}} [r(\BFw,y)]$. The revenue function $r$ might involve two components, denoted as $r_1$ and $r_2$, where $r = r_1 + r_2$. A desired property of the allocation mechanism is to generate consistent payoff, when the two components are considered separately and as a whole. However, the anonymity-proof Shapley value only guarantees this consistency conditioned on the grand coalition but not other coalition and such inconsistency can generate persistent disputes over surplus division, complicating the collaborative learning process.

Therefore, we believe that the anonymity-proof Shapley value, although also immune to dishonest data splitting behavior, is not applicable to the MCFL setting, due to the computational complexity and its failure to satisfy the additivity axiom.

\section{Conclusions}
In conclusion, our study has shed light on the intricate dynamics of collaborative learning in multi-agent systems through the lens of FL technology. By establishing a comprehensive Multi-action Collaborative Federated Learning (MCFL) framework, we study the pitfall of Shapley value in FL mechanism design. Our investigation reveals that while Shapley value based mechanism ensures fair and efficient allocation and guarantees quality decisions among agents through encouraging full participation, they inadvertently introduce significant communication costs during the FL process due to the agents' misconduct through data splitting, highlighting a crucial trade-off between decision quality and operational efficiency. This discovery not only addresses a gap in existing research, but also opens new avenues for exploring mechanism design that balances decision quality with the practicalities of implementation in FL optimization.

Moreover, our work stands as a pioneering effort to systematically explore the interplay between mechanism design and FL performance, offering valuable insights for both theoreticians and practitioners interested in optimizing collaborative learning settings. The identification of Shapley value mechanisms' limitations further enriches the studies in collaborative learning, prompting a re-evaluation of widely accepted practices and encouraging the development of more efficient, cost-effective solutions. 

Several promising directions are as follows. First, extend the MCFL framework to settings where the coalition value $v(\cdot)$ is not strictly concave, such as vertical or hybrid FL in which participants provide complementary feature sets and the characteristic function may exhibit synergy (potentially convex with increasing marginal returns). In this regime, both SV and MC may induce data splitting, so a direct question is whether it is possible to design fair allocation mechanisms that are robust to data splitting under synergistic values. Second, characterize business-specific objective functions arising from concrete operational decisions (e.g. newsvendor problem, portfolio optimization, pricing) and study how their structure affects FL adoption and performance within the MCFL framework.



\bibliographystyle{informs2014} 



\bibliography{bibliography}

%
%
%
\newpage
\begin{APPENDICES}
\newpage 
\section{Examples of Collaborative Learning in Business}
\label{app:example}

Here, we present two examples of collaborative learning in specific business contexts. 

\begin{manualexample}{1.1}
{\textup{\textbf{Collaborative Newsvendor Optimization in Customer-to-Manufacturer (C2M) Systems: Background}}} 

The Customer-to-Manufacturer (C2M) model connects consumers and manufacturers through digital platforms \citep{mak2021triple}. Platforms like Amazon, Walmart, and JD.com collect granular data to improve supply chain decisions \citep{qi2020data}, and some have begun sharing such data with retailers to enhance coordination \citep{Masters2019, arora2023data}. In a Customer-to-Manufacturer (C2M) system, agents are retailers, each holding local inventory and demand data. Their shared goal is to optimize ordering decisions under demand uncertainty. However, regulatory constraints limit direct data sharing. To overcome this, the platform enables collaborative, data-driven optimization via federated learning, allowing joint demand forecasting and better ordering decision-making while preserving data privacy. \end{manualexample}

\begin{manualexample}{2.1}{\textup{\textbf{ Collaborative Portfolio Optimization in Online Financial Services: Background}}}

FL technology has already been adopted in collaborative learning in online financial services. For example, FedAI, an AI group from WeBank, leverages federated
learning frameworks to enhance financial services like risk management and anti-money laundering through collaborative learning across multiple small and micro enterprises \citep{fedai_fate_2024_aml,fedai_fate_2024}. Specifically, in collaborative portfolio optimization, financial agents seek to improve portfolio recommendations by utilizing diverse investment return data. However, direct data sharing is constrained by regulatory and competitive concerns. To address this, platforms such as WeBank employ federated learning (FL) to enable joint learning across financial agents while preserving data privacy.
\end{manualexample}

\newpage
\section{Justification for Assumption \ref{assump:pi}} 
\label{app:pi}

In this section, we consider the data-driven decision-making objectives, which are prevalent in business applications. If agents face a common decision-making problem under uncertainty, the performance measure  $\pi(\hat{\boldsymbol{\theta}}, \boldsymbol{\theta}^*)$ is the business objective, and is determined by a decision variable $\boldsymbol{w}$ and a random outcome $\BFy \in \mathcal{Y}$ whose distribution depends on observed contextual information $\boldsymbol{x} \in \mathbb{R}^p$ and an unknown true parameter $\boldsymbol{\theta}^* \in \Theta$. The agent solves the following contextual data driven optimization problem:
\[
\max_{\boldsymbol{w} \in \mathcal{C}} \ R(\boldsymbol{w}, \BFy) := \mathbb{E}_{\BFy\sim F_{\BFtheta^*,\BFx} } \left[ r(\boldsymbol{w}, \BFy) \right|\BFx],
\]
where $\boldsymbol{w} \in \mathcal{C} \subseteq \mathbb{R}^d$ is the decision variable constrained to a feasible set $\mathcal{C}$, and  $r: \mathcal{C} \times \mathcal{Y} \rightarrow \mathbb{R}$ is the objective/reward function. The expectation above is with respect to the conditional distribution of $\BFy$ given $\BFx$, under true model $h_{\BFtheta^*}$. We denote such a conditional distribution of $\BFy$ as $F_{\BFtheta^*,\BFx}$, which is the true data generating process.

Given an estimator $\hat{\boldsymbol{\theta}}$ and contextual information $\BFx$, let $F_{\hat{\BFtheta},\BFx}$ denote the conditional distribution of $\BFy$ given $\BFx$ under the estimated model $h_{\hat{\BFtheta}}$, the agent makes the following data-driven decision : 
\[
\boldsymbol{w}^*(\hat{\boldsymbol{\theta}}, \boldsymbol{x}) := \arg\max_{\boldsymbol{w} \in \mathcal{C}} \ \mathbb{E}_{ \BFy\sim F_{\hat{\BFtheta},\BFx}} \left[ r(\boldsymbol{w}, \BFy)|\BFx \right].
\]

Let $\mathbb{P}_{\BFx}$ denote the contextual distribution. $\pi(\hat{\boldsymbol{\theta}}, \boldsymbol{\theta}^*)$, under agent decision $\boldsymbol{w}^*(\hat{\boldsymbol{\theta}},\BFx)$ is thus given by:
\[
\pi(\hat{\boldsymbol{\theta}}, \boldsymbol{\theta}^*) := \mathbb{E}_{\BFx\sim\mathbb{P}_{\BFx}}\left[\mathbb{E}_{\BFy \sim F_{\boldsymbol{\theta}^*, \boldsymbol{x}}} \left[ r(\boldsymbol{w}^*(\hat{\boldsymbol{\theta}},\BFx), \BFy) |\BFx\right]\right].
\]

Assume (i) $\mathcal{C}$ is a convex set; and (ii) $r(\BFw, \BFy)$ is Lipschitz continuous in $\mathcal{C}\times\mathcal{Y}$ and strictly concave in $\BFw$ for all $\BFy \in \mathcal{Y}$, then $\pi(\hat{\boldsymbol{\theta}}, \boldsymbol{\theta}^*)$ is Lipschitz continuous in $\hat{\boldsymbol{\theta}}$, and there exists a unique maximizer to $\max_{\hat{\boldsymbol{\theta}}} \ \pi(\hat{\boldsymbol{\theta}}, \boldsymbol{\theta}^*)$, with $\arg\max_{\hat{\boldsymbol{\theta}}} \ \pi(\hat{\boldsymbol{\theta}}, \boldsymbol{\theta}^*):=\BFtheta^*$. This is a standard result in, e.g., \cite{qi2021integrated}.  

Below, we present examples of the decision-making problem that satisfies the previous assumption for newsvendor problem and portfolio optimization. 

\begin{manualexample}{1.2}\textbf{Collaborative  Newsvendor Problem: Performance Measure} 
In the Newsvendor problem, the random demand is denoted by $y \in \mathbb{R}$, and the decision variable $w \in \mathbb{R}$ represents the order quantity. The decision-maker aims to minimize the cost $r(w, y) = h(w - y)^+ + b (y - w)^+$, where $h$ and $b$ represents the unit overstock and understock cost, respectively.  In the contextual setting, the demand $y$ depends on observable features $\boldsymbol{x} \in \mathbb{R}^p$ through a linear model:
\(y = \boldsymbol{x}^\top \boldsymbol{\theta}^* + \epsilon,\)
where $\boldsymbol{\theta}^* \in \mathbb{R}^p$ is an unknown parameter vector and $\epsilon$ is a zero-mean noise term, independent from contextual feature distribution. Given an estimator $\hat{\boldsymbol{\theta}}$ and contextual information $\BFx$, the estimated conditional distribution of $y$ given $\boldsymbol{x}$ is denoted by $F_{\hat{\boldsymbol{\theta}},\boldsymbol{x}}$. The optimal decision based on $\hat{\boldsymbol{\theta}}$ is the quantile decision rule:
\[
\boldsymbol{w}^*(\hat{\boldsymbol{\theta}}, \boldsymbol{x}) = F^{-1}_{\hat{\boldsymbol{\theta}},\boldsymbol{x}} \left( \frac{b}{b+h} \right),
\]
where the quantile is taken with respect to the conditional distribution of $y$ given $\boldsymbol{x}$ under the model parameterized by $\hat{\boldsymbol{\theta}}$, and $\pi(\hat{\BFtheta},\BFtheta^*):=\mathbb{E}_{\BFx\sim\mathbb{P}_\BFx}\mathbb{E}_{y\sim F_{{\BFtheta^*,\BFx}}}[r(w^*(\hat{\BFtheta},\BFx),y)]$.
\label{example: NV}
\end{manualexample}

\begin{manualexample}{2.2}\textbf{Collaborative  Portfolio Optimization: Performance Measure}

In the portfolio optimization problem, the decision-maker allocates capital across $d$ assets. The decision variable is denoted by $\boldsymbol{w} = [w_0; \boldsymbol{w}_{\boldsymbol{y}}] \in \mathbb{R}^{d+1}$, where $w_0$ is a baseline benchmark and $\boldsymbol{w}_{\boldsymbol{y}} \in \mathbb{R}^d$ represents the allocation weights across assets. The uncertain return vector is $\boldsymbol{y} \in \mathcal{Y} \subseteq \mathbb{R}^d$. The portfolio objective is defined as:
\(r(\boldsymbol{w}, \boldsymbol{y}) := \alpha \left( \sum_{l=1}^d w^l y^l - w_0 \right)^2 - \sum_{l=1}^d w^l y^l,\)
where $\alpha > 0$ is a risk-aversion parameter, and $w^l$, $y^l$ denote the $l$-th components of $\boldsymbol{w}_{\boldsymbol{y}}$ and $\boldsymbol{y}$, respectively.

In the contextual setting, each asset $i \in \{1, \dots, d\}$ is associated with observable features $\boldsymbol{x}_i \in \mathbb{R}^p$, and the return $y_i$ is modeled as:
\(y_i = \boldsymbol{x}_i^\top \boldsymbol{\theta}^* + \epsilon_i,\)
where $\boldsymbol{\theta}^* \in \mathbb{R}^p$ is the unknown market response parameter and $\epsilon_i$ is a zero-mean noise term, independent from feature distribution. The return vector $\boldsymbol{y}$ is thus jointly distributed according to the contextual model defined by $\{\boldsymbol{x}_i\}_{i=1}^d$ and $\boldsymbol{\theta}^*$. Given an estimator $\hat{\boldsymbol{\theta}}$ and contextual information $\boldsymbol{X} := [\boldsymbol{x}_1^\top; \dots; \boldsymbol{x}_d^\top] \in \mathbb{R}^{d \times p}$, the estimated return vector $\hat{\boldsymbol{y}}$ is modeled by its conditional distribution $F_{\hat{\boldsymbol{\theta}}, \boldsymbol{X}}$.

The optimal portfolio decision based on $\hat{\boldsymbol{\theta}}$ and context $\boldsymbol{X}$ is given by:
\[
\boldsymbol{w}^*(\hat{\boldsymbol{\theta}}, \boldsymbol{X}) = \arg\max_{\boldsymbol{w} \in \mathcal{C}} \ \mathbb{E}_{\boldsymbol{y} \sim F_{\hat{\boldsymbol{\theta}}, \boldsymbol{X}}} \left[ r(\boldsymbol{w}, \boldsymbol{y}) \right],
\]
where $\mathcal{C} \subseteq \mathbb{R}^{d+1}$ is a convex feasible region for the portfolio. The platform evaluates the quality of the learned estimator $\hat{\boldsymbol{\theta}}$ via the expected reward under the true return distribution:
\[
\pi(\hat{\boldsymbol{\theta}}, \boldsymbol{\theta}^*) := \mathbb{E}_{\boldsymbol{X} \sim \mathbb{P}_\BFX} \ \mathbb{E}_{\boldsymbol{y} \sim F_{\boldsymbol{\theta}^*, \boldsymbol{X}}} \left[ r(\boldsymbol{w}^*(\hat{\boldsymbol{\theta}}, \boldsymbol{X}), \boldsymbol{y}) \right].
\]
\label{example: portfolio}
\end{manualexample}

\newpage
\section{Justification for Performance Guarantee} 
\label{app:pg}

This appendix motivates the performance guarantee in  \eqref{eq:prob_bound_revenue}.  If the platform deploys an estimator $\hat{\boldsymbol{\theta}}$ under participation profile $\BFtau_{\mathcal{A}}$, with $|\BFtau_{\mathcal{A}}|$ being the total sample size, in place of the true parameter $\boldsymbol{\theta}^*$, then under Assumption~\ref{assump:pi} (Lipschitz continuity of the performance metric in its first argument) we have
\begin{equation}
\label{eq:pi-lip}
v^* - \pi\!\left(\hat{\boldsymbol{\theta}}, \boldsymbol{\theta}^*\right) 
\;\le\; L_{\pi}\,\bigl\|\hat{\boldsymbol{\theta}} - \boldsymbol{\theta}^*\bigr\|.
\end{equation}
Assume that under FL, the loss function is designed such that the expected population loss minimizer is $\BFtheta^*$. Under standard assumptions of supervised learning (see Appendix \ref{app:lm5} for details), there exist constants $C_1>0$, $C_2>0$, such that, for any $\delta\in(0,C_2]$, for  
$\epsilon_L(n,\delta) \;:=\; \sqrt{\frac{1}{C_1 |\BFtau_{\mathcal{A}}|}\,\log\!\Bigl(\frac{C_2}{\delta}\Bigr)}$, the empirical risk minimizer
$\hat{\boldsymbol{\theta}}$ that minimizes the empirical loss function
with $|\BFtau_{\mathcal{A}}|$ i.i.d. data samples satisfies : 
\begin{equation}
\label{eq:theta-hp-delta1}
\mathbb{P}\!\left(\bigl\|\hat{\boldsymbol{\theta}} - \boldsymbol{\theta}^*\bigr\| \ge \epsilon_L(|\BFtau_{\mathcal{A}}|,\delta)\right) \;\le\; \delta.
\end{equation}

Combining \eqref{eq:pi-lip} with \eqref{eq:theta-hp-delta1} yields the following form of performance guarantee : 
\begin{equation}
\label{eq:final-bound}
\mathbb{P}\!\left( v^* - \pi\!\left(\hat{\boldsymbol{\theta}}, \boldsymbol{\theta}^*\right) 
\;\le\; \epsilon\!\left(|\boldsymbol{\tau}_{\mathcal{A}}|,\delta\right) \right) \;\ge\; 1-\delta,
\quad\text{where}\quad
\epsilon\!\left(|\boldsymbol{\tau}_{\mathcal{A}}|,\delta\right)
:= L_{\pi}\,\epsilon_L\!\left(|\boldsymbol{\tau}_{\mathcal{A}}|,\delta\right)
= L_{\pi}\sqrt{\frac{1}{C_1\,|\boldsymbol{\tau}_{\mathcal{A}}|}\log\!\Bigl(\frac{C_2}{\delta}\Bigr)} .
\end{equation}

If the platform could successfully train the ML algorithm with enough data samples, the performance guarantee could be satisfied with high probability, 
providing an incentive for the platform to use the form of performance guarantee in \eqref{eq:prob_bound_revenue}.

\newpage 
\section{Supplementary Materials for Proofs in Section \ref{sec: mechanism}}

\subsection{Proof of Proposition \ref{prop: full-part}}
\begin{proof}{Proof of Proposition \ref{prop: full-part}}
when $v(\cdot)$ is strictly increasing (leading to a non-decreasing multi-choice cooperative game), $\psi^{SV}_{\tau_{\mathcal{A}},k}(v)$ is an increasing function in $\tau^k_{\mathcal{A}}$ \citep{hsiao1993shapley}, Theorem 3. For $\psi^{MC}_{\tau_{\mathcal{A}},k}(v)$, as $\psi^{MC}_{\tau_{\mathcal{A}},k}(v) = v(|\BFtau_{\mathcal{A}}|)-v(|\BFtau_{\mathcal{A}}|-\tau^k_{\mathcal{A}})$, for any fixed $|\BFtau_{\mathcal{A}}|$, when $v(\cdot)$ is strictly increasing, $\psi^{MC}_{\tau_{\mathcal{A}},k}(v)$ is increasing in $\tau^k_{\mathcal{A}}$. Hence, both the MCFL Shapley value and marginal contribution mechanism motivate agents’ full participation.
\Halmos

\end{proof}

\subsection{Proof of Proposition \ref{prop: SV-axiom}}

\begin{proof}{Proof of Proposition \ref{prop: SV-axiom}}
This is a direct implication of Theorem 3.1 from \cite{hsiao2004power} applied to linear weights function and the specific format of MCFL Shapley Value defined in Definition \ref{def: MCLG-SV}.
\Halmos
\end{proof}

\subsection{Proof of Theorem \ref{thm: MC-BB}}

\begin{proof}{Proof of Theorem \ref{thm: MC-BB}}
By definition $\sum_{k\in \mathcal{A}} \psi^{\text{MC}}_{\BFtau_{\mathcal{A}}, k}(v)=\sum^{K}_{i=1}v(|\BFtau_{\mathcal{A}}|)-v(|\BFtau_{\mathcal{A}}|-\tau_{\mathcal{A},i})$. Therefore, showing $\sum_{k\in \mathcal{A}} \psi^{\text{MC}}_{\BFtau_{\mathcal{A}}, k}(v)<v(|\BFtau_{\mathcal{A}}|)$ is equivalent to show
\myeql{
(K-1)v(|\BFtau_{\mathcal{A}}|)<\sum^{K}_{i=1}v(|\BFtau_{\mathcal{A}}|-\tau_{\mathcal{A},i}).
}
In the rest of this proof, we drop $\mathcal{A}$ in the subscript of $\BFtau_{\mathcal{A}}$, and let $\BFtau$ denote $\BFtau_{\mathcal{A}}$ to simplify the notation. Note that, for a strictly real-valued concave function $v(\cdot)$, we have for all $\alpha\in(0,1)$, $x$, $y$
\myeql{
\alpha v(x)+(1-\alpha)v(y)<v(\alpha x+(1-\alpha)y).
}
Let $x = |\BFtau|$, $y=0$ and $\alpha = \frac{|\BFtau|-\tau_i}{|\BFtau|}$, we have,
\myeql{\frac{|\BFtau|-\tau_i}{|\BFtau|}v(|\BFtau|) = \frac{|\BFtau|-\tau_i}{|\BFtau|}v(|\BFtau|)+\frac{\tau_i}{|\BFtau|}v(0)< v(|\BFtau|-\tau_{i}).}
This leads to
\myeql{(K-1)v(|\BFtau|)=\sum^{K}_{i=1} \frac{|\BFtau|-\tau_i}{|\BFtau|}v(|\BFtau|)<\sum^{K}_{i=1}v(|\BFtau|-\tau_{i}).}
Similarly, for a strictly convex function,
\myeql{\frac{|\BFtau|-\tau_i}{|\BFtau|}v(|\BFtau|) = \frac{|\BFtau|-\tau_i}{|\BFtau|}v(|\BFtau|)+\frac{\tau_i}{|\BFtau|}v(0)> v(|\BFtau|-\tau_{i}) ,}
and 
\myeql{(K-1)v(|\BFtau|)=\sum^{K}_{i=1} \frac{|\BFtau|-\tau_i}{|\BFtau|}v(|\BFtau|)>\sum^{K}_{i=1}v(|\BFtau|-\tau_{i}) .}
The equality holds when $v(\cdot)$ is a linear function. 
\Halmos
\end{proof}

\subsection{Proof of Theorem \ref{thm: main-SV}}
\label{subsec: proof-main-thm}

\paragraph{\textbf{Additional Notation.}} In this part, we drop the subscript of $\BFtau_{\mathcal{A}}$ to simplify the notation and let $|\BFm|:=\sum_k m_k$ represent the total number of data samples. We also let 
$2^{\mathcal{N}(K)}$ denote the set of all subsets of $\{1, \dots, K\}$ and 
let $\BFm_{[2:K]}=[m_2,\dots,m_K]$ be the maximum number of data points for agent $k' = 2, \dots, K$. 
$\BFtau_{ [2:K]} = [\tau_2,\dots,\tau_K]$ with $\tau_{k'}\in\{0,\dots,m_{k'}\}$ represents a fixed participation decision profile of agent $k'=2, \dots, K$. Lastly, let $M_{k}(\BFtau)=\{v|\tau_v\neq m_v, v\neq k\}$ denote the set of agents in profile $\BFtau$ who have not reached their maximum exertion effort, excluding agent $k$. With a slight abuse of notation, we let $M(\BFtau)=\{v|\tau_v\neq m_v\}$ denote the set of agents in profile $\BFtau$ who have not reached their maximum exertion effort (note that this notation does not exclude any agent $k$). For simplicity of the notation and to avoid ambiguity, in the analysis, we set $k=1$ without loss of generality, while retaining the explicit dependence on $k$ in the expressions. We also let
$\Tau(\BFm_{[2:K]}) := \{\BFtau: \BFtau_{k'} \in \{0, \dots, m_{k'}\}, \forall k' = 2, \dots, K\} $ denote the set of all possible profiles given $\bmvec$.

For later usage in the proof, we define
\begin{equation*}
\nabla v^{t}_{|\BFtau|}=v(|\BFtau|+t)-v(|\BFtau|+t-1),
\end{equation*} 
and \begin{equation*}
c^k_t(\BFtau)=\sum^{|M_k(\BFtau)|}_{l=0}C^{l}_{|M_k(\BFtau)|} (-1)^{l}\frac{1}{|\BFtau|+l+t}.
\end{equation*}

\medskip

We first introduce the following lemma.

\begin{lemma}
\label{lemma: reform-lemma1}
Consider an agent $k$ contributing $T+T'$ number of data samples in a participation profile $\BFtau := [\tau_1, \dots, \tau_k = T+T', \dots, \tau_K]$. The agent considers splitting into two fake identities $k_1$ and $k_2$ with data samples $T$ and $T'$, respectively, and the participation profile becomes $\BFtau' := [\tau_1, \dots, \tau_{k_1} = T, \tau_{k_2}=T', \dots, \tau_K]$. Then we have 
\begin{equation*}
\psi_{\BFtau', k_1}=\sum^{T}_{t=1}\sum_{\BFtau\in \Tau(\bmvec)}tc_{t}(\BFtau)\nabla v^{t}_{|\BFtau|}+\sum^{T}_{t=1}\left[\sum^{T'}_{t_1=1}\sum_{\BFtau\in \Tau(\bmvec)}t c_{t+t_1}(\BFtau)(\nabla v^{t_1+t}_{|\BFtau|}- \nabla v^{t_1+t-1}_{|\BFtau|} ) \right].
\end{equation*}
Similarly,
\begin{equation*}\psi_{\BFtau', k_2}=\sum^{T'}_{t=1}\sum_{\BFtau\in \Tau(\bmvec)}tc_{t}(\BFtau)\nabla v^{t}_{|\BFtau|}+\sum^{T'}_{t=1}\left[\sum^{T}_{t_1=1}\sum_{\BFtau\in \Tau(\bmvec)}t c_{t+t_1}(\BFtau)(\nabla v^{t_1+t}_{|\BFtau|}-\nabla v^{t_1+t-1}_{|\BFtau|} ) \right].
\end{equation*}
\end{lemma}

\begin{proof}{Proof of Lemma \ref{lemma: reform-lemma1}}

With some algebra, 
\begin{align*}
&\psi_{\BFtau', k_1}\\
& =\sum^{T}_{t=1}\left[\sum^{T'-1}_{t_1=0}\sum_{\BFtau\in \Tau(\bmvec)}t \left(\sum^{|M_k(\BFtau)|+1}_{l=0}C^{l}_{|M_k(\BFtau)|+1} (-1)^{l}\frac{1}{|\BFtau|+l+t+t_1}\right)\nabla v^{t_1+t}_{|\BFtau|} \right.\\
& \quad \left.+ \sum_{\BFtau\in \Tau(\bmvec)}t \left(\sum^{|M_k(\BFtau)|}_{l=0}C^{l}_{|M_k(\BFtau)|} (-1)^{l}\frac{1}{|\BFtau|+l+t+T'}\right) \nabla v^{T'+t}_{|\BFtau|} \right]\\
& =\sum^{T}_{t=1}\left[\sum^{T'-1}_{t_1=0}\sum_{\BFtau\in \Tau(\bmvec)}t \left(\sum^{|M_k(\BFtau)|}_{l=0}C^{l}_{|M_k(\BFtau)|+1} (-1)^{l}\frac{1}{|\BFtau|+l+t+t_1} + (-1)^{|M_k(\BFtau)|+1}\frac{1}{|\BFtau|+(|M_k(\BFtau)|+1)+t+t_1} \right) \right.\\
& \quad \left.  \nabla v^{t_1+t}_{|\BFtau|} + \sum_{\BFtau\in \Tau(\bmvec)}t c_{t+T'}(\BFtau)\nabla v^{t+T'}_{|\BFtau|} \right]\\
&=\sum^{T}_{t=1}\left[\sum^{T'-1}_{t_1=0}\sum_{\BFtau\in \Tau(\bmvec)}t \left(\sum^{|M_k(\BFtau)|}_{l=0}[C^{l-1}_{|M_k(\BFtau)|} + C^{l}_{|M_k(\BFtau)|}  ]  (-1)^{l}\frac{1}{|\BFtau|+l+t+t_1}  \right. \right.\\
& \quad \left. \left. +(-1)^{|M_k(\BFtau)|+1}\frac{1}{|\BFtau|+(|M_k(\BFtau)|+1)+t+t_1} \right)\nabla v^{t_1+t}_{|\BFtau|}  + \sum_{\BFtau\in \Tau(\bmvec)}t c_{t+T'}(\BFtau)\nabla v^{t+T'}_{|\BFtau|} \right],
\end{align*}
where we use the recurrence relation of the binomial coefficient. Then the previous equation equals to 
\begin{align*}
&=\sum^{T}_{t=1}\left[\sum^{T'-1}_{t_1=0}\sum_{\BFtau\in \Tau(\bmvec)}t c_{t+t_1}(\BFtau)\nabla v^{t_1+t}_{|\BFtau|}+ \sum^{T'-1}_{t_1=0}\sum_{\BFtau\in \Tau(\bmvec)}t \left(\sum^{|M_k(\BFtau)|}_{l=1} C^{l-1}_{|M_k(\BFtau)|}(-1)^{l}\frac{1}{|\BFtau|+l+t+t_1} \right. \right.\\
& \quad \left. \left.+(-1)^{|M_k(\BFtau)|+1}\frac{1}{|\BFtau|+(|M_k(\BFtau)|+1)+t+t_1}\right)\nabla v^{t_1+t}_{|\BFtau|}+\sum_{\BFtau\in \Tau(\bmvec)}tc_{t+T'}(\BFtau) \nabla v^{t+T'}_{|\BFtau|} \right].
\end{align*}
The equality is defined by $c_{t+t_1}(\BFtau)$, and by changing the index from $l=1$ to $l=0$, we have
\begin{align*}
& =\sum^{T}_{t=1}\left[\sum^{T'-1}_{t_1=0}\sum_{\BFtau\in \Tau(\bmvec)}t c_{t+t_1}(\BFtau)\nabla v^{t_1+t}_{|\BFtau|}  
+\sum_{\BFtau\in \Tau(\bmvec)}tc_{t+T'}(\BFtau) \nabla v^{t+T'}_{|\BFtau|} \right.\\
& \quad \left.- \sum^{T'-1}_{t_1=0}\sum_{\BFtau\in \Tau(\bmvec)}t \left(\sum^{|M_k(\BFtau)|}_{l=0} C^{l}_{|M_k(\BFtau)|}\frac{(-1)^{l}}{|\BFtau|+l+1+t+t_1}\right)\nabla v^{t_1+t}_{|\BFtau|}   \right] \\
& =\sum^{T}_{t=1}\left[\sum^{T'-1}_{t_1=0}\sum_{\BFtau\in \Tau(\bmvec)}t c_{t+t_1}(\BFtau)\nabla v^{t_1+t}_{|\BFtau|}- \sum^{T'-1}_{t_1=0}\sum_{\BFtau\in \Tau(\bmvec)}t c_{t+t_1+1}(\BFtau)\nabla v^{t_1+t}_{|\BFtau|}  +\sum_{\BFtau\in \Tau(\bmvec)}tc_{t+T'}(\BFtau) \nabla v^{t+T'}_{|\BFtau|} \right].\\
\end{align*}
Similarly, the previous equality comes from the definition of $c_{t+T'}(\BFtau)$ and $c_{t+t_1+1}(\BFtau)$, and changing the index of $t_1=0$ to $t_1=1$, 
\begin{align*}
\psi_{\BFtau', k_1}&=\\
& =\sum^{T}_{t=1}\left[\sum_{\BFtau\in \Tau(\bmvec)}tc_{t}(\BFtau)\nabla v^{t}_{|\BFtau|}+\sum^{T'}_{t_1=1}\sum_{\BFtau\in \Tau(\bmvec)}t c_{t+t_1}(\BFtau)\nabla v^{t_1+t}_{|\BFtau|}- \sum^{T'}_{t_1=1}\sum_{\BFtau\in \Tau(\bmvec)} t c_{t+t_1}(\BFtau)\nabla v^{t_1+t-1}_{|\BFtau|} \right]\\
& =\sum^{T}_{t=1}\left[\sum_{\BFtau\in \Tau(\bmvec)}tc_{t}(\BFtau)\nabla v^{t}_{|\BFtau|}+\sum^{T'}_{t_1=1}\sum_{\BFtau\in \Tau(\bmvec)}t c_{t+t_1}(\BFtau)(\nabla v^{t_1+t}_{|\BFtau|}-\nabla v^{t_1+t-1}_{|\BFtau|} ) \right].
\end{align*}
Similarly, we have
$$\psi_{\BFtau', k_2}=\sum^{T'}_{t=1}\sum_{\BFtau\in \Tau(\bmvec)}tc_{t}(\BFtau)\nabla v^{t}_{|\BFtau|}+\sum^{T'}_{t=1}\left[\sum^{T}_{t_1=1}\sum_{\BFtau\in \Tau(\bmvec)}t c_{t+t_1}(\BFtau)(\nabla v^{t_1+t}_{|\BFtau|}-\nabla v^{t_1+t-1}_{|\BFtau|} ) \right].$$
\Halmos
\end{proof}

We next introduce Lemma \ref{lemma: reform-lemma2} for the future reorganization of $\psi_{\BFtau',k_1}$ and $\psi_{\BFtau',k_2}$.

\begin{lemma}
\label{lemma: reform-lemma2}
For any $K, T, T' $, and any maximum data vector $\bmvec$, we have 
\begin{align*}
&\sum^{T'}_{t=1}\sum_{\BFtau\in \Tau(\bmvec)} \left[t c_{t}(\BFtau)\nabla v^{t}_{|\BFtau|} -(t+T) c_{t+T}(\BFtau)\nabla v^{t+T}_{|\BFtau|}\right] \\
& = \sum_{t=0}^{T-1} \sum^{T'}_{t_1=1}\sum_{\BFtau\in \Tau(\bmvec)} \left[(t_1+t)c_{t_1+t}(\BFtau)\nabla v^{t_1+t}_{|\BFtau|}-(t_1+t+1)c_{t_1+t+1}(\BFtau)\nabla v^{t_1+t+1}_{|\BFtau|}\right].
\end{align*}
\end{lemma}
\begin{proof}{Proof of Lemma \ref{lemma: reform-lemma2}}
For any fixed $t_1 \in [1, \dots, T']$ and $\BFtau \in \Tau(\bmvec)$, 
\begin{align*}
& t_1 c_{t_1}(\BFtau)\nabla v^{t_1}_{|\BFtau|} -(t_1+T) c_{t_1+T}(\BFtau)\nabla v^{t_1+T}_{|\BFtau|} \\
& = \sum_{t'=t_1}^{t_1+T-1}\left[ t' c_{t'}(\BFtau)\nabla v^{t'}_{|\BFtau|} -(t'+1) c_{t'+1}(\BFtau)\nabla v^{t'+1}_{|\BFtau|} \right]\\
& = \sum_{t=0}^{T-1} \left[(t_1+t)c_{t_1+t}(\BFtau)\nabla v^{t_1+t}_{|\BFtau|}-(t_1+t+1)c_{t_1+t+1}(\BFtau)\nabla v^{t_1+t+1}_{|\BFtau|}\right].
\end{align*}
Thus the desired result follows. 
\end{proof}

We then introduce the next lemma to simplify expression of $c^k_t(\BFtau)$. 

\begin{lemma}
\label{lemma: frac-expansion}
For any $n\in \mathbb{N}^{+}$ and $x\in \mathbb{N}^{+}$, we have
\begin{equation*}
\sum_{k=0}^n C_n^k\frac{(-1)^k}{x+k} = \frac{n!}{\Pi_{k=0}^n(x+k)}.
\end{equation*}
\end{lemma}
\begin{proof}{Proof of Lemma \ref{lemma: frac-expansion}}
Note that 
    $\frac{1}{x+k} = \int_0^1 t^{x+k-1}dt$.
Then, interchanging finite sum and integral directly, we have
\begin{equation*}
\sum_{k=0}^n C_n^k(-1)^k \int_0^1 t^{x+k-1}dt = \int_0^1 t^{x-1} \sum_{k=0}^n C_n^k(-t)^k dt = \int_0^1 t^{x-1}(1-t)^n dt = B(x, n+1).
\end{equation*}  
The second equality holds due to applying the Binomial theorem $(a+b)^n = \sum_{k=0}^n C_n^k a^{n-k}b^k$ and the last equality comes from the definition of beta function. The desired result follows from the fact that 
\begin{equation*}
B(x, n+1) = \frac{\Gamma(x)\Gamma(n+1)}{\Gamma(x+n+1)} = \frac{(x-1)!n!}{(x+n)!} = \frac{n!}{\Pi_{k=0}^n(x+k)}.
\end{equation*}

\Halmos
\end{proof}

We further introduce the following useful Lemmas \ref{lemma: sum-x} and \ref{lemma: sum-c}.

\begin{lemma}
\label{lemma: sum-x}
For any vector $\BFm$ and for any $x\in [0,1]$, we have that 
\begin{equation*}
    \sum_{\BFtau \in \Tau(\BFm)}x^{|\BFtau|}(1-x)^{|M(\BFtau)|} =1
\end{equation*}

\end{lemma}
\begin{proof}{Proof of Lemma \ref{lemma: sum-x}}
By definition, $|\BFtau| = \sum_{i=1}^K\tau_i$ and $|M(\BFtau)| = \sum_{i=1}^K \mathbbm{1}(\tau_i<m_i) $, then we have
\begin{align*}
&\sum_{\BFtau \in \Tau(\BFm)}x^{|\BFtau|}(1-x)^{|M(\BFtau)|}  = \sum_{\BFtau \in \Tau(\BFm)}\Pi_{i=1}^K x^{\tau_i}(1-x)^{\mathbbm{1}(\tau_i<m_i)} \\
= &\sum_{\tau_1 =0}^{m_1}\sum_{\tau_2 =0}^{m_2}\cdots\sum_{\tau_K =0}^{m_K}  \Pi_{i=1}^K x^{\tau_i}(1-x)^{\mathbbm{1}(\tau_i<m_i)} = \Pi_{i=1}^K \sum_{\tau_i \leq m_i } x^{\tau_i}(1-x)^{\mathbbm{1}(\tau_i<m_i)}\\
= & \Pi_{i=1}^K (\sum_{\tau_i =0} ^{m_i-1}x^{\tau_i}(1-x) + x^{m_i}) =\Pi_{i=1}^K \left((1-x)\frac{1-x^{m_i}}{1-x} +x^{m_i}\right) = \Pi_{i=1}^K 1 = 1.
\end{align*}
Note that the second equation holds because it is a sum over cartesian products with independent coordinates. 
\Halmos
\end{proof}

\begin{lemma}
\label{lemma: sum-c}
For any vector $\BFm$, there is $$\sum_{\BFtau \in \Tau(\BFm)} c_t(\BFtau)=\frac{1}{t}.$$
\end{lemma}

\begin{proof}{Proof of Lemma \ref{lemma: sum-c}}
By definition and Lemma \ref{lemma: frac-expansion}, 
\begin{equation*}
c_t(\BFtau)=\sum^{|M(\BFtau)|}_{l=0}C^{l}_{|M(\BFtau)|} (-1)^{l}\frac{1}{|\BFtau|+l+t} = \frac{|M(\BFtau)|!}{\Pi_{k=0}^{|M(\BFtau)|} (|\BFtau|+t+k)}.
\end{equation*}
Note that by definition of Beta function, we have,
\begin{equation}
\int_0^1 x^{A-1}(1-x)^{B-1}\,dx = \frac{(A-1)!\,(B-1)!}{(A+B-1)!}. 
\label{eq: frac-exp-1}
\end{equation}
Then 
\begin{align*}
\sum_{\BFtau \in \Tau(\BFm)} c_t(\BFtau) & = \sum_{\BFtau \in \Tau(\BFm)} \int_0^1 x^{A-1}(1-x)^{B-1} dx \nonumber\\
& = \sum_{\BFtau \in \Tau(\BFm)} \int_0^1 x^{|\BFtau|+t-1}(1-x)^{|M(\BFtau)|} dx \nonumber\\
& = \int_0^1 \sum_{\BFtau \in \Tau(\BFm)} x^{|\BFtau|+t-1}(1-x)^{|M(\BFtau)|} dx \nonumber\\
& = \int_0^1 x^{t-1} \sum_{\BFtau \in \Tau(\BFm)} x^{|\BFtau|}(1-x)^{|M(\BFtau)|}dx.  
\end{align*}
Since $|\BFtau| = \sum_{i=1}^K\tau_i$ and $|M(\BFtau)| = \sum_{i=1}^K \mathbbm{1}(\tau_i<m_i)$,  by Lemma \ref{lemma: sum-x},
\begin{align*}
\sum_{\BFtau \in \Tau(\BFm)}x^{|\BFtau|}(1-x)^{|M(\BFtau)|} = 1.
\end{align*}
Note that the second equation holds because $\BFtau$ forms a grid where each coordinate $\tau_i < m_i$, then it is equivalent to swap sum and multiplication. Thus, 
\begin{equation*}
 \sum_{\BFtau \in \Tau(\BFm)} c_t(\BFtau) = \int_0^1 x^{t-1}dx = \frac{1}{t}.
\end{equation*}
\Halmos
\end{proof}

\begin{lemma}
\label{lemma: concave-sum}
For any vector $\BFm$, any positive integer $t \in \mathbb{N}^+$, and a strictly decreasing function $u(\cdot): \mathbb{R}\rightarrow \mathbb{R}^+$, and provided there exists at least one coordinate with $m_i\ge 1$, we have,
\begin{equation*}
    S := \sum_{\BFtau\in \Tau(\BFm)} [tc_t(\BFtau) -(t+1)c_{t+1}(\BFtau)]u(|\BFtau|) > 0.
\end{equation*}
\end{lemma}
\begin{proof}{Proof of Lemma \ref{lemma: concave-sum}}
By the Beta integral $\int_0^1 x^{A-1}(1-x)^{B-1}\,dx = \frac{(A-1)!\,(B-1)!}{(A+B-1)!}$, similar to the proof of Lemma \ref{lemma: sum-c}, we have
\begin{equation*}
    S =  \int_0^1 \left((tx^{t-1}-(t+1)x^t)\cdot  \left(\sum_{\BFtau\in \Tau(\BFm)}x^{|\BFtau| }(1-x)^{|M(\BFtau)|}u(|\BFtau|)\right)\right)dx. 
\end{equation*}
Let $f(x) := tx^{t-1} - (t+1)x^t$ and $g(x):= \sum_{\BFtau\in \Tau(\BFm)}x^{|\BFtau| }(1-x)^{|M(\BFtau)|}u(|\BFtau|)$.
Then we have
\begin{align*}
    S & = \int_0^{\frac{t}{t+1}} f(x)g(x) dx + \int_{\frac{t}{t+1}}^1 f(x)g(x) dx \\
    & > \int_0^{\frac{t}{t+1}} f(x)g\left(\frac{t}{t+1}\right) dx + \int_{\frac{t}{t+1}}^1 f(x)g\left(\frac{t}{t+1}\right) dx  \\
   & = \int_0^{1} f(x)g\left(\frac{t}{t+1}\right) dx \\
   &= g\left(\frac{t}{t+1}\right) \left(\int_0^1 tx^{t-1}dx - \int_0^1 (t+1)x^tdx\right) \\
   & = g\left(\frac{t}{t+1}\right) (1-1) = 0
\end{align*}
The strict inequality holds because $f(x) <0$ for $1>x>\frac{t}{t+1}$ and $f(x) >0$ for $0<x<\frac{t}{t+1}$, and that $g(x)$ is a non-negative decreasing function according to Lemma \ref{lemma: g-decreasing}. 
\Halmos
\end{proof}

\begin{lemma}
\label{lemma: g-decreasing}
 For any $x\in [0,1]$ and any $\BFm$, the function $g$, defined as $g(x):= \sum_{\BFtau\in \Tau(\BFm)}x^{|\BFtau| }(1-x)^{|M(\BFtau)|}u(|\BFtau|)$, is a non-negative, strictly decreasing function.
\end{lemma}
\begin{proof}{Proof of Lemma \ref{lemma: g-decreasing}}
Consider $P_x(\BFtau) := x^{|\BFtau|}(1-x)^{|M(\BFtau)|}$. Note that according to Lemma \ref{lemma: sum-x}, $\sum_{\BFtau\in \Tau(\BFm)} P_x(\BFtau) =1$. Together with the fact that $P_x(\cdot) \geq0$ for any $\BFtau$, we view $P_x(\BFtau)$ as the probability mass function of a distribution with the support on $\Tau(\BFm)$. We decompose the distribution into each coordinate of $\BFtau$,
\begin{equation*}
    P_x(\BFtau) = \Pi_{i=1}^K Q_{x, m_i}(\tau_i),
\end{equation*}
where $Q_{x, m_i}(\tau_i) :=  x^{\tau_i} (1-x)^{\mathbbm{1}(\tau_i<m_i)}$. Note that we still have 
\begin{equation*}
    \sum_{\tau_i =0}^{m_i} Q_{x, m_i}(\tau_i) = \sum_{\tau_i =0}^{m_i-1} x^{\tau_i} (1-x)  + x^{m_i} = (1-x)\cdot\frac{1-x^{m_i}}{1-x}+ x^{m_i}=  1. 
\end{equation*}
Therefore, the coordinates are independent, each following distribution $Q_{x, m_i}(\tau_i)$. For each coordinate $\tau_i$, the tail probability is $\Pr(\tau_i \geq n) = \sum_{j=n}^{m_i-1}x^j(1-x) +x^{m_i} = x^n$, for all $1\leq n \leq m_i$, and is strictly increasing in $x$. Therefore, for each coordinate, $\tau_i$, is stochastically increasing in $x$.

Then we show that, suppose that $\sum_{i=1}^{k-1}\tau_i$ is stochastically (strictly) increasing in $x$, then $\sum_{i=1}^{k}\tau_i$ is also stochastically (strictly) increasing in $x$. For any $x' >x$, fix $n, j$, we have
    \begin{align*}
        & \Pr_{P_{x'}}(\sum_{i=1}^{k-1}\tau_i + \tau_k \geq n) - \Pr_{P_x}(\sum_{i=1}^{k-1}\tau_i + \tau_k \geq n)\\
        &= \sum_{j=0}^{m_k} Q_{x', m_k}(j) \Pr_{P_{x'}}(\sum_{i=1}^{k-1}\tau_i \geq n-j) - \sum_{j=1}^{m_k} Q_{x, m_k}(j) \Pr_{P_{x}}(\sum_{i=1}^{k-1}\tau_i \geq n-j)\\
        & = \sum_{j=0}^{m_k} Q_{x, m_k}(j) \left[\Pr_{P_{x'}}(\sum_{i=1}^{k-1}\tau_i \geq n-j) - \Pr_{P_{x}}(\sum_{i=1}^{k-1}\tau_i \geq n-j)\right] + \sum_{j=0}^{m_k} \left[Q_{x', m_k} (j) - Q_{x, m_k} (j)\right]\Pr_{P_{x'}}(\sum_{i=1}^{k-1}\tau_i \geq n-j) 
    \end{align*}
The first term on the right hand side is greater than zero due to the inductive hypothesis. The second term is also greater than zero because $Q_{x', m_k}$ stochastically dominates $Q_{x, m_k}$ and $\Pr_{P_{x'}}(\sum_{i=1}^{k-1}\tau_i \geq n-j)$ is an increasing function of $j$. 
Therefore, it follows that \( g(x)=\mathbb{E}_{P_x}[u(|\BFtau|)] \) is decreasing in \(x\), and is strictly decreasing whenever at least one coordinate has \(m_i\ge 1\). 
\Halmos
\end{proof}

With all of the above lemmas, we are now ready to prove the main theorem.

\begin{proof}{Proof of Theorem \ref{thm: main-SV}}
We start by considering an agent $k$ contributing $T+T'$ number of data samples in a participation profile $\BFtau := [\tau_1, \dots, \tau_k = T+T', \dots, \tau_K]$. The agent considers splitting into two fake identities $k_1$ and $k_2$ with data samples $T$ and $T'$, respectively, and the participation profile becomes $\BFtau' := [\tau_1, \dots, \tau_{k_1} = T, \tau_{k_2}=T', \dots, \tau_K]$.

By Lemma \ref{lemma: reform-lemma1},  we have 
\begin{align*}
\psi_{\BFtau', k_1}+\psi_{\BFtau', k_2} &= \sum^{T}_{t=1}\sum_{\BFtau\in \Tau(\bmvec) }tc_{t}(\BFtau)\nabla v^{t}_{|\BFtau|}+\sum^{T}_{t=1}\left[\sum^{T'}_{t_1=1}\sum_{\BFtau\in \Tau(\bmvec)}t c_{t+t_1}(\BFtau)(\nabla v^{t_1+t}_{|\BFtau|}- \nabla v^{t_1+t-1}_{|\BFtau|} ) \right] \\
& \quad + \sum^{T'}_{t=1}\sum_{\BFtau\in \Tau(\bmvec)}tc_{t}(\BFtau)\nabla v^{t}_{|\BFtau|}+\sum^{T'}_{t=1}\left[\sum^{T}_{t_1=1}\sum_{\BFtau\in \Tau(\bmvec)}t c_{t+t_1}(\BFtau)(\nabla v^{t_1+t}_{|\BFtau|}-\nabla v^{t_1+t-1}_{|\BFtau|} ) \right] \\
& = \sum^{T}_{t=1}\sum_{\BFtau\in \Tau(\bmvec)}tc_{t}(\BFtau)\nabla v^{t}_{|\BFtau|} + \sum^{T'}_{t=1}\sum_{\BFtau\in \Tau(\bmvec)}tc_{t}(\BFtau)\nabla v^{t}_{|\BFtau|}\\
& \quad + \sum^{T}_{t=1}\left[\sum^{T'}_{t_1=1}\sum_{\BFtau\in \Tau(\bmvec)}(t+ t_1) c_{t+t_1}(\BFtau)(\nabla v^{t_1+t}_{|\BFtau|}- \nabla v^{t_1+t-1}_{|\BFtau|} ) \right].
\end{align*}
Now, without splitting among fake identities, we have the allocated surplus as
\begin{align*}
\psi_{\BFtau,k}&=\sum^{T+T'}_{t=1}\sum_{\BFtau\in \Tau(\bmvec)} t c_{t}(\BFtau)\nabla v^{t}_{|\BFtau|} \\
& = \sum^{T}_{t=1}\sum_{\BFtau\in \Tau(\bmvec)} t c_{t}(\BFtau)\nabla v^{t}_{|\BFtau|} + \sum^{T'}_{t=1}\sum_{\BFtau\in \Tau(\bmvec)} (t+T) c_{t+T}(\BFtau)\nabla v^{t+T}_{|\BFtau|}. 
\end{align*}
Then to compare the two options,
\begin{align}
\psi_{\BFtau', k_1}+\psi_{\BFtau', k_2}-\psi_{\BFtau,k} &= \sum^{T}_{t=1}\left[\sum^{T'}_{t_1=1}\sum_{\BFtau\in \Tau(\bmvec)}(t+ t_1) c_{t+t_1}(\BFtau)(\nabla v^{t_1+t}_{|\BFtau|}- \nabla v^{t_1+t-1}_{|\BFtau|} ) \right] \nonumber\\
& \quad + \sum^{T'}_{t=1}\sum_{\BFtau\in \Tau(\bmvec)} \left[t c_{t}(\BFtau)\nabla v^{t}_{|\BFtau|} -(t+T) c_{t+T}(\BFtau)\nabla v^{t+T}_{|\BFtau|}\right] \nonumber\\
& = \sum^{T-1}_{t=0}\left[\sum^{T'}_{t_1=1}\sum_{\BFtau\in \Tau(\bmvec)}(t+ t_1+1) c_{t+t_1+1}(\BFtau)(\nabla v^{t_1+t+1}_{|\BFtau|}- \nabla v^{t_1+t}_{|\BFtau|} ) \right] \nonumber\\
& \quad + \sum_{t=0}^{T-1} \sum^{T'}_{t_1=1}\sum_{\BFtau\in \Tau(\bmvec)} \left[(t_1+t)c_{t_1+t}(\BFtau)\nabla v^{t_1+t}_{|\BFtau|}-(t_1+t+1)c_{t_1+t+1}(\BFtau)\nabla v^{t_1+t+1}_{|\BFtau|}\right] \label{eq: thm1-1}\\
& =  \sum_{t=0}^{T-1} \sum^{T'}_{t_1=1}\sum_{\BFtau\in \Tau(\bmvec)} \left[(t_1+t)c_{t_1+t}(\BFtau)-(t_1+t+1)c_{t_1+t+1}(\BFtau)\right]\nabla v^{t_1+t}_{|\BFtau|} \label{eq: thm1-2}.
\end{align}
where \eqref{eq: thm1-1} holds according to Lemma \ref{lemma: reform-lemma2}.

We first prove (ii). If $v(\cdot)$ is a linear function, there exists a constant $v_0$ such that $\nabla v^{t}_{|\BFtau|} = v_0$ for any $t$ and $\BFtau$. Therefore, we have 
\begin{align}
\eqref{eq: thm1-2} &=  \sum_{t=0}^{T-1} \sum^{T'}_{t_1=1}\sum_{\BFtau\in \Tau(\bmvec)} \left[(t_1+t)c_{t_1+t}(\BFtau)-(t_1+t+1)c_{t_1+t+1}(\BFtau)\right]v_0 \nonumber \\
& =  \sum_{t=0}^{T-1} \sum^{T'}_{t_1=1}\left[(t_1+t)\sum_{\BFtau\in \Tau(\bmvec)}c_{t_1+t} (\BFtau)-(t_1+t+1)\sum_{\BFtau\in \Tau(\bmvec)} c_{t_1+t+1}(\BFtau)\right]v_0 \nonumber\\
& = \sum_{t=0}^{T-1} \sum^{T'}_{t_1=1}\left[(t_1+t)\frac{1}{t_1+t}-(t_1+t+1)\frac{1}{t_1+t+1}\right]v_0  \label{eq: thm1-3} \\
& = 0\nonumber.
\end{align}
where \eqref{eq: thm1-3} holds due to Lemma \ref{lemma: sum-c}.
Now we prove (i). If $v(\cdot)$ is strictly concave and increasing, then $\nabla v^{t_1+t}_{|\BFtau|}$  is a non-negative, strictly decreasing function in $|\BFtau|$. According to Lemma \ref{lemma: concave-sum}, considering $u(|\BFtau|)$ to be $\nabla v^{t_1+t}_{|\BFtau|}$ , we have that, for any $t \in \{0, \dots, T-1\}$ and any $ t_1 \in \{1, \dots, T'\}$, 
\begin{equation*}
    \sum_{\BFtau\in \Tau(\bmvec)} \left[(t_1+t)c_{t_1+t}(\BFtau)-(t_1+t+1)c_{t_1+t+1}(\BFtau)\right]\nabla v^{t_1+t}_{|\BFtau|}  > 0.
\end{equation*}
Therefore, $\eqref{eq: thm1-2} > 0$.
\Halmos
\end{proof}

\subsection{Proof of Corollary \ref{cor: equi-SV}}
\begin{proof}{Proof of Corollary \ref{cor: equi-SV}} According to Proposition \ref{prop: full-part}, the MCFL Shapley value incentives agents to contribute all of their data samples. However, it also encourages data splitting.
Recursively applying Theorem \ref{thm: main-SV} to each (fake) identity until the data samples have been fully split to as many fake identities as possible. Therefore, the equilibrium of agent participation for the real agent $k$ with $m_k$ samples becomes a full split across $m_k$ fake identities, each with one data sample.
Applying this conclusion to each real agent leads to the desired result.
\Halmos
\end{proof}

\subsection{Proof of Corollary \ref{cor:impossible}}
\begin{proof}{Proof of Corollary \ref{cor:impossible}} Corollary \ref{cor:impossible} is a direct result from Corollary \ref{cor: equi-SV} and Proposition \ref{prop: SV-axiom}.
\Halmos
\end{proof}
\subsection{Proof of Theorem \ref{thm: main-MC}}
\begin{proof}{Proof of Theorem \ref{thm: main-MC}}
Let $m=|\BFm_{[2:K]}|$ and we follow the notation in the proof of Theorem \ref{thm: main-SV}. $\psi^{MC}_{\BFtau,k}(v)=v(m+T+T')-v(m)$, $\psi^{MC}_{\BFtau',k_1}(v)+\psi^{MC}_{\BFtau',k_2}(v)=v(m+T+T')-v(m+T')+v(m+T+T')-v(m+T)$. Comparing $\psi^{MC}_{\BFtau,k}(v)$ and 
$\psi^{MC}_{\BFtau',k_1}(v)+\psi^{MC}_{\BFtau',k_2}(v)$ is similar to comparing $v(m+T)-v(m)$ and $v(m+T'+T)-v(m+T')$. Hence, $v(m+T)-v(m)>v(m+T'+T)-v(m+T')$ when the function satisfies a strict decreasing difference, and $\psi^{MC}_{\BFtau,k} > \psi^{MC}_{\BFtau',k_1}(v)+\psi^{MC}_{\BFtau',k_2}(v)$. Similarly, $v(m+T)-v(m)<v(m+T'+T)-v(m+T')$ when the function has a strict increasing difference, and 
$\psi^{MC}_{\BFtau,k} < \psi^{MC}_{\BFtau',k_1}(v)+\psi^{MC}_{\BFtau',k_2}(v)$. When $v(\cdot)$ is a linear function with $v(x)=ax+b$, $v(m+T)-v(m)=a T=v(m+T'+T)-v(m+T')=aT$ and splitting makes no difference. \Halmos
\end{proof}

\subsection{Proof of Corollary \ref{cor: equi-MC}}
\begin{proof}{Proof of Corollary \ref{cor: equi-MC}}
 Proposition \ref{prop: full-part} shows that the marginal contribution mechanism induces full participation of all agents with all possessed data samples. According to Theorem \ref{thm: main-MC}, the agents have no incentive to split the data among fake identities. 
 \Halmos
\end{proof}

\newpage 
\section{Supplementary Material for FL Algorithm}
\label{app:FL_alg}
\begin{algorithm}[htb!]
\KwIn{$t=0$, $\Phi=\{K^{FL}, \rho, \BFtheta^0, T, H\}$, $\BFtheta^{0}_{k}=\BFtheta^0$  for all $k$, $S_0\leftarrow$ random set of $K^{FL}$ agents}
\KwOut{Training weights $\hat{\BFtheta}^{FL}_{|\BFtau_\mathcal{A}|}$ as average of $\BFtheta^{t}, t\in\{H,2H,\dots,T-1\}$}
\While{$t < T$}{
        \eIf{$t+1\in\{H,2H,\dots,T-1\}$}{
            $\BFtheta^{t}=\frac{1}{K^{FL}}{\sum_{i\in S_{t}}\BFtheta^{t}_{i}}$ \tcp*{Synchronization}
            $S_t\leftarrow$ random set of $K^{FL}$ agents \tcp*{Client Selection}
            $\BFtheta^{t+1}_k=\BFtheta^{t+1}$ for $k\in S_{t}$ \tcp*{Broadcasting}

        }{   
        \For{$k \in S_t$ in Parallel}{
            $\BFtheta^{t+1}_{k}=\BFtheta^{t}_{k}-\rho\nabla L_{k}(\BFtheta^{t}_{k})$ \tcp*{Local Training}}
        $S_t\leftarrow S_{t-1}$
        }
    $t = t + 1$
}
\caption{Federated learning approach with Local Gradient Descent for solving (\ref{eq:FL_general}) (e.g. \cite{mangasarian1995parallel,khaled2019first})}
\label{alg: FL}
\end{algorithm}

\newpage
\section{Proof for Lemma \ref{lemma: high-prob-consistent}} 
\label{app:lm5}
\begin{proof}{Proof for Lemma \ref{lemma: high-prob-consistent}}
Under supervised learning, assume the platform minimizes the empirical loss function given by $(\ref{eq:FL_general})$ based on i.i.d.\ samples $\{\boldsymbol{z}_i\}_{i=1}^{|\BFtau_{\mathcal{A}}|}$, and suppose:
\begin{enumerate}[label=(\roman*), leftmargin=1.3em]
\item The population risk $\mathcal{L}(\boldsymbol{\theta}) := \mathbb{E}_z[\ell(\boldsymbol{\theta};\boldsymbol{z})]$ is uniquely minimized at $\boldsymbol{\theta}^*\in\Theta$;
\item $\mathcal{L}$ is  strongly convex;
\item $\ell(\boldsymbol{\theta};\boldsymbol{z})$ is non-negative, bounded, and Lipschitz in $\boldsymbol{\theta}$ with respect to the $l_2$-norm for all $\BFz$.
\end{enumerate}

Under these conditions, the empirical risk minimizer
$\hat{\boldsymbol{\theta}} := \arg\min_{\boldsymbol{\theta}\in\Theta} L(\BFtheta)$
satisfies the following high-probability parameter error bound (e.g., Theorem 4.8. \citep{Ma2022CS229MNotes}) : there exist constants $C_1>0$, $C_2>0$, and a radius $r>0$ such that for all $\varepsilon \le r$,
\begin{equation}
\label{eq:theta-hp}
\mathbb{P}\!\left(\bigl\|\hat{\boldsymbol{\theta}} - \boldsymbol{\theta}^*\bigr\| \ge \varepsilon\right) 
\;\le\; C_2\,\exp(-C_1 |\BFtau_{\mathcal{A}}| \varepsilon^2).
\end{equation}
Equivalently, for any $\delta\in(0,C_2]$, if
\(
|\BFtau_{\mathcal{A}}|\;\ge\;\frac{1}{C_1 r^2}\log\!\Bigl(\frac{C_2}{\delta}\Bigr),
\) so that $\epsilon_L(|\BFtau_{\mathcal{A}}|,\delta)\le r$, defining
\begin{equation}
\label{eq:epsL}
\epsilon_L(n,\delta) \;:=\; \sqrt{\frac{1}{C_1 |\BFtau_{\mathcal{A}}|}\,\log\!\Bigl(\frac{C_2}{\delta}\Bigr)},
\end{equation}
we have
\begin{equation}
\label{eq:theta-hp-delta}
\mathbb{P}\!\left(\bigl\|\hat{\boldsymbol{\theta}} - \boldsymbol{\theta}^*\bigr\| \ge \epsilon_L(|\BFtau_{\mathcal{A}}|,\delta)\right) \;\le\; \delta.
\end{equation}

\Halmos
\end{proof}
\section{Proof for Theorem \ref{thm:convergence}}
\label{app:proof-convergence}

\begin{proof}{Proof for Theorem \ref{thm:convergence}}In this proof, we omit the dependency of $\BFtau_{\mathcal{A}}$ in $\hat{\BFtheta}^{FL}_{\BFtau_{\mathcal{A}}}$, and use
$\hat{\BFtheta}_{FL}$ to denote $\hat{\BFtheta}^{FL}_{\BFtau_{\mathcal{A}}}$.
We denote
\(
L_k({\BFtheta})
:=\frac{|\mathcal{A}|}{|\BFtau_{\mathcal{A}}|}\sum_{j=1}^{\tau_{\mathcal{A},k}} l({\boldsymbol{\theta}}; \mathbf{z}_j)
\)
to be the local loss function for agent identity $k$.
Under SV, $L_k(\BFtheta)=l(\BFtheta;\BFz_k)$ since $\BFtau=[1,\dots,1]\in \mathbb{R}^{|\BFm|\times 1}$. We further define local variation of loss function. Let $\hat{\BFtheta}^*_{FL}:=\arg\min_{\BFtheta}L(\BFtheta)$. Under SV, define
\[
\sigma^2
:=\frac{1}{|\mathcal{A}|}\sum_{k\in\mathcal{A}}\|\nabla L_k(\hat{\BFtheta}^*_{FL})\|^2
=\frac{1}{|\BFm|}\sum_{i=1}^{|\BFm|}\|\nabla l(\hat{\BFtheta}^*_{FL};\BFz_i)\|^2.
\]
Under MC with $K$ participants, let $m_k$ be the number of samples assigned to participant $k$ with samples 
$\{\BFz_{k,j}\}_{j=1}^{m_k}$, $\sum_{k=1}^K m_k=|\BFm|$. Assume the MC partition is $\eta$-balanced: $m_k \le (1+\eta)\frac{|\BFm|}{K}$ for all $k$. Then
\[
\sigma_K^2
:=\frac{1}{K}\sum_{k=1}^K \|\nabla L_k(\hat{\BFtheta}^*_{FL})\|^2
=\frac{1}{K}\sum_{k=1}^K\left\|\frac{K}{|\BFm|}\sum_{j=1}^{m_k}\nabla l(\hat{\BFtheta}^*_{FL};\BFz_{k,j})\right\|^2
\]
To compare $\sigma^2$ and $\sigma^2_K$, by Jensen's inequality,
\[
\left\|\frac{K}{|\BFm|}\sum_{j=1}^{m_k} \nabla l(\hat{\BFtheta}^*_{FL};\BFz_{k,j})\right\|^2
\le \frac{K^2}{|\BFm|^2}\,m_k\sum_{j=1}^{m_k}\|\nabla l(\hat{\BFtheta}^*_{FL};\BFz_{k,j})\|^2.
\]
Hence,
\[
\sigma_K^2
\le \frac{K m_{\max}}{|\BFm|}\,\sigma^2\leq (1+\eta)\sigma^2.
\]

Lastly, define \myeql{\underline{T}=C_1^{2}{\bigl[\log\!\bigl(\tfrac{C_2}{\delta}\bigr)\bigr]^{-2}}{\Bigl(\sqrt{{64\nu\|\BFtheta_0-\BFtheta^*\|^2}/{\mu}
+\tfrac{12(1+\eta)\sigma^2}{\nu\mu\,\min\{C_{\mathrm{SV}}^{2},C^2_{MC}\}}}
+\tfrac{(1+\eta)L}{4\nu\,\min\{C_{\mathrm{SV}},C_{MC}\}}\Bigr)^{4}}\frac{|\mathbf m|^2}{K},\label{eq-barT}} where $C_{SV}$ and $C_{MC}$ are two constant FL parameter configuration in SV and MC. Fix the total epoch $T$ in both FL algorithms with $T\geq \underline{T}$.
\paragraph{\textbf{Analysis for convergence under SV mechanism.}}
By Assumption \ref{assump: asmp_Lsmooth_convex}(a), for some $\mu>0$ and all $\BFx,\BFy$,
\[
L(\BFy)\ge L(\BFx)+\nabla L(\BFx)^T(\BFy-\BFx)+\frac{\mu}{2}\|\BFy-\BFx\|^2.
\]

Hence, for $\hat{\BFtheta}^*_{FL}=\arg\min_{\BFtheta}L(\BFtheta)$,
\begin{equation}
\label{eq:proof_strong_convex}
L(\hat{\BFtheta}_{FL})-L(\hat{\BFtheta}^*_{FL})\le \epsilon
\ \Rightarrow\
\|\hat{\BFtheta}^*_{FL}-\hat{\BFtheta}_{FL}\|\le \left(\frac{2\epsilon}{\mu}\right)^{1/2}.
\end{equation}

We now consider a sequence of auxiliary training results $\hat{\BFtheta}^{a}_{FL}=\frac{1}{T}\sum^{T-1}_{t=0}\frac{1}{|\BFm|}\sum^{|\BFm|}_{k=1}\BFtheta^{t}_{k}$ (which is not actually computed in algorithm, but is useful for analysis). Let $\hat{\BFtheta}_{FL}=\frac{1}{T}\sum^{T-1}_{t=0}\BFtheta^t$, with $\BFtheta^t=\BFtheta^{t-1}$ if $t\notin\{0,H, 2H,\dots,\}$\footnote{Here we assume $T=\mathbb{Z}H$ for some positive integer $\mathbb{Z}$, and we use the average of synchronization as the grand output of FL algorithm, which, although differs from the original FL algorithm, is widely used in Federated optimization analysis \cite{stich2018local,khaled2019first}. }. \myeqmodel{\label{eq:bound_grad} \|\hat{\BFtheta}^{a}_{FL}-\hat{\BFtheta}_{FL}\|& =\|\frac{1}{T}\sum^{T}_{t=1}\left(\frac{1}{|\BFm|}\sum^{\BFm}_{k=1}\BFtheta^{t}_{k}-\BFtheta^t\right)\|\\ &\leq \frac{1}{T}\sum^{T}_{t=1}\|\frac{1}{|\BFm|}\sum^{|\BFm|}_{k=1}\BFtheta^{t}_{k}-\BFtheta^t\|\\ &= \frac{1}{T}\sum_{t'\in\{0,H,2H,\dots\}}\sum^{H-1}_{j=1}\|\frac{1}{|\BFm|}\sum^{\BFm}_{k=1}\BFtheta^{t'+j}_{k}-\BFtheta^{t'}\|\\ &= \frac{1}{T}\sum_{t'\in\{0,H,2H,\dots\}}\sum^{H-1}_{j=1}\|\frac{1}{|\BFm|}\sum^{\BFm}_{k=1}(\BFtheta^{t'+j-1}_{k}-\rho\nabla L_k(\BFtheta^{t'+j-1}_{k}))-\BFtheta^{t'}\|\\ &\leq \frac{1}{T}\frac{T}{H}\sum^{H-1}_{j=1}j\rho L\leq \frac{\rho H L }{2}. } In \eqref{eq:bound_grad}, the first equality is by the definitions of $\hat{\BFtheta}^{a}_{FL}$ and $\hat{\BFtheta}_{FL}$, expand the difference of their time averages. The first inequality is by applying the triangle/Jensen inequality. For the second equality, we partition time into synchronization blocks of length $H$ by writing $t=t'+j$ with $t'\in\{0,H,2H,\dots,T-1\}$ and $j\in\{1,\dots,H-1\}$, and use $\theta^t=\theta^{t'}$ for $t$ inside the block. For the third equality, we expand one local gradient step on each client: $\theta^{t'+j}_{k}=\theta^{t'+j-1}_{k}-\rho\,\nabla L_k(\theta^{t'+j-1}_{k})$, and substitute. For the second inequality, we bound within-block drift by cumulative step lengths and the uniform Lipschitz (gradient) bound $\|\nabla L_k(\cdot)\|=\|\nabla l(\BFtheta;\BFz)\|\le L$ under SV, and $\|\theta^{t'+j}_{k}-\theta^{t'}\|\le \sum_{\ell=1}^{j}\rho\|\nabla L_k(\theta^{t'+\ell-1}_{k})\|\le j\rho L$. Averaging over clients does not increase the bound. Summing $j=1$ to $H-1$ and averaging over the $\tfrac{T}{H}$ blocks yields $\frac{1}{T}\cdot\frac{T}{H}\sum_{j=1}^{H-1} j\rho L=\frac{\rho L}{2}(H-1)\le \frac{\rho L H}{2}$. And the final bound uses $H-1\le H$.

Following Theorem~1 of \cite{khaled2019first},
\myeql{\label{eq:convergence}
L(\hat{\BFtheta}^{a}_{FL})-L({\BFtheta}^{*}_{FL})
\le \frac{2\|\BFtheta_0-\BFtheta^*\|^2_2}{\rho T}+24\rho^2\sigma^2H^2\nu.
}

In order to achieve the best balanced leading-order dependence on both $T$ and $|\BFm|$ to $N_{sync}$ (for large $|\BFm|$ this is required and optimal for FL), fix constant $C_{\mathrm{SV}}\in(0,T^{1/4}|\BFm|^{-3/4})$ to ensure $H\geq1$ and choose
\[
H := \frac{1}{C_{\mathrm{SV}}}\,T^{1/4}|\BFm|^{-3/4},
\qquad
\rho := \frac{\sqrt{|\BFm|}}{4\nu\sqrt{T}}.
\]
Then the number of synchronizations is
\[
N^{SV}_{\mathrm{sync}}=C_{\mathrm{SV}}(T|\BFm|)^{3/4}.
\]
Substituting the above $(H,\rho)$ into \eqref{eq:convergence} yields
\begin{equation}
\label{eq:Lconvergence}
L(\hat{\BFtheta}^{a}_{FL})-L({\BFtheta}^{*}_{FL})
\le \left(8\nu\|\BFtheta_0-\BFtheta^*\|^2+\frac{3\sigma^2}{2\nu\,C_{\mathrm{SV}}^{2}}\right)\frac{1}{\sqrt{T|\BFm|}}.
\end{equation}

Combining \eqref{eq:proof_strong_convex} and \eqref{eq:Lconvergence}, we have
\[
\|\hat{\BFtheta}^*_{FL}-\hat{\BFtheta}^{a}_{FL}\|
\le
\left(\frac{16\nu\|\BFtheta_0-\BFtheta^*\|^2}{\mu}
+\frac{3\sigma^2}{\nu\mu\,C_{\mathrm{SV}}^{2}}\right)^{1/2}(T|\BFm|)^{-1/4}.
\]
Moreover, by \eqref{eq:bound_grad} and the chosen $(H,\rho)$,
\[
\|\hat{\BFtheta}^{a}_{FL}-\hat{\BFtheta}_{FL}\|
\le \frac{\rho H L}{2}
= \frac{L}{8\nu\,C_{\mathrm{SV}}}(T|\BFm|)^{-1/4}.
\]
Hence,
\[
\|\hat{\BFtheta}^*_{FL}-\hat{\BFtheta}_{FL}\|
\le
\left(
\left(\frac{16\nu\|\BFtheta_0-\BFtheta^*\|^2}{\mu}
+\frac{3\sigma^2}{\nu\mu\,C_{\mathrm{SV}}^{2}}\right)^{1/2}
+\frac{L}{8\nu\,C_{\mathrm{SV}}}
\right)(T|\BFm|)^{-1/4}.
\]

To guarantee
\(
P(\|\hat{\BFtheta}_{FL}-\BFtheta^*\|\ge \varepsilon_L(|\BFm|,\delta))\le \delta
\),
it suffices that
\(
P(\|\hat{\BFtheta}^*_{FL}-\BFtheta^*\|\ge \varepsilon_L(|\BFm|,\delta)/2)\le \delta
\)
and
\[
\|\hat{\BFtheta}_{FL}-\hat{\BFtheta}^*_{FL}\|
\le
\left(
\left(\frac{16\nu\|\BFtheta_0-\BFtheta^*\|^2}{\mu}
+\frac{3\sigma^2}{\nu\mu\,C_{\mathrm{SV}}^{2}}\right)^{1/2}
+\frac{L}{8\nu\,C_{\mathrm{SV}}}
\right)(T|\BFm|)^{-1/4}
\le \frac{\varepsilon_L(|\BFm|,\delta)}{2}.
\]

Equivalently, letting
\[
A_{\mathrm{SV}}
:=
\left(\frac{16\nu\|\BFtheta_0-\BFtheta^*\|^2}{\mu}
+\frac{3\sigma^2}{\nu\mu\,C_{\mathrm{SV}}^{2}}\right)^{1/2}
+\frac{L}{8\nu\,C_{\mathrm{SV}}},
\]
a sufficient condition is
\[
T|\BFm|\ \ge\ \left(\frac{2A_{\mathrm{SV}}}{\varepsilon_L(|\BFm|,\delta)}\right)^4.\]
Using $\varepsilon_L(|\BFm|,\delta)=\frac{1}{L_\pi}\epsilon(|\BFm|,\delta)$ and
$\epsilon(|\BFm|,\delta)= L_\pi \sqrt{\frac{1}{C_1|\BFm|}\log\left(\frac{C_2}{\delta}\right)}$, it is straightforward that $T$ satisfies the sufficient condition for FL convergence under SV for all $T\geq\underline{T}$.

\paragraph{\textbf{Analysis for convergence under MC mechanism.}}
Under MC, the local objective is
$
L_k(\BFtheta)=\frac{K}{|\BFm|}\sum_{j=1}^{m_k}l(\BFtheta;\BFz_{k,j}).
$
By Assumption~\ref{assump: asmp_Lsmooth_convex}(b), for all $\BFtheta\in\Theta$,
\[
\|\nabla L_k(\BFtheta)\|
=
\left\|\frac{K}{|\BFm|}\sum_{j=1}^{m_k}\nabla l(\BFtheta;\BFz_{k,j})\right\|
\le \frac{K}{|\BFm|}m_k\,L
\le (1+\eta)L.
\]
Hence the drift bound \eqref{eq:bound_grad} continues to hold with $L$ replaced by $(1+\eta)L$ (With some abuse of notation, here define $\hat{\BFtheta}^{a}_{FL}=\frac{1}{T}\sum^{T-1}_{t=0}\frac{1}{K}\sum^{K}_{k=1}\BFtheta^{t}_{k}$), i.e.,
\begin{equation}\label{eq:bound_grad_MC}
\|\hat{\BFtheta}^{a}_{FL}-\hat{\BFtheta}_{FL}\|
\le \frac{\rho H (1+\eta)L}{2}.
\end{equation}

Define the MC local-variation term at $\hat{\BFtheta}^*_{FL}$ as
\[
\sigma_K^2
:=\frac{1}{K}\sum_{k=1}^K \|\nabla L_k(\hat{\BFtheta}^*_{FL})\|^2
=\frac{1}{K}\sum_{k=1}^K\left\|\frac{K}{|\BFm|}\sum_{j=1}^{m_k} \nabla l(\hat{\BFtheta}^*_{FL}; \BFz_{k,j})\right\|^2.
\]
Then Theorem~1 of \cite{khaled2019first} gives the analogue of \eqref{eq:convergence}:
\begin{equation}\label{eq:convergence_MC}
L(\hat{\BFtheta}^{a}_{FL})-L({\BFtheta}^{*}_{FL})
\leq \frac{2\|\BFtheta_0-\BFtheta^*\|^2_2}{\rho T}+24\rho^2\sigma_K^2H^2\nu.
\end{equation}

Fix a constant $C_{\mathrm{MC}}\in(0,T^{1/4}K^{-3/4})$ to ensure $H\geq1$, and choose
\[
H := \frac{1}{C_{\mathrm{MC}}}\,T^{1/4}K^{-3/4},
\qquad
\rho := \frac{\sqrt{K}}{4\nu\sqrt{T}}.
\]
Then the synchronization count is
\[
N^{MC}_{\mathrm{sync}}:=\frac{T}{H}=C_{\mathrm{MC}}(TK)^{3/4}.
\]
Substituting the above $(H,\rho)$ into \eqref{eq:convergence_MC} yields the explicit constant dependence
\begin{equation}\label{eq:Lconvergence_MC}
L(\hat{\BFtheta}^{a}_{FL})-L({\BFtheta}^{*}_{FL})
\leq \left(8\nu\|\BFtheta_0-\BFtheta^*\|^2+\frac{3\sigma_K^2}{2\nu\,C_{\mathrm{MC}}^{2}}\right)\frac{1}{\sqrt{TK}}.
\end{equation}

Combining \eqref{eq:proof_strong_convex} with \eqref{eq:Lconvergence_MC} gives
\[
\|\hat{\BFtheta}^*_{FL}-\hat{\BFtheta}^{a}_{FL}\|
\le
\left(\frac{16\nu\|\BFtheta_0-\BFtheta^*\|^2}{\mu}
+\frac{3\sigma_K^2}{\nu\mu\,C_{\mathrm{MC}}^{2}}\right)^{1/2}(TK)^{-1/4}.
\]
Moreover, by \eqref{eq:bound_grad_MC} and the chosen $(H,\rho)$,
\[
\|\hat{\BFtheta}^{a}_{FL}-\hat{\BFtheta}_{FL}\|
\le \frac{\rho H (1+\eta)L}{2}
= \frac{(1+\eta)L}{8\nu\,C_{\mathrm{MC}}}(TK)^{-1/4}.
\]
Hence,
\[
\|\hat{\BFtheta}^*_{FL}-\hat{\BFtheta}_{FL}\|
\le
\left(
\left(\frac{16\nu\|\BFtheta_0-\BFtheta^*\|^2}{\mu}
+\frac{3\sigma_K^2}{\nu\mu\,C_{\mathrm{MC}}^{2}}\right)^{1/2}
+\frac{(1+\eta)L}{8\nu\,C_{\mathrm{MC}}}
\right)(TK)^{-1/4}.
\]
Therefore, a sufficient condition for
$P(\|\hat{\BFtheta}_{FL}-\BFtheta^*\|\geq \varepsilon_L(|\BFm|,\delta))\leq \delta$
is
\[
\left(
\left(\frac{16\nu\|\BFtheta_0-\BFtheta^*\|^2}{\mu}
+\frac{3\sigma_K^2}{\nu\mu\,C_{\mathrm{MC}}^{2}}\right)^{1/2}
+\frac{(1+\eta)L}{8\nu\,C_{\mathrm{MC}}}
\right)(TK)^{-1/4}
\le \frac{\varepsilon_L(|\BFm|,\delta)}{2}.
\]
Equivalently, letting
\[
A_{\mathrm{MC}}
:=
\left(\frac{16\nu\|\BFtheta_0-\BFtheta^*\|^2}{\mu}
+\frac{3\sigma_K^2}{\nu\mu\,C_{\mathrm{MC}}^{2}}\right)^{1/2}
+\frac{(1+\eta)L}{8\nu\,C_{\mathrm{MC}}}.
\]
a sufficient condition is       
\[TK \ge \left(\frac{2A_{\mathrm{MC}}}{\varepsilon_L(|\BFm|,\delta)}\right)^4.\]

Using $\varepsilon_L(|\BFm|,\delta)=\frac{1}{L_\pi}\epsilon(|\BFm|,\delta)$ and
$\epsilon(|\BFm|,\delta)=L_\pi\sqrt{\frac{1}{C_1|\BFm|}\log\left(\frac{C_2}{\delta}\right)}$, it is also straightforward that $T$ satisfies the sufficient condition for convergence of FL under MC for all $T\geq\underline{T}$. \Halmos
\end{proof}
 
\section{Proof for Corollary \ref{cor:system_analysis_MS}}

\begin{proof}{Proof for Corollary \ref{cor:system_analysis_MS}}For the first claim, note that
\[
\Pi^{sys}(MC)-\Pi^{sys}(SV)
\ge
\sum_{k=1}^{K}\bigl(v(|\BFm|)-v(|\BFm|-m_k)\bigr)-v(|\BFm|)
\;+\;
c\Bigl(C_{SV}(T|\BFm|)^{3/4}-C_{MC}(TK)^{3/4}\Bigr).
\]
Since $\sum_{k=1}^{K}(v(|\BFm|)-v(|\BFm|-m_k))=K\,v(|\BFm|)-\sum_{k=1}^{K}v(|\BFm|-m_k)$, the condition
\[
c>\frac{\sum_{k=1}^{K}v(|\BFm|-m_k)-(K-1)v(|\BFm|)}
{T^{3/4}\bigl(C_{SV}|\BFm|^{3/4}-C_{MC}K^{3/4}\bigr)}
\]
implies $\Pi^{sys}(MC)-\Pi^{sys}(SV)>0$, i.e., \eqref{ineq:sys_eff}.

For the second claim, fix any $c>0$ and define $v(x)=x^{1-t}$ with some $t\in(0,1)$, so $v$ is concave on $\mathbb{R}_+$.
By concavity (first-order upper bound), for each $k$,
\[
(|\BFm|-m_k)^{1-t}
\le
|\BFm|^{1-t}-(1-t)|\BFm|^{-t}m_k.
\]
Summing over $k$ and using $\sum_{k=1}^{K}m_k=|\BFm|$ yields
\[
\sum_{k=1}^{K}(|\BFm|-m_k)^{1-t}
\le
K|\BFm|^{1-t}-(1-t)|\BFm|^{1-t}
=(K-1+t)|\BFm|^{1-t}.
\]
Therefore,
\begin{align*}
\Pi^{sys}(MC)-\Pi^{sys}(SV)
&\ge
cT^{3/4}\bigl(C_{SV}|\BFm|^{3/4}-C_{MC}K^{3/4}\bigr)
-\sum_{k=1}^{K}(|\BFm|-m_k)^{1-t}
+(K-1)|\BFm|^{1-t}\\
&\ge
cT^{3/4}\bigl(C_{SV}|\BFm|^{3/4}-C_{MC}K^{3/4}\bigr)
-t|\BFm|^{1-t}.
\end{align*}
Since $C_{SV}|\BFm|^{3/4}>C_{MC}K^{3/4}$, the first term is strictly positive. Choose $t\in(0,1)$ sufficiently small so that
\[
t|\BFm|^{1-t}
<
cT^{3/4}\bigl(C_{SV}|\BFm|^{3/4}-C_{MC}K^{3/4}\bigr),
\]
which implies $\Pi^{sys}(MC)-\Pi^{sys}(SV)>0$. The lower bound above does not depend on the particular allocation
$(m_1,\dots,m_K)$ beyond $\sum_k m_k=|\BFm|$, hence \eqref{ineq:sys_eff} holds for all such profiles. \Halmos
\end{proof}
\section{Proof of Proposition \ref{prop_opt_participation}}

  \begin{proof}{Proof of Proposition \ref{prop_opt_participation}}
  We prove each part in turn. Without loss of generality, label agents so that $m_1 \geq m_2 \geq \cdots \geq m_K$.

  \medskip
  \noindent\textbf{Step 1: Full data provision is optimal for any participating agent.}

  Fix any set of participating agents $\mathcal{A}$. If agent $k \in \mathcal{A}$ contributes $\hat{m}_k < m_k$, then the profile with
  $\tau_k = m_k$ (and all other unchanged) produces strictly higher surplus: $v(\cdot)$ strictly increases while $|\mathcal{A}|$ is unchanged,
  so $N_{\text{sync}}$ is unchanged. Hence, any optimal profile has $\tau_k = m_k$ for all $k \in \mathcal{A}$.

  \medskip
  \noindent\textbf{Step 2: Optimal selection of participants is a threshold policy (Part (i)).}

  By Step 1, we may restrict attention to profiles where each agent participates with either full data or abstains. For any subset
  $\mathcal{A} \subseteq \{1, \dots, K\}$, the social surplus is
  \[
  \Pi^{\text{social}}(\mathcal{A}) = v\!\left(\sum_{k \in \mathcal{A}} m_k\right) - C|\mathcal{A}|^{3/4}.
  \]

  \noindent\emph{Claim: For any fixed coalition size $\hat{K}$, the optimal coalition of size $\hat{K}$ is $\{1, \dots, \hat{K}\}$.}

  Since $N_{\text{sync}}$ depends only on $|\mathcal{A}| = \hat{K}$, and $v(\cdot)$ is strictly increasing, the surplus is maximized by
  maximizing $\sum_{k \in \mathcal{A}} m_k$, which is achieved by selecting the $\hat{K}$ agents with the largest $m_k$, i.e., agents $\{1,
  \dots, \hat{K}\}$.

  \noindent\emph{Claim: The optimal coalition size $\bar{k}$ induces a threshold policy.}

  Define
  \[
  \bar{k} := \argmax_{0 \leq j \leq K} \; \Pi^{\text{social}}(\{1, \dots, j\}),
  \]
  where $\Pi^{\text{social}}(\emptyset) = 0$. According to the previous claim, the overall optimal coalition is $\mathcal{A}^* = \{1, \dots,
  \bar{k}\}$, and the threshold is $m^* = m_{\bar{k}}$. All agents with $m_k \geq m^*$ participate; all with $m_k < m^*$ remain out.

  \medskip
  \noindent\textbf{Step 3: MC with $v^S$ induces the optimal profile (Part (ii)).}

  Under the marginal contribution mechanism with the characteristic function $v^S(\BFtau_{\mathcal{A}}) = v(|\BFtau_{\mathcal{A}}|) -
  C|\mathcal{A}|^{3/4}$, the payoff to agent $k$ in coalition $\mathcal{A}$ is
  \[
  \psi^{MC}_k(v^S) = v^S(\BFtau_{\mathcal{A}}) - v^S(\BFtau_{\mathcal{A} \setminus \{k\}}) = \bigl[v(|\BFtau_{\mathcal{A}}|) -
  C|\mathcal{A}|^{3/4}\bigr] - \bigl[v(|\BFtau_{\mathcal{A}}| - m_k) - C(|\mathcal{A}|-1)^{3/4}\bigr].
  \]

  By Theorem~\ref{thm: main-MC}, the MC mechanism is robust to data splitting when $v^S$ is concave in $|\BFtau_{\mathcal{A}}|$ (which we
  verify below), so each agent participates with full data or stays out of the FL system.

 The agent $k$ participates if and only if $\psi^{MC}_k(v^S) \geq 0$, i.e.,
  \[
  v(|\BFtau_{\mathcal{A}}|) - v(|\BFtau_{\mathcal{A}}| - m_k) \geq C\bigl[|\mathcal{A}|^{3/4} - (|\mathcal{A}|-1)^{3/4}\bigr].
  \]
  Consider the optimal coalition $\mathcal{A}^* = \{1, \dots, \bar{k}\}$. For agent $k \leq \bar{k}$: since $v$ is concave and $m_k \geq
  m_{\bar{k}}$, the left-hand side is at least as large as agent $\bar{k}$'s marginal learning contribution. Meanwhile, the right-hand side
  is the marginal synchronization cost of adding one agent to a coalition. The optimality of $\bar{k}$ (from Step 2) ensures that each agent
   $k \leq \bar{k}$ has non-negative MC payoff in the optimal coalition.

  For agent $k > \bar{k}$: adding agent $k$ to $\mathcal{A}^*$ would decrease $\Pi^{\text{social}}$ (by definition of $\bar{k}$), so
  \[
  v\!\left(\sum_{i=1}^{\bar{k}} m_i + m_k\right) - v\!\left(\sum_{i=1}^{\bar{k}} m_i\right) < C\bigl[(\bar{k}+1)^{3/4} -
  \bar{k}^{3/4}\bigr].
  \]
  Hence $\psi^{MC}_k(v^S) < 0$, and agent $k$ voluntarily abstains.

  Therefore, the MC equilibrium under $v^S$ is exactly $\mathcal{A}^* = \{1, \dots, \bar{k}\}$ with full data provision, matching the social
   optimum.

  \medskip
  \noindent\textbf{Step 4: SV with $v^S$ leads to data splitting (Part (iii)).}

  By Theorem~\ref{thm: main-SV}, the MCFL Shapley value induces data splitting whenever the characteristic function is strictly concave. It
  remains to verify that $v^S(\BFtau_{\mathcal{A}}) = v(|\BFtau_{\mathcal{A}}|) - C|\mathcal{A}|^{3/4}$ is strictly concave as a function
  of $|\BFtau_{\mathcal{A}}|$ when agents can split data.

 In data splitting, an agent with $m_k$ samples who creates $m_k$ fake single-sample identities increases $|\mathcal{A}|$ while keeping
  $|\BFtau_{\mathcal{A}}|$ fixed. The SV splitting incentive from Theorem~\ref{thm: main-SV} applies to the $v(\cdot)$ component: since $v$
  is strictly concave, each agent strictly benefits from splitting under SV. Moreover, the splitting also inflates $|\mathcal{A}|$, which
  increases the cost term $C|\mathcal{A}|^{3/4}$, further reducing social surplus. Hence, SV under $v^S$ leads to data splitting and a
  socially inefficient outcome. 
  \Halmos
  \end{proof}

\newpage
\section{Implementation Details for Federated Learning with Neural Networks}
\label{app:flwr-parameter}

We build a simple federated image classifier with Flower (TensorFlow/Keras). The dataset is partitioned by clinical center; each center is split into several client identities. Every client trains the same small CNN locally, and the server aggregates model weights with standard FedAvg. We report accuracy on each client’s held-out data. Images are loaded once, shuffled per center, and then divided into fixed sub-partitions to form clients. Each client receives one sub-partition and internally splits it into train/test.

Each RGB image is resized to \(224\times224\) and scaled to \([0,1]\) (float32). Batches are materialized as NumPy arrays for Keras. On each client, RGB skin images are preprocessed by resizing to \(224\times224\) and normalizing pixel values to \([0,1]\). The local model is a lightweight CNN: two conv–ReLU–max-pool blocks, followed by flattening, dropout, and a softmax layer producing probabilities over eight lesion classes. Each client trains with the local optimizer as Adam \citep{kingma2014adam}, while the server coordinates rounds with standard federated averaging (FedAvg , \citep{mcmahan2017communication}).

When an agent $k$ with $n_k$ samples splits into $s$ identities, the local dataset is evenly partitioned into $n_k/s$ sub-samples. Each sub-sample is divided into a $80\%\!/\!20\%$ train/test split. Clients train with Adam as the optimizer and the sparse categorical crossentropy loss as the loss function. After local training, the server performs federated averaging (FedAvg) over client weights.

We summarize the key parameters in the following table :  
\medskip
\begin{center}
\begin{tabular}{ll}
\toprule
\textbf{Parameter} & \textbf{Default Setting} \\
\midrule
Number of actual centers & \(6\) \\
Split per Center & \texttt{SPLIT\_PER\_CENTER}\(\in\{1,2,10\}\) \\
Total clients & \(6 \times \texttt{SPLIT\_PER\_CENTER}\) \\
Shuffle seed & \(42\) \\
Train/Test split & \(80/20\) (per client) \\
Image size & \((224,224)\) \\
Normalization & divide pixels by \(255.0\) \\
Conv filters & \(32,\,64\) \\
Kernel / Pool size & \(3\times3\) conv, \(2\times2\) max-pool \\
Dropout & \(0.5\) \\
Classes & \(8\) (softmax) \\
Optimizer & Adam (Keras default) \\
Loss Function & sparse categorical cross-entropy \\
\bottomrule
\end{tabular}
\end{center}

\newpage
\section{Other Mechanisms and Resulting Equilibrium}
\label{sec: extensions_mechanism}

In this section, we discuss other commonly adopted mechanisms, including equal share (ES), proportional share (PS). We will first present the mechanism division rule, then show what axioms those mechanisms violate, and finally present the resulting participation equilibrium under those mechanisms.

\subsection{Equal Share}
\begin{definition}[Equal Share in MCFL] Suppose there are $K$ agents participating in MCFL with participation profile $\BFtau$ with the characteristic function $v(|\BFtau|)$. Then the payoff allocated to agent $k$ is defined as
\begin{equation*}
\psi^{ES}_k(v) = \frac{v(|\BFtau|)}{K}
\end{equation*}
\end{definition}

Equal share mechanism violates the null player axiom. Consider a null player where $v([\tau_1,\dots,\tau_i,\dots,\tau_K])=v([\tau_1,\dots,0,\dots,\tau_K])$, under equal share mechanism, that player receives $\psi^{ES}_k(v) = \frac{v(|\BFtau|)}{K}>0$. Having the mechanism satisfies null player axiom is important to the platform, as there would not be free-riding problems for the players who add no value to learning.

As $v(|\BFtau|)$ is an increasing function, all agents are willing to fully participate. We further show that under equal share mechanism, the resulting equilibrium is to participate with full splitting on data. 

\begin{proposition}
\label{thm: ES}
For any $K, T, T' $, and any maximum data vector $\bmvec$, under equal share mechanism,
$\psi^{T+T'}<\psi^{T}_{T,T'}+\psi^{T'}_{T,T'}$.

In other words,  an agent tends to create a duplicated identity and split her original data set for higher benefit allocation. 
\end{proposition}

\begin{proof}{Proof of proposition\ref{thm: ES}} The proof is direct. Let $m=|\BFm_{[2:K]}|$ be the number of data samples provided by other participants excluding the first agent. $\psi^{T+T'}=\frac{v(m+T+T')}{K}<\frac{v(m+T+T')}{K+1}+\frac{v(m+T+T')}{K+1}$ for $K>1$.\Halmos
\end{proof}

\subsection{Proportional Share}

\begin{definition}[Proportional Share in MCFL] Suppose there are $K$ agents participating MCFL with participation profile $\BFtau$ with the characteristic function $v(|\BFtau|)$. Then the payoff allocated to agent $k$ is defined as
\begin{equation*}
\psi^{PS}_{\tau_k,k}(v) = \frac{\tau_k}{|\BFtau|}v(|\BFtau|),
\end{equation*}
\end{definition}

Proportional share mechanism also violates the null player axiom. Similarly, consider a null player where $v([\tau_1,\dots,\tau_i,\dots,\tau_K])=v([\tau_1,\dots,0,\dots,\tau_K])$, under proportional share mechanism, that player receives $\psi^{ES}_{\tau_k,k}(v) = \frac{\tau_k}{K}v(|\BFtau|)>0$.  

Similarly, under proportional share, all agents are still willing to fully participate. We further show that under proportional share mechanism, the agent is indifferent between splitting or not splitting.

\begin{proposition}
\label{thm: PS}
For any $K, T, T' $, and any maximum data vector $\bmvec$, under proportional share mechanism,
$\psi^{T+T'}=\psi^{T}_{T,T'}+\psi^{T'}_{T,T'}$.
In other words,  an agent is indifferent between splitting her original data set, or participating with her original data set. 
\end{proposition}

\begin{proof}{Proof of Proposition \ref{thm: PS}}Similarly, let $m=|\BFm_{[2:K]}|$ be the number of data samples provided by other participants excluding the first agent. A direct proof of $\psi^{T+T'}=\frac{T+T'}{|\BFtau|}v(m+T+T')=\frac{T}{|\BFtau|}v(m+T+T')+\frac{T'}{|\BFtau|}v(m+T+T')$ gives this result. \Halmos

\end{proof}

\end{APPENDICES}

\end{document}